\RequirePackage{lineno} 
\documentclass[12pt]{article}

\usepackage[svgnames]{xcolor}
\usepackage{pgf}
\usepackage{tikz}
\usepackage{pgfplots}

\usepackage{caption}
\usepackage{subcaption}
\usepackage{amsmath,enumerate}
\usepackage{amssymb}
\usepackage{soul}
\usepackage{fancyhdr}
\usepackage{graphicx}
\usepackage{eso-pic}
\usepackage{setspace}
\usepackage{natbib}
\usepackage{authblk}
\usepackage{geometry}

\usepackage[colorlinks,linkcolor=Blue,citecolor=DarkGreen,urlcolor=Blue]{hyperref}
\usepackage{cleveref}
\crefname{assumption}{assumption}{assumptions}
\crefname{lemma}{lemma}{lemmas}
\crefname{example}{example}{examples}
\crefname{algorithm}{algorithm}{algorithms}
\crefname{equation}{equation}{equations}
\crefname{conjecture}{conjecture}{conjectures}
\crefname{figure}{figure}{figures}
\crefname{eqnarray}{equation}{equations}

\newcommand{\convp}{\stackrel{p}{\rightarrow}}

\newtheorem{theorem}{Theorem}

\newtheorem{assumption}{Assumption}
\newtheorem{proposition}{Proposition}

\newtheorem{lemma}{Lemma}
\newtheorem{example}{Example}
\newtheorem{remark}{Remark}
\newtheorem{algo}{Algorithm}

\numberwithin{algo}{section}

\numberwithin{equation}{section}
\numberwithin{lemma}{section}
\numberwithin{assumption}{section}
\numberwithin{definition}{section}

\usepackage{threeparttablex}
\usepackage{longtable}

\usepackage{comment}

\usepackage{fancyhdr}
\newenvironment{proof}[1][Proof]{\noindent\textbf{#1.} }{\ \rule{0.5em}{0.5em}}
\usepackage[color=Orange]{todonotes}

\newcommand{\qed}{\textcolor{red}{\text{$\pmb{\Box}$}}}

\newcommand{\argmin}{\mathop{\mathrm{argmin}}}
\newcommand{\argmax}{\mathop{\mathrm{argmax}}}

\usepackage{pgf}
\usepackage{tikz}

\begin{document}

\title{ A Sharp Test for the Judge Leniency Design\footnote{We have greatly benefited from insightful comments from Bocar Ba. We also thank William Arbour, Kory Kroft, and Brigham Frandsen for their valuable discussions. 
Mourifi\'e and Wan thank the SSHRC Insight Grant \#43520190500. Wan thanks the SSHRC Insight Grant \#43520240428. Hsu gratefully acknowledges the research support from the National Science and
Technology Council of Taiwan (NSTC112-2628-H-001-001), the Academia Sinica Investigator Award of
the Academia Sinica, Taiwan (AS-IA-110-H01), and from the Center for Research in Econometric Theory and Applications of National Taiwan University (Grant No.\ 113L8601).}}

\author{\vspace{-0.0cm} Mohamed Coulibaly\footnote{Department of Applied Economics, HEC Montr\'eal. Email: mohamed.coulibaly@hec.ca.}\quad Yu-Chin Hsu\footnote{Institute of Economics, Academia Sinica; Department of Finance, National Central University; Department of Economics, National Chengchi University; CRETA, National Taiwan University. E-mail: ychsu@econ.sinica.edu.tw.}\quad Ismael Mourifi\'e\footnote{Corresponding author. Department of Economics, Washington University in St. Louis, One Brookings Drive
St. Louis, MO 63130-4899, USA. E-mail: ismaelm@wustl.edu.}\quad Yuanyuan Wan\footnote{Department of Economics, University of Toronto. E-mail: yuanyuan.wan@utoronto.ca.}}

\date{\vspace{-0.0cm}  \small \text{August 4, 2025}}
\maketitle
\vspace{-0.5cm}

\begin{abstract}
\noindent
We propose sharp testable implications and tests to jointly assess the random assignment, exclusion, and monotonicity assumptions in judge leniency designs. Our procedures accommodate various data scenarios in which the number of defendants handled by a judge may be either small or large, and allow for discrete or continuous instrumental variables. When the validity of the design is rejected, a variant of the marginal treatment effect can be identified under weaker assumptions. We apply our test to the Philadelphia court data studied by \cite{stevenson2018distortion} and demonstrate that it outperforms non-sharp joint tests by significant margins in simulation studies. 
\end{abstract}

\begin{spacing}{1.35}
\section{Introduction}
We propose a novel sharp test to assess the validity of the judge leniency design, which has emerged as a prominent instrumental variable (IV) approach in recent years, particularly in empirical research exploring causal effects within the criminal justice system. This design has proven beneficial in investigating the impacts of various interactions with the legal system, such as pretrial detentions and incarcerations, on subsequent outcomes, including recidivism rates, conviction probabilities, and employment prospects. What sets the judge leniency design apart is its distinctive feature of randomly assigning judges to different cases, with each judge handling a significant number of cases while having discretion over the final decision. The random assignment of judges enhances the credibility of this IV approach and has led to its increasing popularity among researchers \citep[][]{kling2006incarceration,di2013criminal,aizer2015juvenile,mueller2015criminal}.\footnote{\cite{kling2006incarceration} exploits randomized judge assignment along with judge propensities to instrument for incarceration length, aiming to investigate the causal impact of incarceration on labor market outcomes.} Importantly, the judge leniency design's random assignment feature extends beyond the context of criminal justice, making it a valuable methodology in diverse research contexts, including medicine, patents and startups, bankruptcy protection, evictions, and access to foster care \citep[see][]{doyle2015measuring,farre2020patent,dobbie2017consumer,gross2022temporary}.\footnote{For example, \cite{doyle2015measuring} employs the judge leniency design in the medical context to examine the impact of ambulance companies on patients in emergencies, relying on the pseudo-random assignment of ambulance companies to patients. Similarly, \cite{dobbie2017consumer} uses the leniency of randomly assigned bankruptcy judges as an instrument to study the implications of Chapter 13 bankruptcy protection on future financial events.}

However, in addition to the random assignment, an instrumental variable must adhere to two additional crucial conditions: (i) an exclusion restriction, which means that judges' actions should only influence the treatment and should not have any direct influence on the defendant's future outcomes; and (ii) a monotonicity restriction, which means that judges should consistently exhibit more or less leniency. This means that if a defendant was treated (detained) by one judge, she would always be treated (detained) by a less lenient judge. Trial decisions (treatment) are often multidimensional, including incarceration, fines, community service, sentence length, and others \citep[][]{johnson2014judges}. These decisions impact future outcomes. Because different judges may have varying attitudes on these decisions, the exclusion restriction can be violated if some of the decisions are unobserved or uncontrolled. Furthermore, \cite{abrams2012judges} and \cite{stevenson2018distortion} argue there is considerable heterogeneity in how judges rank defendants when considering various types of offenses. If this heterogeneity is not observed, then it is possible that judges exhibit varying levels of leniency under different circumstances, and the monotonicity assumption would be violated. These observations align with \cite{mogstad2019identification}, who demonstrate that, in general, monotonicity effectively requires homogeneous choice behavior for economic agents when there are multiple instruments. Therefore, offering a statistical test to evaluate the validity of the judge leniency design becomes a highly relevant empirical question.

In this paper, we characterize the \textit{sharp testable implications} of the judge leniency design as a set of inequality restrictions on the distribution of the observed data. Our result is novel and contributes to the testable implications derived in the seminal work of \cite{heckman2005structural} in two ways. First, our implications belong to a tractable subset of the constraints of \cite{heckman2005structural} and are easier to implement in practice. Second, we establish the sharpness of our testable implication, that is, they possess the unique quality of exploiting all available information within the data distribution that is useful to refute the validity of the judge leniency design. 

Numerous efforts have been made to test the judge leniency design in the existing literature. A common approach involves providing separate evidence for the validity of the individual assumptions made in the judge leniency design. For instance, to assess the random assignment of judges, \cite{dobbie2018intergenerational} examines whether a measure of judge stringency (the instrumental variable) correlates with baseline cases and family characteristics of criminal defendants. Regarding the monotonicity assumption, they test an implication that requires the first-stage estimates to be non-negative for all subsamples. \cite{bhuller2018incarceration} and \cite{norris2021effects} employ similar individual testing approaches. Assessing the assumptions individually is effective in empirical scenarios where researchers know which assumption to test and have prior knowledge that other assumptions hold. Our approach adds to the existing body of knowledge by introducing a test that does not depend on prior information. In fact, the three key assumptions may collectively impose certain constraints on the observable data-generating process (DGP), which could not be detected by examining only the testable implications of each assumption in isolation.

Unlike individually testing each assumption, \cite{frandsen2023judging} proposes a joint test for all assumptions underlying the judge leniency design. Their test leverages the property that, in the judge leniency design, the average outcome at the judge level should exhibit a smooth relationship with the propensity score (or the judge-level treatment probability). It ought to have a bounded slope, where the bounds depend on the limits of the outcome variable’s support. Although \cite{frandsen2023judging}'s testable implication has the desirable property that it assesses all the assumptions simultaneously, we show there is still significant relevant information in the data distribution essential for evaluating the judge leniency design's validity, but not used in \cite{frandsen2023judging}'s testable implication. This difference is also demonstrated by numerical examples and empirical studies reported in \Cref{Section: MTESharp-Test,section: simulation}.

To the best of our knowledge, our test is the only sharp test available for assessing the validity of the judge leniency design. In other words, our testable implications exhaust all the information in the observed data distribution. As seen in previous methods, non-sharp tests have practical virtue when there is no easily tractable characterization of the sharp testable implications of a model's assumptions. If a non-sharp rejects, it conveys an informative result that the assumptions should be rejected. However, there are also important trade-offs to consider. First, a non-sharp test can have no power against certain violations since it does not consider all possible constraints on the data distribution. Second, different non-sharp tests can lead to discordant empirical results and potentially misleading interpretations of the estimand of interest \citep[see][]{li2024discordant}. For instance, two different non-sharp tests may produce conflicting results because they consider different aspects of the observed data distribution. Our sharp test addresses both issues as it is a consistent test built upon sharp testable implications and, therefore, a useful complement to the existing literature. 

We construct valid and consistent semi-nonparametric and semiparametric tests based on these tractable testable implications. Our asymptotic tests support a diverse range of data structures. For example, we can apply our tests to the empirical context in which each judge handles a large number of defendants, and the number of judges can be either large or small (as in our empirical application). As we will further elaborate in \Cref{section: testing procedure}, our asymptotic tests are also applicable when the number of defendants for each judge is small, as long as the data regime permits a root-n estimation of the propensity score. 

We also provide an easy-to-compute finite sample test for cases involving a small number of judges and a small number of defendants per judge. To the best of our knowledge, ours and the finite sample test of \cite{frandsen2023judging} are the only finite sample specification tests in the judge leniency design literature. Like \cite{frandsen2023judging}, our finite sample test also focuses on binary outcomes. Unlike \cite{frandsen2023judging}'s test, which ensures the finite sample validity by computing a ``least favorable p-value" via a high-dimensional nonlinear optimization routine, we use Bonferroni correction. The computation for our test is very light and requires little more than simulating Bernoulli random variables. Therefore, it serves as a useful addition to the existing finite sample tests.

As a potential alternative to the existing non-sharp tests, one may consider testing the validity of the judge leniency design employing some of the existing sharp tests developed for the Local Average Treatment Effect (LATE) framework, i.e., \cite{kitagawa2015}, \cite{huber2015testing}, and \cite{MW2017}. However, it is worth noting that these tests may over-reject since they are based on a priori direction in the monotonicity assumption and are not directly applicable in the context of judge leniency design. For instance, in the judge leniency design, the number of judges can be quite large, and in some cases, it might even be infinite, especially when judges' types are continuous. In such scenarios, the number of potential directions to consider becomes large, possibly infinite. Imposing a specific ex-ante direction in the judge leniency design is therefore overly restrictive, and considering all possible directions might be impractical or impossible. Furthermore, imposing an incorrect a priori direction bears an additional risk of model misspecification. These issues highlight the need for a more flexible testing approach, like the one proposed in this paper, which is free from making overly restrictive assumptions on the direction of monotonicity.

While our test is primarily motivated by testing judge leniency designs, it can also be applied to assess the identifying assumptions in a general Marginal Treatment Effect framework with continuous or discrete instrument variables, which has been applied to various empirical settings. See \cite{carneiro2011estimating,kowalski2016doing, Brinch2017}, among many others. In the context of judge leniency designs, this also means that our test does not require observing a judge's identity and accommodates continuous judge types. Finally, motivated by \cite{mogstad2019identification}, we propose to relax monotonicity and exclusion assumptions to partial monotonicity and partial exclusion, respectively, when our test rejects the null hypothesis. 

We organize the rest of the paper as follows.  \Cref{Section: MTESharp-Test} presents the analytical framework and the sharp testable implications of the judge leniency design. \Cref{section: testing procedure} presents the testing procedures. In \Cref{section: simulation}, we show the results of the simulations and discuss our empirical illustration. In \Cref{Section: After the test}, we explore approaches to salvage the judge leniency design when its sharp testable implications are violated. The last section concludes the paper, and the proofs are collected in the online supplementary materials.

\section{Model and Sharp Testable Implications}\label{Section: MTESharp-Test}
 
We adopt the potential outcomes framework. Let the observed treatment indicator be $D\in \{0,1\}$. For example, in the judge leniency design, the unit of observation is defendants. Hence, $D=1$ indicates that a defendant is incarcerated. Let $Z\in \mathcal Z \subseteq \mathbb R^{d_z}$ be the type of the judge assigned to the defendant. $Y_{d}(z)\in\mathcal Y \subseteq \mathbb R$ denotes the potential outcome of interest (e.g., recidivism) when the treatment and the judge's type are externally set to $D=d$, and $Z=z$, respectively. Similarly, $D_z$ denotes the potential treatment when the judge's type is externally set to $Z=z$. Let $Y = Y_{1}(Z)D + Y_{0}(Z)(1-D)$ be the observed outcome. For the moment, we omit observed defendant and case covariates $X$ (such as time and courtroom of the trial) for ease of notation. The identification analysis in this section can be extended by conditioning on $X$. We will also discuss the implementation of our test in the presence of $X$ in \Cref{section: add covariates}.

In our setting, $Z$ can be multidimensional, continuous, discrete, or a combination of both. For example, if there is a group of judges $\mathcal J$, and if their identities are observed, then $Z \in \mathcal J$ can be chosen as the identity of the judge assigned to the defendant. This is the instrumental variable that \citet[][FLL hereafter]{frandsen2023judging} consider. On the other hand, we allow scenarios in which the judge's identity is unobserved but with observed characteristics. In this case, $Z$ may contain a set of continuous or discrete variables, such as the judge's experience, gender, and race. 

The literature mainly relies on the following assumptions to evaluate the causal effects of treatment $D$ on outcome $Y$.
\begin{assumption}[Random assignment of judges]\label{ass: MTEI}
$Z\perp (Y_{0}(z),Y_{1}(z), D_z; z\in \mathcal Z)$.
\end{assumption}


\begin{assumption}[Exclusion restriction] \label{ass: Exclusion}
There is no direct effect of judges' type on the potential outcomes. For $d \in \{0,1\}$, $Y_{d}(z)=Y_d$ for all $z \in \mathcal Z$.
\end{assumption}

\begin{assumption}[Monotonicity]\label{ass: Vyt-Mon}
For any pair $(z,z') \in \mathcal Z\times \mathcal Z$ either $D_z\geq D_{z'}$ for all defendants or $D_z\leq D_{z'}$ for all defendants. 
\end{assumption}

A particular feature of the judge leniency design is that judges are usually randomly assigned to different cases, making the random assignment assumption likely to hold in practice. However, \Cref{ass: Exclusion,ass: Vyt-Mon} are usually less credible. \Cref{ass: Exclusion} means the effect of judges on the potential outcomes must necessarily transit through their effect on treatment assignment. \Cref{ass: Vyt-Mon} requires that any defendants treated (incarcerated) by a more lenient judge be also treated if assigned to a less lenient one. \cite{heckman2005structural} refers to the monotonicity assumption as a uniformity condition since it restricts that the treatment on all the defendants must vary in a uniform direction when externally assigned to another judge.    \cite{vytlacil2002independence} provides an equivalent characterization of the monotonicity assumption, which can be stated as follows:
\begin{assumption}[Single Threshold-Crossing: STC]\label{ass: STC}
The judge treatment assignment mechanism is governed by the following threshold crossing model $D=1\{\nu(Z)\geq  U\}$ for some measurable and non-trivial function $\nu$, where the distribution of $ U$ is absolutely continuous.
\end{assumption}

Under \Cref{ass: MTEI,ass: STC}, we can rewrite the threshold crossing model without loss of generality as follows: 
\[
D = 1\left\{F_{U}(\nu(Z))\geq F_{U}(U)\right\} \equiv 1\left\{P(Z)\geq V\right\}, 
\]
where $F_{U}(\cdot)$ is the distribution function of $U$, $P(\cdot)\equiv F_{U}(\nu(\cdot))$ is identified from the observed variables $(D,Z)$ by $P(z) = \mathbb P(D=1|Z=z)$, and $V\equiv F_{U}(U)\sim Uniform[0,1]$. Hereafter, we will write $P(Z)$ as $P$ when it causes no confusion. Let $\mathcal P \subseteq [0,1]$ denote the support of $P(Z)$.
It is worth noting that the STC does not impose a priori direction in $z$ in the monotonicity condition since \Cref{,ass: STC} is equivalent to  \Cref{ass: Vyt-Mon} \citep[][]{vytlacil2002independence}. 
Under \Cref{ass: MTEI,ass: Exclusion,ass: STC}, the judge leniency design model can be equivalently written as:
\begin{equation} \label{eq: model}
  \begin{cases}
  Y = Y_{1}D + Y_{0}(1-D),  \\
  D=1\left\{P(Z)\geq V\right\}.
\end{cases}
\end{equation}

\Cref{ass: MTEI,ass: Exclusion,ass: STC} (equivalently \Cref{ass: MTEI,ass: Exclusion,ass: Vyt-Mon}) 
impose some restrictions on the joint distribution of the observed variables $(Y,D,P(Z))$, which we will characterize in \Cref{Thm: sharpMTE}. But before stating the theorem, we will discuss the intuition of the testable implications. Let $g: \mathcal Y\rightarrow \mathbb R^{+}$ be a nonnegative real integrable function such that $\mathbb E \vert g(Y_d)\vert < \infty$. Taking $d=0$ as an illustration. For any pair $(p,p') \in \mathcal P \times \mathcal P $ such that $p \leq p'$, we have:
\begin{multline*} 
\mathbb E [g(Y)(1-D) \vert P=p]=\mathbb E[g(Y_0)1\{ V\geq P\}\vert P=p]
=\mathbb E[g(Y_0)1\{ V\geq p\}]
\\ \geq \mathbb E[g(Y_0)1\{ V\geq p'\}]=\mathbb E[g(Y_0)1\{ V\geq P\}\vert P=p']= \mathbb E[g(Y)(1-D)\vert P=p'].
\end{multline*}
The first and fourth equalities hold by \Cref{ass: STC} (STC) and \Cref{ass: Exclusion} (exclusion); the second and third equalities hold because of \Cref{ass: MTEI} (random assignment),  and the inequality holds because $p\leq p'$. Intuitively, under the assumptions of the judge leniency design, if a defendant is released by judge $p'$, then he/she would necessarily be released by judge $p$ since judge $p$ is more lenient than judge $p'$. On the other hand, there can exist a set of defendants who were released by a type $p$ judge, but not by a type $p'$ judge: a group of ``compliers". Because $g(Y_0)$ is nonnegative, the average $g(Y_0)$ for this group of compliers is also nonnegative, delivering the inequality we see from the displayed equation above.  
The discussion is formalized in the following theorem. 
\begin{theorem}[Sharp characterization of the Judges' IV design assumptions] \label{Thm: sharpMTE}
 Let the collection of variables $(Y,D,Y_1,Y_0,P(Z))$ define a potential outcome model $Y=Y_1D+Y_0(1-D)$.
 
 (i) If \Cref{ass: MTEI,ass: Exclusion,ass: STC} (equivalently \Cref{ass: MTEI,ass: Exclusion,ass: Vyt-Mon})  hold, then for all $y, y' \in \mathcal Y$,  $\mathbb P(y<Y \leq y' , D=1 \vert P=p)$ and $-\mathbb P(y<Y \leq y' , D=0 \vert P=p)$ are non-decreasing in $p$ for all $p \in \mathcal P$. 
 
 (ii)  If for all $y, y' \in \mathcal Y$, $\mathbb P(y<Y \leq y' , D=1 \vert P=p)$ and $-\mathbb P(y<Y \leq y' , D=0 \vert P=p)$ are non-decreasing in $p$ for all $p \in \mathcal P$, there exists a joint distribution of $(\tilde V,\tilde{Y}_1,\tilde{Y}_0,P(Z))$ such that \Cref{ass: MTEI,ass: Exclusion,ass: STC} hold, and $(\tilde Y, \tilde D, P(Z))$ has the same distribution as $(Y,D,P(Z))$.
\end{theorem}

The proof of \Cref{Thm: sharpMTE} is collected in \Cref{proof: sharpMTE}. The testable implications in \Cref{Thm: sharpMTE}(i) are a subset of the implications previously derived in \citet[][Appendix A]{heckman2005structural}, who show for any non-negative integrable function, i.e. $g(\cdot): \mathcal Y \rightarrow \mathbb R^{+}$, $\mathbb E [g(Y)D \vert P=p]$ and $-\mathbb E [g(Y)(1-D) \vert P=p]$ are non-decreasing in $p$ under \Cref{ass: MTEI,ass: Exclusion,ass: STC}. 
The contribution of \Cref{Thm: sharpMTE}-(i) is that it shows we do not need to visit every single non-negative measurable function. It is sufficient to restrict our attention to a tractable subclass of these functions to screen all possible observable violations. This tractable characterization provides a basis for constructing a formal statistical test to verify the validity of the assumptions.\footnote{ We note that use the the half-interval class $g(Y)=1\{Y\leq y\}, y\in\mathcal Y$ will result in loss of power. To see this, suppose the support is finite, that is, $\mathcal Y=\{y_1,y_2,\cdots, y_K\}$, then it is without loss of information to consider the class of singletons: $g(Y)=1\{Y=y_k\}$, $k=1,2,\cdots, K$.  However, if one considers $g(Y)=1\{Y\leq y_k\}$, then it is possible that both $\mathbb P(Y\leq y_1, D=1 \vert P=p)$ and $\mathbb P(Y\leq y_2, D=1 \vert P=p)$ are non-decreasing function, but $\mathbb P(Y=y_2, D=1 \vert P=p)$ is not.}

The second part of \Cref{Thm: sharpMTE} is new, and it shows that the testable implications in \Cref{Thm: sharpMTE}(i) are the most informative way to detect all observable violations of the random assignment, the exclusion restriction, and the monotonicity assumption (without an ex-ante imposed direction). These testable implications cannot be strengthened without making additional assumptions. Various tests or testable implications are used in the literature to screen violations of the judge leniency design assumptions; for instance, \cite{dobbie2018intergenerational,bhuller2018incarceration,norris2021effects,frandsen2023judging}. However, to the best of our knowledge, only \Cref{Thm: sharpMTE} provides sharp testable implications without imposing an a priori direction in the monotonicity assumption. 

Tests based on sharp testable implications have empirical virtue. In practice, one may use tests developed from non-sharp testable implications for the sake of traceability. However, as recently discussed in \cite{li2024discordant}, non-sharp tests can lead to discordant empirical results and misleading interpretations of the estimand of interest. It is possible that for the same data, two different non-sharp tests may generate contradictory results, as they may use different sets of information from the same observed DGP to screen violations of the model assumptions. Thus, the conclusion may largely depend on which test the empirical researcher implements. 

Moreover, after implementing a specification test and obtaining a non-rejection result, one often proceeds and provides a causal interpretation of the estimand. For example, in judge leniency designs, the 2SLS or Local IV (LIV) estimand is interpreted as the LATE or MTE, respectively. However, since a non-sharp test only uses part of the observable information in the data and fails to reject the model when it is misspecified, we must be cautious about interpreting the 2SLS or the LIV estimand as identifying the LATE/MTE solely based on the result of a non-sharp test. Therefore, using a sharp test must be viewed not only as a theoretical exercise, but also as having an important empirical relevance. A sharp test provides the most informative way to detect all observable violations of a given model's assumptions and is more robust to possible misleading interpretations and discordant results.

\subsection{Connection to existing tests}

\subsubsection{\cite{kitagawa2015}, and  \cite{MW2017} testable implications}
Inspired by \citet[][Appendix A]{heckman2005structural}, \cite{kitagawa2015} and  \cite{MW2017} derive a set of sharp testable implications assuming an a priori direction in the monotonicity assumption. When judges' types are binary, i.e. $ Z\in \mathcal \{0,1\}$, there are only two potential directions, so it is not restrictive to assume the direction of the monotonicity. However, when the cardinality of the judges' types is large (or even infinite when the judges' types are continuous), imposing a specific ex-ante direction is extremely restrictive because the number of possible directions to consider can be rather large (or even infinite). One could implement their test by visiting all the possible directions, but this can be cumbersome or even computationally impossible if $Z$ takes many values.

One significant difference between the testable implication of \cite{kitagawa2015} and  \cite{MW2017} and ours is we do not assume a prior direction. To illustrate this point, suppose $\mathcal Z=\{z_1,...,z_K\}$ and suppose we assume one of the $K!$ potential directions as: 
$$D_{z_{K}}\geq D_{z_{K-1}} \geq...\geq D_{z_1}$$ meaning that type $z_{K}$ judge is less lenient than type $z_{K-1}$ judge, which, in turn, is less lenient than $z_{K-2}$, $z_{K-3}$, $\cdots, z_{1}$ judge.
Given this imposed ordering, \Cref{ass: MTEI,ass: Exclusion,ass: Vyt-Mon} imply the following testable implications studied in \cite{sun2023instrument}:
\begin{align*}
\mathbb P(y<Y \leq y' , D=1 \vert Z=z_{k}) &\leq \mathbb P(y<Y \leq y' , D=1 \vert Z=z_{k+1}), \\
-\mathbb P(y<Y \leq y' , D=0 \vert Z=z_{k}) &\leq -\mathbb P(y<Y \leq y' , D=0 \vert Z=z_{k+1}),\\
&~~~ \text{ for all } k \in \{1,...,K-1\}  \text{ and }  y,y' \in \mathcal Y. \nonumber
\end{align*}
A key point to note is that the above implications restrict $F_{Y,D|Z}(y,d|z)$ while the testable implications in \Cref{Thm: sharpMTE}(i) instead restrict  $F_{Y,D|P}(y,d|p)$. In the first case, the induced direction of inequalities is with respect to the observed judge type $Z$, while in our case, the inequalities are with respect to the propensity score $P$, which is obtained without imposing a prior direction. Also, noteworthy is that if one takes $y=-\infty$ and $y'=\infty$, the testable implications in \Cref{Thm: sharpMTE}(i) no longer have any empirical content. But, the testable implications with an ex-ante monotonicity direction still restrict the propensity scores and the judges' types, i.e., $P$, and $Z$, such that
\begin{align*}
\mathbb P(D=1 \vert Z=z_{k}) \leq \mathbb P(D=1 \vert Z=z_{k+1}), \text{ for all } k \in \{1,...,K-1\}.
\end{align*}
Therefore, implementing the testing approaches of \cite{kitagawa2015} and  \cite{MW2017} may reject the judge leniency design assumptions even if \Cref{ass: Exclusion,ass: MTEI,ass: Vyt-Mon} hold, but just the ex-ante imposed direction of monotonicity is wrong.

\subsubsection{\cite{frandsen2023judging}'s test}

FLL proposes a set of testable implications for  \Cref{ass: MTEI,ass: Exclusion,ass: Vyt-Mon}. Their testable implication has sound features of not relying on the ex-ante specified direction of monotonicity and assessing all the assumptions jointly. Their testable implication, however, is not sharp and can fail to screen some {\color{blue} non-negligible} observable violations of the judge leniency design. To see this, consider any integrable function $g(\cdot): \mathcal Y \rightarrow \mathbb R$, and let $p\neq p' \in \mathcal P$. Under \Cref{ass: MTEI,ass: Exclusion,ass: Vyt-Mon}, we can derive the following equality:
\begin{multline*}
W(g(Y),p,p')\equiv \frac{\mathbb E[g(Y)|P=p']-\mathbb E[g(Y)|P=p]}{p'-p}\\=\mathbb E[g(Y_1)-g(Y_0)|p<V\leq p']1\{p<p'\}+\mathbb E[g(Y_1)-g(Y_0)|p'<V\leq p]1\{p<p'\}.
\end{multline*}
If we denote by $L_g$ and $U_g$ the known lower bound and upper bound of the support of $g(Y)$, the latter equality implies:
\begin{eqnarray}\label{eq: Frand-Test}
L_g-U_g\leq W(g(Y),p,p') \leq U_g-L_g,
\end{eqnarray}
where the inequality in \eqref{eq: Frand-Test} is the main testable implication used by FLL (see Theorem 1 and Equation (2) therein) to implement their test. However, under \Cref{ass: MTEI,ass: Exclusion,ass: Vyt-Mon}, we should also have:
\begin{align*}
W(g(YD),p,p')&=\mathbb E[g(Y_1)|p<V\leq p']1\{p<p'\}+\mathbb E[g(Y_1)|p'<V\leq p]1\{p>p'\},
\\
W(g(Y(1-D)),p,p')&=-\mathbb E[g(Y_0)|p<V\leq p']1\{p<p'\}-\mathbb E[g(Y_0)|p'<V\leq p]1\{p>p'\},
\end{align*}
where those two latter equalities lead to the following observable restrictions:
\begin{align}
L_g&\leq W(g(YD),p,p') \leq U_g,\label{eq:SF1}\\
-U_g&\leq W(g(Y(1-D)),p,p') \leq -L_g, \label{eq:SF2}
\end{align}
One can easily observe that the testable restrictions in \eqref{eq:SF1} and \eqref{eq:SF2} could be violated, whereas the restriction used by FLL, i.e. inequality \eqref{eq: Frand-Test} still holds. Hence, implementing FLL's statistical testing procedure based on inequalities \eqref{eq:SF1} or \eqref{eq:SF2} could provide a different result compared to their test based on inequality \eqref{eq: Frand-Test} alone. These discordant implications confirm the concern about developing a statistical test based on non-sharp restrictions. \Cref{example: FFL fails} provides a concrete numerical example.

\begin{example}\label{example: FFL fails}
    Consider the potential outcome model:
    \[
  \begin{cases}
  Y = Y_{1}D + Y_{0}(1-D),  \\
  D=1\left\{P\geq V\right\}.
\end{cases}
    \]
    Suppose $V$ is independent of $(Y_1,Y_0,P)$. However, $(Y_1,Y_0)$ and $P$ are dependent:
    \[
      \begin{cases}
  Y_1|P=\tilde p\sim \text{ degenerate at 1 }, &\text{ if } \tilde p<\frac{1}{2}  \\
  Y_1|P=\tilde p\sim Bernoulli(\tilde p),&\text{ if } \tilde p\geq \frac{1}{2}
\end{cases}
    \]
        \[
      \begin{cases}
  Y_0|P=\tilde p\sim \text{ degenerate at 0 }, &\text{ if } \tilde p<\frac{1}{2}  \\
  Y_0|P=\tilde p\sim Bernoulli(\tilde p),&\text{ if } \tilde p\geq \frac{1}{2}
\end{cases}
    \]
    Therefore, the randomization assumption is violated, but the monotonicity and exclusion conditions are met. In this case, $Y_d$ is binary and $U_g=1$ and $L_g=0$, so we also take $g(\cdot)$ to be the identity function without loss of generality. 

    For this DGP, we can show that for any $p'\in(0,1)$ and $p\in(0,1)$, $ W(g(Y),p,p')=1$. Hence, FLL's testable implication (inequality \ref{eq: Frand-Test} above) always holds and has no power to detect the violation. There is missing information. For example, when $p'>p>\frac{1}{2}$, we can verify that $W(g(YD),p,p') =p'+p > 1\equiv U_g$. Therefore, condition (\ref{eq:SF1}) is violated. Please see derivation details in \Cref{section: numerical example}.
    
     On the other hand, our testable implication can capture such a violation. To see this, note 
    \[
    \mathbb E[YD|P=p] = \mathbb E[Y_1|P=p]p = \begin{cases}
  p \;\; \text{ if } p <\frac{1}{2},  \\
 p^2\;\; \text{ if } p\geq \frac{1}{2}.   \\
\end{cases}.
    \]
    It is apparent that $\mathbb E[YD|P=p]$ is not a monotone function of $p$, and therefore violates our testable implication. \qed
\end{example}

The intuition behind \Cref{example: FFL fails} is not pathological and is reflected in the derivation in \Cref{section: numerical example}. Because $\mathbb E[Y|P=p]=\mathbb E[Y_1D|P=p]+\mathbb E[Y_0(1-D)|P=p]$, it is possible the violations on the $\mathbb E[Y_1D|P=p]$ and $\mathbb E[Y_0(1-D)|P=p]$ ``cancel'' out. As a consequence, the quantity $\mathbb E[Y|P=p]$ provides no power to detect violations in these cases.

Another evident reason why FLL's implications cannot exhaust all violations of the judge leniency design is that they only focus on $g(Y)=Y$, whereas the inequality in \eqref{eq: Frand-Test} should hold for any integrable function $g$ and for any pair $p\neq p' \in \mathcal P$. $g(Y)=Y$ is not a sufficient class of functions to screen all violations of the model.

Finally, we note our testable implications in \Cref{Thm: sharpMTE} do not rely on the known support of $g(Y)$, whereas to test inequality (\ref{eq: Frand-Test}), one needs to know the bounds of the support $(U_g,L_g)$. If the support of $g(Y)$ is unbounded, i.e., $U_g=+\infty$ and $L_g=-\infty$, then the testable implication in (\ref{eq: Frand-Test}) holds trivially and FLL's test does not have any power in detecting violations to the identification assumptions.

In the next section, we propose a testing procedure based on the sharp testable implications of \Cref{Thm: sharpMTE}. We will show that in large samples, our test is consistent against all the violations of our testable implication and is, therefore, more powerful asymptotically than the existing ones. 

\section{Testing Procedures} \label{section: testing procedure}
In this section, we construct tests based on \Cref{Thm: sharpMTE}. For the defendant $i\in\{1,2,\cdots, n$\}, researchers observed a vector $(Y_i,D_i,Z_i,X_i)$, where $Y_i$, $D_i$, $Z_i$, and $X_i$ represent his/her observed outcome, observed treatment status, the vector of characteristics of the judge that $i$ was assigned to, and the vector of additional control variables, respectively.

We first present our baseline semi-nonparametric test in \Cref{section: nonparametric test} without the presence of control variables $X_i$. For this test, we make no functional form or distributional assumptions about potential outcomes. We do need to estimate the propensity score $P(z)\equiv \mathbb P(D_i=1|Z_i=z)$ first, for which our procedure can accommodate different data scenarios. If $Z_i$ contains continuous variables, we follow the common practice in the literature to employ a parametric model so that $P(z)=P(z,\theta_0)$ for all $z\in\mathcal{Z}$ and for a finite-dimensional parameter vector $\theta_0\in\Theta$. Popular choices include the Probit or Logit model with a linear index $z'\theta_0$ \citep[see, for instance,][among many others]{carneiro2011estimating,kowalski2016doing}.\footnote{When $Z$ is continuous, the rejection result of our semi-nonparametric test can be interpreted as rejecting the joint assumption of the judge leniency design and the parametric form imposed on the propensity score. In our simulation studies, we always keep the propensity score correctly specified. In these studies, therefore, the rejection shows the power of our test to reject false judge leniency assumptions.} When $Z_i$ only contains discrete variables, such as judge's gender, we can estimate $\mathbb P(D_i=1|Z_i=z)$ by the sample averages of $D$ conditioning on each possible value of $z\in\mathcal Z$. In this case, our test is indeed nonparametric. 

We should emphasize that, in both cases above, we do not require any knowledge of the identity of judges, nor do we need the number of defendants handled by each judge to diverge to infinity. For example, when $Z$ is gender, we only need the number of defendants for each judge's gender to go to infinity. This can happen when the number of judges is large, but each judge handles a finite number (or even one) of defendants. Therefore, our test can also be applied to other empirical contexts than the judge leniency design. There is another scenario in which $Z_i$ is the judge's identity. Suppose the number of defendants handled by each judge is large, as in our empirical application. In this case, we can also consistently estimate judge $j$'s propensity score $\mathbb P (D=1|Z_i=j)$ by the sample frequency estimator for each judge $j$, regardless of whether the number of judges is small or large. 

In practice, the number of defendants that a judge handles can be small. In this case, one can not estimate $\mathbb P (D=1|Z_i=j)$ consistently without additional assumptions and conventional inference methods can be invalid. This phenomenon has received attention from the literature, see discussions in \cite{jochmans2023many}, \cite{ren2024extrapolating}, \cite{sithole2024locally}, and \cite{yap2024inference}. These papers, however, focus on inference on the parameters instead of testing model specification. To account for data scenarios with small numbers of defendants per judge, we design a test that does not require a consistent estimator for the propensity score and controls the size at any sample sizes for the case of a binary outcome. To the best of our knowledge, this test and FLL's finite sample test are the only ones for testing judge leniency design specification with finite samples, and both focus on binary outcome variables. Our test uses upper bounds of the null distribution to calculate the critical values, and hence is very easy to implement. It only requires simulating Bernoulli random variables, and no nonlinear optimization is involved. For the purpose of exposition, we collect the finite sample test in \Cref{section: finite sample test} and focus on the cases in which the propensity score can be consistently estimated in this section. 

In practice, researchers may observe a set of defendant and case covariates $X$ and assume the randomization and monotonicity hold conditioning on $X$ (see \Cref{ass: MTEI with X,ass: STC with X} below). In the presence of covariates, researchers can use the semi-nonparametric test introduced in \Cref{section: nonparametric test} when the dimension of covariates is small or the number of support points in $\mathcal X$ is not large; please see \Cref{remark: small x} below. In other cases, the semi-nonparametric test may encounter challenges associated with the curse of dimensionality. To address this concern, we introduce an alternative semiparametric test designed to accommodate situations with a large (but fixed) number of covariates in \Cref{section: add covariates}. 

\subsection{A Semi-nonparametric test}  \label{section: nonparametric test}

For the convenience of the exposition, we restate the testable implications as the null hypothesis $H_0$. That is, for all $p_1\geq p_2 \text{ with }  p_1,p_2\in \mathcal P  \text{ and all }  y,y' \in \mathcal Y$, 
\begin{align}
    \label{MTE00} &\mathbb P(y<Y \leq y' , D=1 \vert P= p_1) \geq \mathbb P(y<Y \leq y' , D=1 \vert P=p_2), \\ 
    \label{MTE01} &\mathbb P(y<Y \leq y' , D=0 \vert P= p_1) \leq \mathbb P(y<Y \leq y' , D=0 \vert P=p_2).
\end{align}
The alternative hypothesis $H_1$ is then inequality (\ref{MTE00}) or (\ref{MTE01}) fails to hold for some $(p_1,p_2)$ and $(y,y')$. Without loss of generality, we assume the support of $Y$ is $[0,1]$.\footnote{We can always apply a transformation to ensure the support of $Y$ is $[0,1]$.  If $Y$ has a finite support $[a,b]$, we can apply an affine transformation $\tilde Y=(Y-a)/(b-a)$.  If $Y$'s support is the whole real line, we can apply standard normal CDF after rescaling and re-centering: $\tilde Y= \Phi\left(\frac{Y-\bar Y}{\hat{std}(Y)}\right)$, where $\bar Y$ is the sample average and $\hat{std}(Y)$ is the sample standard deviation.} Testing inequalities (\ref{MTE00}) and (\ref{MTE01}) involves two features; first, it is a set of inequality restrictions defined on conditional moments where the conditioning variable is possibly continuous. We deal with the first difficulty by employing the method of \cite{hsu2019testing} to transform them into an equivalent set of restrictions on unconditional moments. The second feature is that the conditioning variable $P$ is not directly observed from the data. We derive the new influence functions and show that the first-stage estimation error is properly accounted for.

To be more specific, we define a collection of functions $\{\nu_d(\ell):\ell\in \mathcal L, d=0,1\}$ as follows:
\begin{align}
&\nu_1(\ell) \equiv \mathbb{E}[D 1\{y \leq Y\leq y+r_y\}  1\{p_2 \leq P\leq p_2+r_p\}]
\cdot   \mathbb{E}[1\{p_1 \leq P\leq p_1+r_p\}]\nonumber\\
&~~-\mathbb{E}[D 1\{y \leq Y\leq y+r_y\}  1\{p_1 \leq P\leq p_1+r_p\}]
\cdot   \mathbb{E}[\{p_2 \leq P\leq p_2+r_p\}],\label{eq: nu1}
\end{align}
and
\begin{align}
&\nu_0(\ell) \equiv  \mathbb{E}[(D-1) 1\{y \leq Y\leq y+r_y\}  1\{p_2 \leq P\leq p_2+r_p\}]
\cdot   \mathbb{E}[1\{p_1 \leq P\leq p_1+r_p\}]\nonumber\\
&~~-\mathbb{E}[(D-1) 1\{y \leq Y\leq y+r_y\}  1\{p_1 \leq P\leq p_1+r_p\}]
\cdot   \mathbb{E}[1\{p_2 \leq P\leq p_2+r_p\}],\label{eq: nu0}
\end{align}
where the index $\ell \in \mathcal L$ is defined as
\begin{align*}
& \ell = (\ell_y',\ell_p')', \quad\ell_y=(y,r_y)',\quad\ell_p=(p_1,p_2,r_p)', \quad\mathcal L=\mathcal L_Y\otimes\mathcal L_P,\\
&{\mathcal L}_{Y}=\left\{(y,r_y):~r_y=q_y^{-1},~\text{ } q_y\cdot y\in\{0,1,2,\cdots,(q_y-1)\}\text{ for } q_y=1,2,\cdots,\right\}.\\
&{\mathcal L}_{P}=\left\{(p_1,p_2,r_p):~r_p=q_p^{-1},~\text{ } q_p\cdot (p_1,p_2)\in\{0,1,2,\cdots,(q_p-1)\}^2, p_1\geq p_2 \text{ for } q_p=1,2,\cdots,\right\}.
\end{align*}
Then, following the same calculation as in \cite{hsu2019testing}, we can formulate the null hypothesis in inequalities (\ref{MTE00}) and (\ref{MTE01}) as the following:
\begin{equation}\label{eq: null}
H_0:    \nu_d(\ell) \leq 0,\quad \text{ for all } \ell\in\mathcal L \text{ and } d=0,1,
\end{equation}
against the alternative hypothesis $H_1$ that inequality (\ref{eq: null}) fails to hold for some $\ell\in \mathcal L$ and for $d=0$ or $d=1$. Consequently, testing the original sharp implication in \Cref{Thm: sharpMTE} is equivalent to testing the set of inequalities indexed by $\ell\in\mathcal L$, a class of cubes. There is no loss of information for such transformation \citep[see][]{AndrewsShi2013}. Under $H_0$, we expect to see $T \equiv \sum_{d=0,1} \sum_{\ell\in\mathcal L} \max\{\nu_d(\ell),0\}^2 \Omega(\ell) = 0$, where $\Omega(\cdot)$ is a positive weighting function. On the other hand, $T>0$ under $H_1$. Our test statistics are based on the appropriately rescaled and standardized sample analog of $T$. 

In the expression of $\nu_d(\ell)$, the propensity score $P(Z_i)$ is unknown, but can be replaced by its root-n consistent estimate $\hat P_i$. When we estimate the propensity score by a parametric model, we denote it as $\hat P_i \equiv P(Z_i,\hat\theta)$, where $\hat\theta$ is the MLE. When $Z_i$ is the judge's identity and the number of defendants for each judge is large, we simply use the frequency estimator $\hat P_i=\frac{\sum_{j=1}^n D_j1\{Z_j=Z_i\}}{\sum_{j=1}^n 1\{Z_j=Z_i\}}$. \Cref{algorithm: critical value} below summarizes the semi-nonparametric test's implementation procedure. Please see \Cref{app: implementation} for detailed equations and expressions.

\begin{algo} \label{algorithm: critical value}
This algorithm shows the steps for constructing the test statistics and critical value. 
\begin{enumerate}
    \item Specify integers $Q_Y$ and $Q_P$, and create a coarser version $\mathcal L_Q$ of $\mathcal L$ set by limiting $q_y=1,2,\cdots, Q_Y$ and $q_p=1,2,\cdots,Q_P$. 
    \item Compute the estimator for the propensity score $\hat P_i$, as detailed in \Cref{app: implementation}.
    \item For each $\ell\in\mathcal L_Q$, construct estimates $\hat\nu_1(\ell)$ and $\hat\nu_0(\ell)$ as sample analogs of \Cref{eq: nu1,eq: nu0}, as detailed in \Cref{eq: m functoin estimates,eq: w functoin estimates}. 
    \item Choose a positive integer $B$ (as the number of bootstrap iterations), and for each $b=1,2,\cdots, B$,
    \begin{enumerate}
        \item Draw $W_1^b,W_2^b,\cdots, W_n^b$ as a sequence of independent random variables with both mean and variance equal to one and are independent of the original sample. 
        \item Estimate propensity score for each bootstrap iteration $\hat P_i^{b}$, defined in \Cref{eq: boot MLE}. 
        \item Obtain $\hat\nu_d^b(\ell)$, $d=0,1$, for each bootstrap iteration using \Cref{eq: nu0estimation boot,eq: nu1estimation boot}. 
    \end{enumerate}
    \item Compute the normalization factor, denoted by  $\hat\sigma_d(\ell)$, as:
    \begin{equation}\label{eq: variance estimator}
 \hat{\sigma}^2_{d}(\ell)
 =\frac{n}{B}\sum_{b=1}^B \big(\hat{\nu}^{b}_d(\ell)-
 \overline{\hat{\nu}^{b}_d}(\ell)\big)^2,\text{ where }\overline{\hat{\nu}}^{b}_d(\ell)=\frac{1}{B}\sum_{b=1}^B
\hat{\nu}^{b}_d(\ell).
\end{equation}
Choose a constant $\epsilon>0$, and let $\hat\sigma^2_{d,\epsilon}(\ell)=\max\{\hat\sigma^2_d(\ell),\epsilon\}$.
    \item Choose the weighting function $\Omega$ over $\mathcal{L}$ such that $\Omega(\ell)>0$ for all  $\ell \in \mathcal{L}$ and $\sum_{\ell\in\mathcal{L}}\Omega(\ell)<\infty$. Calculate the test statistics as 
    \begin{equation} \label{eq: test statistics}
\widehat{T}_n= \sum_{d=0,1}\sum_{\ell\in\mathcal{L}_Q} 
\max\Big\{ \sqrt{n}\frac{\hat{\nu}_{d}(\ell)}{\hat{\sigma}_{d,\epsilon}(\ell)},
0 \Big\}^2  \Omega(\ell).\footnote{To be specific, for $q_y$ and $q_p$, we suggest to set $\Omega(\ell)= q_y^{-3} \cdot \frac{q_p^{-2}}{ q_p (q_p-1)}$.}
    \end{equation}
    \item Let $a_n$ and $B_n$ be positive deterministic sequences.\footnote{See \cite{AndrewsShi2013} for the rate condition of $a_n$ and $B_n$ and they suggest to set $a_n=\sqrt{0.3 \ln n}$ and $B_n=\sqrt{{0.4\ln n}/{\ln\ln n}}$. Here, we propose $a_n=0.15\ln n$ and $B_n={0.85\ln n}/{\ln\ln n}$, as in \cite{hsu2019testing}.} Calculate the generalized moment selection (GMS) terms as 
    \[
    \hat\psi_{d}(\ell) = - B_n\cdot 1\left\{\frac{\sqrt{n}\hat\nu_d(\ell)}{\hat\sigma_{d,\epsilon}(\ell)}<-a_n\right\}.
    \]
    \item For $b=1,2,\cdots, B$, calculate the quantity 
        \[
        \widehat T^b = \sum_{d\in\{0,1\},\ell\in\mathcal L_Q}\max \left\{\frac{\hat\Phi^b_d(\ell)}{\hat\sigma_{d,\epsilon}(\ell)} +\hat\psi_d(\ell)\right\} ^2 \Omega(\ell),
        \]
        where
        \begin{equation}
            \Phi^{b}_d(\ell) = \sqrt{n} \left(\hat\nu_d^b(\ell) - \hat\nu_d(\ell)\right).
        \end{equation}
    \item Let $\hat c = \hat q(1-\alpha +\eta) +\eta$, where $\hat q(\tau)$ is the $\tau$-th empirical quantile of $\left\{\widehat T^b\right\}_{b=1}^B$ and $\eta$ is a small positive constant, e.g. $\eta=10^{-6}$.\footnote{$\eta$ is the infinitesimal constant which is introduced mainly for the sake of proof; see for instance \cite{AndrewsShi2013}. Our simulation exercises set it to $10^{-6}$. } 
    \item Define the test to be $\phi_n = 1\{\widehat T\geq \hat c\}$. That is, we reject the null hypothesis if $\widehat T\geq \hat c$.
\end{enumerate} 
\end{algo}

\Cref{thm: validity} shows that the test $\phi_n$ has its size controlled asymptotically and is consistent. The proof for \Cref{thm: validity} is collected in \Cref{proof: validity} of the online supplementary material. We also list all the technical assumptions, such as conditions that ensure the first-stage estimator converges at a sufficiently fast rate, in that section for the sake of exposition. 

\begin{theorem}\label{thm: validity}
    Suppose \Cref{ass: continuity,ass: theta influence,ass: wb theta influence,ass: iid} in \Cref{proof: validity} are satisfied. Let $\alpha\in(0,1/2)$ be the pre-chosen significance level. 
    \begin{enumerate}
        \item[(i)] Under the $H_0$ in characterized by inequalities (\ref{eq: null}), we have
\begin{equation}
    \limsup_{n\rightarrow \infty}\mathbb P(\phi_n=1|H_0)\leq \alpha.
\end{equation}
\item[(ii)] Under $H_1$,
\begin{equation}
    \limsup_{n\rightarrow \infty}\mathbb P(\phi_n=1|H_1)=1.
\end{equation}
    \end{enumerate}
\end{theorem}

\subsection{A semiparametric test with covariates dimension reduction}\label{section: add covariates} 

In this section, we introduce a semiparametric test in the presence of covariates $X$. We begin by introducing the following assumptions. 

\begin{assumption}[Conditional Random Assignment of Judges]\label{ass: MTEI with X}
$Z\perp (Y_{0}(z),Y_{1}(z), D_z; z\in \mathcal Z)|X=x$ for all $x\in\mathcal X$.
\end{assumption}

\begin{assumption}[Single Threshold-Crossing with Covaraites: STC]\label{ass: STC with X}
The judge treatment assignment mechanism is governed by the following threshold crossing model $D=1\{\nu(Z,X)\geq  U\}$ for some measurable and non-trivial function $\nu$, where the distribution of $ U$ is absolutely continuous.
\end{assumption}

When \Cref{ass: MTEI with X,ass: Exclusion,ass: STC with X} hold, the testable implications can be written as follows. For all $x\in\mathcal X$, $p_1,p_2\in\mathcal P$ and $p_1\geq p_2$, and all $y,y'\in\mathcal Y$
\begin{align}
    \label{MTE00-x} &\mathbb P(y<Y \leq y' , D=1 \vert P= p_1, X=x) \geq \mathbb P(y<Y \leq y' , D=1 \vert P=p_2, X=x), \\ 
    \label{MTE01-x} &\mathbb P(y<Y \leq y' , D=0 \vert P= p_1, X=x) \leq \mathbb P(y<Y \leq y' , D=0 \vert P=p_2, X=x).
\end{align}

\begin{remark} \label{remark: small x}
    If $ X$ is discrete and $\mathcal X$ only contains a relatively small number of values, or $X$ contains a small number of continuous variables, we can also follow the same procedure as in \Cref{section: nonparametric test} but add cubes for $X$. Under the null hypothesis, we should expect $T \equiv \sum_{d=0,1} \sum_{x\in\mathcal X}\sum_{\ell\in\mathcal L} \max\{\nu_d(\ell,x),0\}^2 \Omega(\ell,x) = 0$, where $\Omega(\ell,x) $ is a positive weighting function chosen by researchers, and
\begin{multline*}
\nu_1(\ell,x) \equiv \mathbb{E}[D 1\{y \leq Y\leq y+r_y\} 1\{x \leq X\leq x+r_x\} 1\{p_2 \leq P\leq p_2+r_p\}] \\ 
\times   \mathbb{E}[1\{x \leq X\leq x+r_x\}1\{p_1 \leq P\leq p_1+r_p\}]-  \mathbb{E}[1\{x \leq X\leq x+r_x\}\{p_2 \leq P\leq p_2+r_p\}]\\
\times \mathbb{E}[D 1\{y \leq Y\leq y+r_y\}1\{x \leq X\leq x+r_x\}  1\{p_1 \leq P\leq p_1+r_p\}],
\end{multline*}
and 
\begin{multline*}
\nu_0(\ell,x) \equiv  \mathbb{E}[(D-1) 1\{y \leq Y\leq y+r_y\} 1\{x \leq X\leq x+r_x\}1\{p_2 \leq P\leq p_2+r_p\}]\\
\times   \mathbb{E}[1\{x \leq X\leq x+r_x\}1\{p_1 \leq P\leq p_1+r_p\}] -   \mathbb{E}[1\{x \leq X\leq x+r_x\}1\{p_2 \leq P\leq p_2+r_p\}]\\
\times {E}[(D-1) 1\{y \leq Y\leq y+r_y\} 1\{x \leq X\leq x+r_x\} 1\{p_1 \leq P\leq p_1+r_p\}],
\end{multline*}
and $r_x$ is similarly defined as $r_p$ and $r_y$. The implementation follows analogously from \Cref{algorithm: critical value}. 
\end{remark}

When the dimension of $X$ is high, an alternative approach is to include the covariates parametrically, as in \citet[][Assumptions A.4 and A.5]{carr2021testing}, which we state below:
\begin{assumption} \label{ass: functional form x}
    (i) For $d=0,1$, then potential outcomes take the form of $Y_d = \alpha_d+ X'\beta_d + U_d$, where $(\alpha_d,\beta_d)$ are constants, and (ii) the residual terms $( U_0, U_1)$ satisfy $( U_0, U_1,V)\perp (X,Z)$.
\end{assumption}

\citet[][Proposition 2]{carr2021testing} show if \Cref{ass: MTEI with X} is strengthened to \Cref{ass: functional form x}, then the testable implications in (\ref{MTE00-x}) and (\ref{MTE01-x}) can be characterized as
\begin{align}
    \label{MTE00-u} &\mathbb P(y<\tilde Y \leq y' , D=1 \vert P= p_1) \geq \mathbb P(y<\tilde Y \leq y' , D=1 \vert P=p_2), \\ 
    \label{MTE01-u} &\mathbb P(y<\tilde Y \leq y' , D=0 \vert P= p_1) \leq \mathbb P(y<\tilde Y \leq y' , D=0 \vert P=p_2),
\end{align}
for $y,y'\in \mathcal Y$, and 
\[
\tilde Y= D(U_1+\alpha_1) + (1-D)(U_0+\alpha_0) = D(Y_1 - X'\beta_1)+(1-D)(Y_0-X'\beta_0) = Y - X'(D\beta_1 +(1-D)\beta_0).
\]
The advantage of using (\ref{MTE00-u}) and (\ref{MTE01-u}) is that both inequalities are only conditional on the scalar-valued propensity score. The effect of covariates has been filtered out by constructing a new outcome variable $\tilde Y$. \Cref{ass: functional form x} is a common assumption made in the literature for estimating the MTE, see for instance \cite{CarneiroLee2009,carneiro2010evaluating,kowalski2016doing}. Nevertheless, we do acknowledge it is subject to the potential risk of model mis-specification. Under the null hypothesis of the model being correctly specified, parameters $\beta_0$ and $\beta_1$ can be estimated by partial linear regression of $Y$ on $X$ and propensity score $P$ separately for the sample of $D=1$ and $D=0$: 
\[
\mathbb E[Y|X=x,P=p,D=d]=x'\beta_d + K_d(p),\quad d\in\{0,1\},
\]
where $K_d(p) = \mathbb E[\alpha_d+U_d|X=x,D=d,P=p]$ only depends on $p$ under \Cref{ass: functional form x}-(ii). 
The following algorithm summarizes the steps for implementation. 
\begin{algo} \label{algorithm: critical value with X} 
   \begin{enumerate}\setlength{\itemsep}{8pt}
\item The procedure starts with estimated propensity score $\hat P_i = P(Z_i,X_i,\hat\theta)$ using \Cref{eq: MLE with X}.
\item Choosing the subsample with $D=d$, and within this subsample,
\begin{enumerate}
\item Estimate $\mathbb E[Y|P]$ nonparametrically,\footnote{One can consider local polynomial estimation as in \cite{CarneiroLee2009} or do global estimation as in \cite{kowalski2016doing}. Since we do not need to estimate the derivative $K'(p)$ in our paper, we use global polynomial regression in \cite{kowalski2016doing} for our simulation and empirical applications.} and calculate the residual $e_i^P\equiv Y_i-\hat{\mathbb E}[Y_i|\hat P_i]$. 
\item Estimate $\mathbb E[X|P]$ nonparametrically, and calculate the residual $e_i^X\equiv X_i-\hat{\mathbb E}[X_i|\hat P_i]$.
\item Regress $e_i^P$ on $e_i^X$ and obtain the OLS estimates, denoted by $\hat\beta_d$. 
\end{enumerate}
\item Once $\hat\beta_1$ and $\hat\beta_0$ are obtained, one can construct estimates for ${\widetilde Y}_i= Y_i - X_i'(D_i\hat\beta_1+(1-D_i)\hat\beta_0)$ 
\item Follow the rest of steps in \Cref{algorithm: critical value} with $Y$ being replaced by ${\widetilde Y}$.
\end{enumerate} 
\end{algo}

\section{Simulation and Empirical Application} \label{section: simulation}

\subsection{Simulation}

In this subsection, we provide two sets of simulations to assess the size and power properties of our sharp test under various DGPs in finite samples. Throughout this section, we ran $1000$ replications for each simulation design, and the bootstrap sample size is chosen to be $B=800$. We set $a_n=0.15\ln n$ and $B_n={0.85\ln n}/{\ln\ln n}$, as in \cite{hsu2019testing}. We choose $Q_P=5$ and $Q_Y=5$ (for continuous $Y$) or  $Q_Y=2$ (for binary $Y$). We set the infinitesimal constant $\eta=10^{-6}$ and the constant $\epsilon=10^{-6}$ (see the definition of $\hat\sigma^2_{d,\epsilon}(\ell)$ in \Cref{algorithm: critical value}-4).

\subsubsection{Binary outcome}
The first set of simulations is based on a DGP introduced in FLL (online appendix, page 22). In this set of simulations, we mimic the random assignment of $n$ defendants to a pool of $J$ judges, ensuring an equitable distribution of $\frac{n}{J}$ defendants to each judge. As in FLL, the severity probability of each judge $j$ is set as follows: 
$$
 p_{j} = p_{a} + \frac{ j-1}{J-1}(1-p_{a} - p_{n}) 
$$
Here, $p_{a}$ and $p_{n}$ stand for the fraction of always and never treated defendants, respectively. FLL consider a binary outcome model where the outcome $Y \in \{0,1\}$ satisfies the following condition:
$$ 
 \mathbb{E} \left[ Y \mid p_{j}  \right]  = \frac{1 - (1-\lambda)(p_n + p_a)}{1 - (p_n + p_a)} p_j - \frac{\lambda}{1-(p_n + p_a)} p_a.
 $$
The parameter $\lambda$ dictates the extent of deviation from the exclusion restriction assumption. When $\lambda = 0$, there is no violation of the judge leniency design assumptions. Consequently, for $\lambda = 0$, the simulations aim at assessing the size property of the two different tests. On the other hand, $\lambda > 0$ signifies a departure from the judge leniency design assumption, with higher (absolute) values indicating a more pronounced deviation. Like in the original paper, we adopt the parametrization for the fraction of always and never treated $p_n = p_a = 0.2$. Meanwhile, we vary the value of $\lambda$ within the range of 0 to 1. Note that the parameter $\lambda$ directly governs the shape of the function $\mathbb E[Y|P=p]$. The nonzero value of $\lambda$ can potentially be generated by violations of one of the three assumptions (or their combinations).

 \begin{singlespace}
\begin{figure} [!htbp]     
     \centering
     \begin{subfigure}[a]{0.49\textwidth}
         \centering
        \includegraphics[trim=10 240 10 200,scale=0.4]{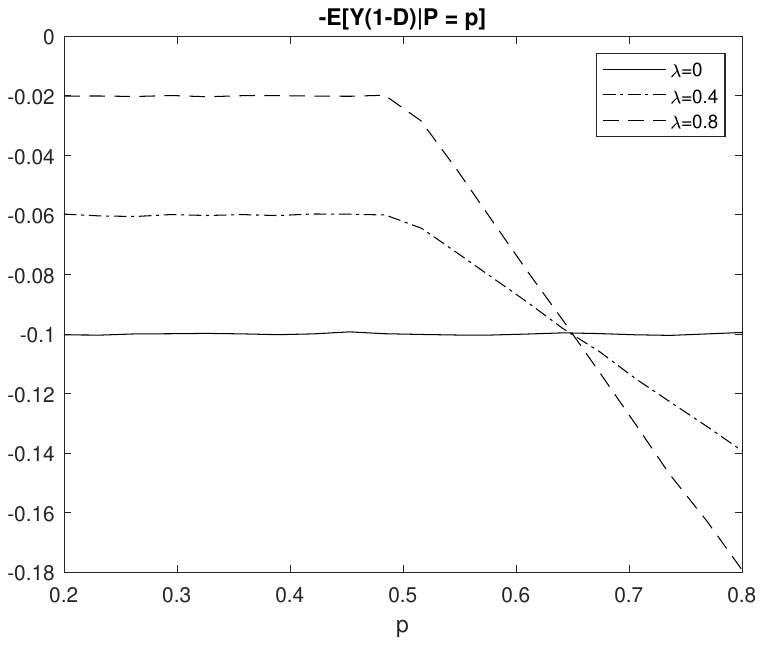} 
        \caption{$-\mathbb E[Y(1-D)|P=p]$}
    \end{subfigure}
    \hfill
   	 \begin{subfigure}[a]{0.49\textwidth}
         \centering
         \includegraphics[trim=10 240 10 200,scale=0.4]{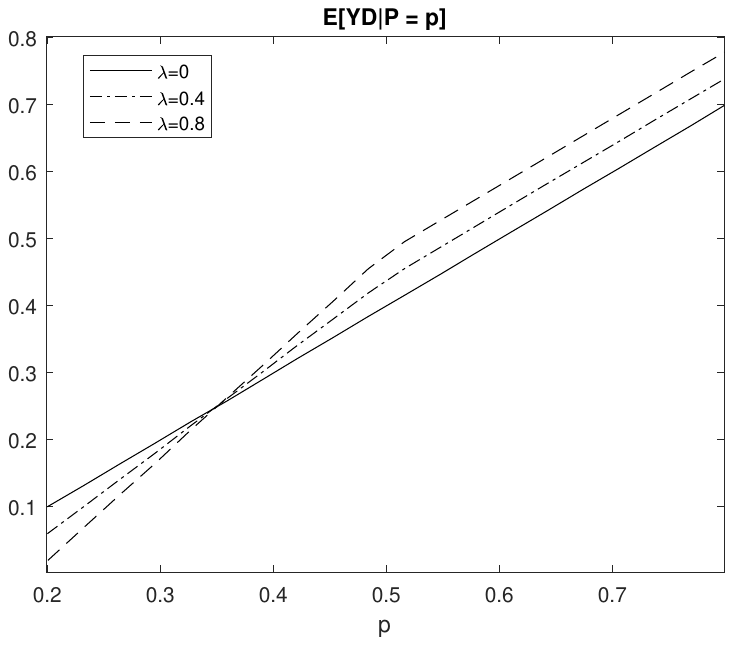} 
         \caption{$\mathbb E[YD|P=p]$}
     \end{subfigure}
     \hfill
        \caption{Testable restrictions by degree of violations of exclusion restriction}
        \label{fig1}
\end{figure}
\end{singlespace}

\Cref{fig1} visually illustrates our testable implications of the judge leniency design for the specific function $g(Y) = 1\{0<Y\leq 1\} = Y$ (because $Y$ is binary). The left and right panels of the figure, respectively, depict $ \mathbb E [-Y(1-D) \vert P=p]$ and $ \mathbb E [YD \vert P=p]$. These population quantities are approximated by a large number of defendants (1 million) for each judge. Intuitively, it is expected that $ \mathbb E [YD \vert P=p]$ and $\mathbb E [-Y(1-D) \vert P=p]$ should be non-decreasing when the judge leniency design holds. When the exclusion restriction holds, as shown in both figures with $\lambda=0$, $ \mathbb E [YD \vert P=p]$ and $ \mathbb E [-Y(1-D) \vert P=p]$ behave as expected. However, for a violation of the exclusion restriction ($\lambda = 0.4$ or $\lambda=0.8$), despite that $\mathbb E[YD|P=p]$ remains to be increasing, the other function $\mathbb E [-Y(1-D) \vert P=p]$ decreases for higher values of the propensity score. This discrepancy starkly contrasts with the implications of the judge leniency design assumptions.

\begin{figure} [h]     
     \centering
   	 \begin{subfigure}[a]{0.49\textwidth}
         \centering
         \includegraphics[trim=10 220 10 260,scale=0.3]{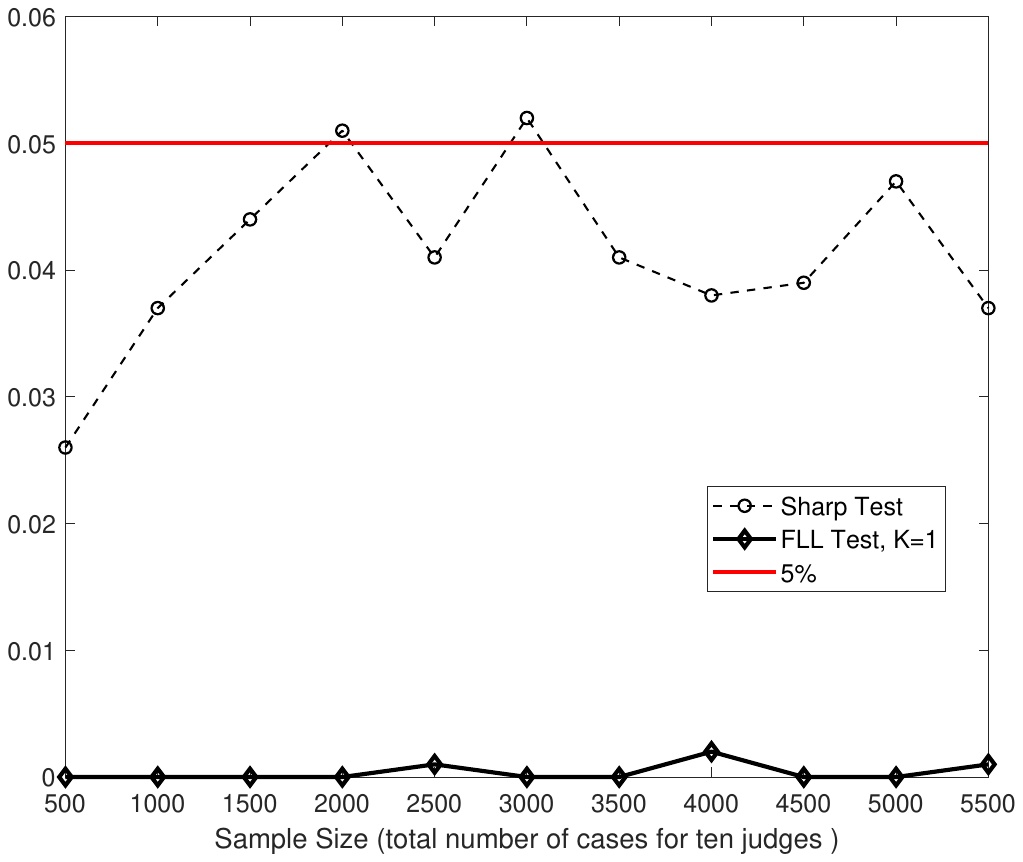} 
         \caption{Rejection rate when $\lambda = 0$}
     \end{subfigure}
     \hfill
   	 \begin{subfigure}[a]{0.49\textwidth}
         \centering
         \includegraphics[trim=10 220 10 260,scale=0.3]{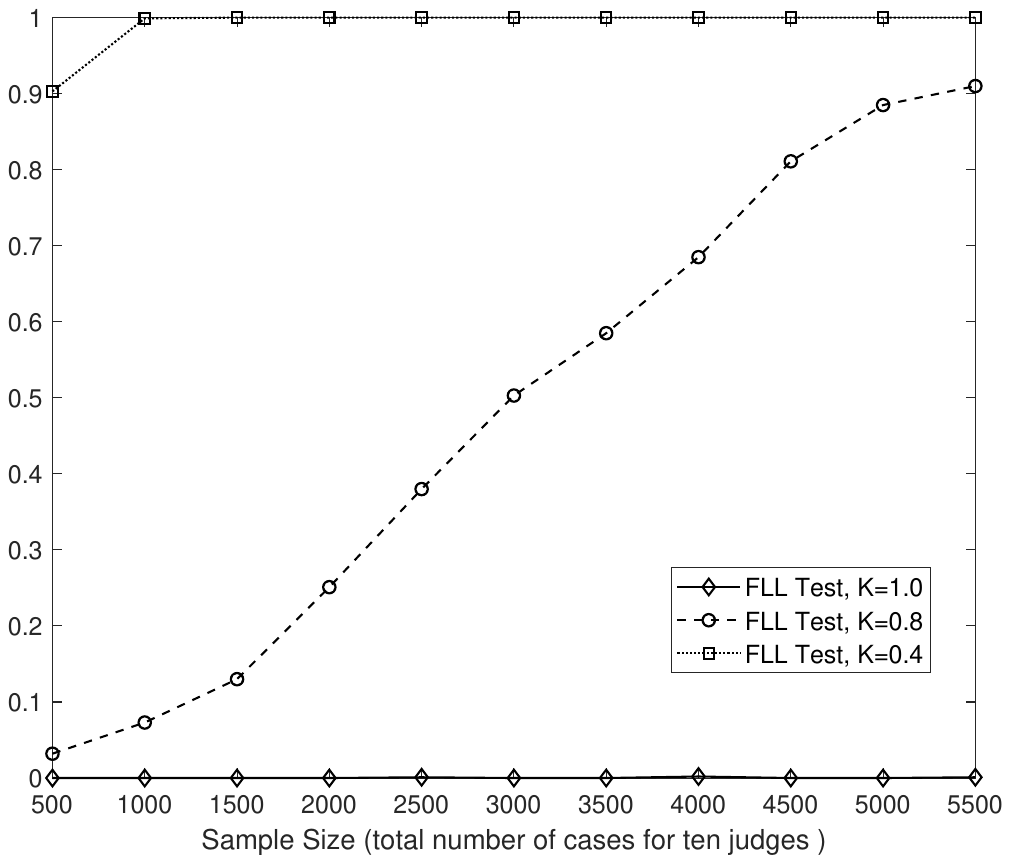} 
         \caption{Rejection rate when $\lambda = 0$}
     \end{subfigure}
     \hfill
         \begin{subfigure}[a]{0.8\textwidth}
         \centering
        \includegraphics[trim=10 280 10 260,scale=0.45]{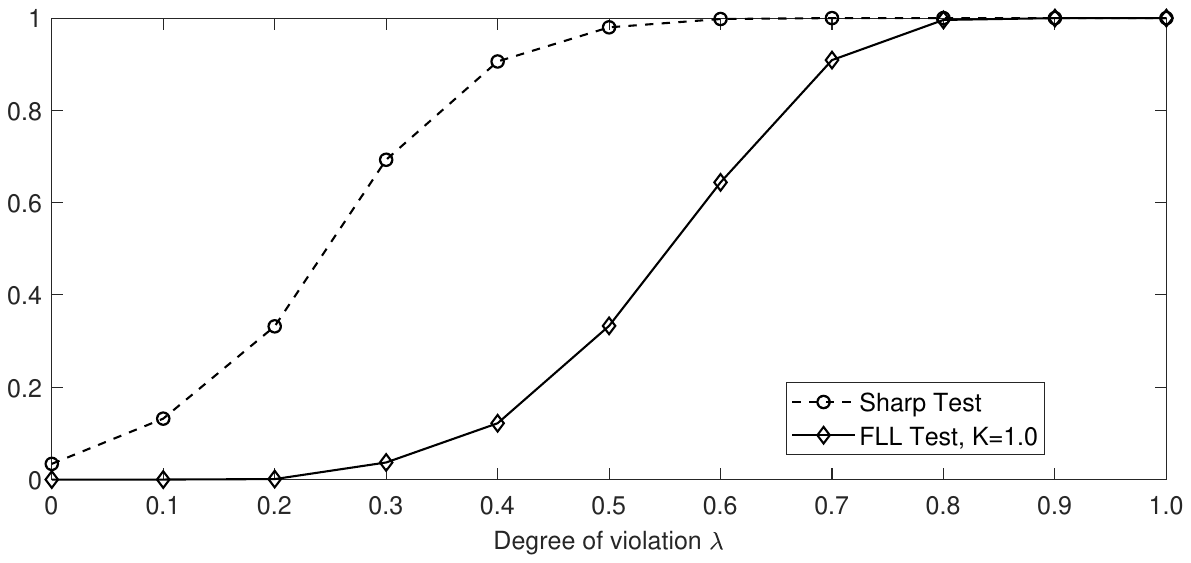} 
        \caption{Rejection rate with varying degree of violation ($n=1000$)}
    \end{subfigure}
    \hfill
        \caption{Rejection rates in FLL's DGP}
        \label{fig2}
\end{figure}

In \Cref{fig2}-(a), we report the size property for our sharp test and FFL's test at 5\% significance level (when $\lambda = 0$). The simulation designs involve twenty judges and varying sample sizes, ranging from 500 defendants (equivalent to 50 defendants per judge) to 5500 defendants (equivalent to 550 defendants per judge). The plot reveals that both tests control size well in the aforementioned DGP. Specifically, it is evident from the graph that the rejection rate of our sharp test is controlled by and close to the nominal level of 5\%. Conversely, the nonparametric test proposed in FLL consistently yields rejection rates close to zero when setting the tuning parameter $K=1$.\footnote{Recall the outcome variable is binary; hence, the largest possible absolute value for the treatment effect is $1$. These results correspond to Figures 9 and 10 in the online appendix of FLL, where the rejection probabilities are nearly zero for various sample sizes when $\lambda=0$.}

 \begin{singlespace}
\end{singlespace}

FLL discuss how one can improve the power of their testing methodology by considering more stringent upper bounds on the largest possible treatment effects (i.e., using a smaller value of $K$). For instance, in their empirical application of a binary outcome model--where the maximum treatment effect is set at 1--they advocate exploring smaller permissible maximum treatment effect values. However, if $K$ is set to be too small, then FLL's test can have server size distortion. Indeed, \Cref{fig2}-(b) graphically represents this situation by plotting the rejection rate associated with FLL's nonparametric test under two additional cases: when the maximum allowable treatment effect $K$ is set at 0.8 and 0.4, respectively. The striking observation is that the conclusions drawn from these scenarios can be misleading, as they suggest an excessive over-rejection of the assumptions even when those assumptions are indeed satisfied. For example, if one sets $K=0.4$, then the rejection rate is always $100\%$ whenever the sample size is greater or equal to $1000$. As a matter of fact, the rejection we observe from \Cref{fig2}-(b) reflects that the ad-hoc imposed magnitude of the treatment effect is not correct, but the underlying exclusion restriction holds. Our test is immune to this problem since it does not require pre-specifying the magnitude order of the unknown treatment effect. 

To assess and compare the power property of the two nonparametric tests, \Cref{fig2}-(c) plots the rejection rate as a function of $\lambda $ for 10 judges and 1000 defendants (100 defendants per judge). The solid line is the rejection rate of the FLL test, which is nearly the same as what is plotted in FLL (Appendix, Figure 10). The rejection rate achieved by our sharp test consistently surpasses that of the FLL test across the entire spectrum of exclusion restriction violations, as indicated by varying degrees of $\lambda$. As shown, the power improvement can be substantial.

\subsubsection{Continuous outcome} \label{section: continuous outcome simulation}
The second set of simulations aims to show the performance of our test in detecting violations of the judge leniency design when the outcome is continuous and unbounded. Let $(U_0,U_1,U,Z^*)\sim N(\pmb{\mu},\Sigma)$, where $\pmb{\mu} =(\mu_0 ,\mu_1 ,\mu_U ,\mu_Z)'$ is a vector of means, and $\Sigma $ is a covariance matrix. For generic random variables $A$ and $B$, let $\sigma_A^2$ be the variance of $A$ and $\rho_{A,B}$ be the correlation coefficient between $A$ and $B$. In this design, we set $\sigma_A=1$ for all $A\in\{U_1,U_0, U, Z^*\}$. We let $\rho_{U_0,U}=-0.5$, $\rho_{U_1,U}=0.5$, $\rho_{U,Z}=0$, $\rho_{U_1,U_0}=0$, $\rho_{U_1,Z}=\delta_1$, and $\rho_{U_0,Z}=\delta_1$. To create discrete judges or IV, we set 
\[
Z=F_{Z^*}^{-1}\left(\frac{\ell(Z^*)}{L}\right), \quad \ell(Z^*) = \argmin_{\ell=1,2,\cdots,L-1} \left|F_{Z^*}(Z^*) - \frac{\ell}{L}\right|.
\]
That is, we divide the support of $Z^*$ by $L$ equal-probability intervals and concentrate the mass over each interval to its nearest cutoff points. Let the potential outcomes and treatment assignment be 
\begin{align*}
    D =& 1\{\nu(X,Z) >U\}\times1\{ \delta_2 =0 \}\\
    &~~+ 
    \left[ 1\{\nu(X,Z) >U\} 1\{ U \geq U_0\} + 1\{1- \nu(X,Z) >U\} 1\{ U < U_0\}  \right]\times 1\{ \delta_2 \neq 0 \}  ,
\end{align*}
and 
\[
Y_{d}(z) = \alpha_d+X\beta_d + \delta_3 z + U_d,\quad Y_d =\sum_{z\in \mathcal Z} Y_{d}(z) 1\{Z=z\}.
\]
where $X\sim N(0,1)$ is independent of all the other random variables. We let $\nu(x,z)=z$ and set $\alpha_0=0$, and $\alpha_1=1$.The $\delta$ parameters, however, are set to be different values to capture different violations of the judge leniency design. More specifically,  
\begin{enumerate}
    \item when $\delta_1=\delta_2=\delta_3=0$, the assumptions of the judge leniency design hold;
    \item $\delta_1\neq 0$ denotes violation of the independence assumption;
    \item $\delta_{2}\neq 0$ denotes violation of the monotonicity assumption;
    In this case, the selection equation becomes
    \[
    D = 1\{Z >U\} 1\{ U \geq U_0\} + 1\{1- Z >U\} 1\{ U < U_0\},
    \]
    which indicates that there are two groups of judges, each with distinct skills (or preferences) in assigning treatment. This is in clear violation of the monotonicity assumption, which requires all judges to have the same skill \citep[][]{chan2022selection}.  
    \item $\delta_3\neq 0$ denotes violation of the exclusion restriction.
\end{enumerate}

\begin{figure}
     \centering
        \includegraphics[trim=10 260 10 280,scale=0.7]{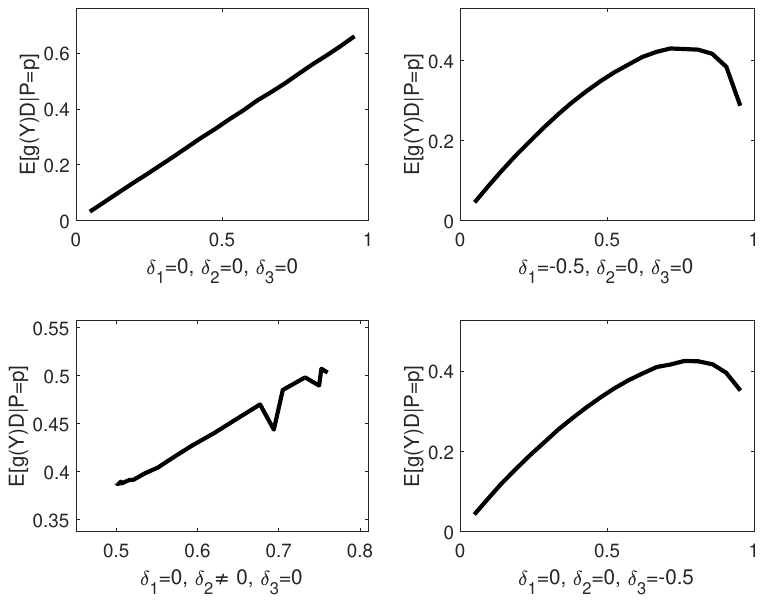} 
        \caption{Sharp Testable Restrictions for Different DGPs}
        \label{fig: novel_violations}
\end{figure}

\Cref{fig: novel_violations} plots $\mathbb E[g(Y)D|P=p]$ as a function of $p$ when $g(Y)=1\{Y\geq 0.5\}$ and 20 judges for a simple illustration. The graphs were simulated with a large sample size (over three million) and approximated the population quantity. The function is non-decreasing when all assumptions are met, as shown in the upper-left panel. In contrast, $\mathbb E [g(Y)D \vert P=p]$ deviates from the expected pattern when the judge leniency design assumptions are violated in different ways.

\Cref{fig: FLL_no_violation}, on the other hand, plots the testable implication used in FLL. The left side panels plot $\mathbb E[Y|P=p]$ for each of the $p\in\{p_1,p_2,\cdots, p_{20}\}$ (sorted in increasing order) for each of the four designs. The right panels plot the ``numerical derivative'' of the form $\frac{\mathbb E[Y|P=p_j]-\mathbb E[Y|P=p_{j-1}]}{p_{j}-p_{j-1}}$ against $\{p_2,\cdots, p_{20}\}$. The FLL testable implications require that the curves in the right-hand side panels be bounded between $[-K, K]$, where $K$ again is the difference between the upper and lower bounds of the support. Note that in this example, the outcomes have unbounded support and, therefore, $K=+\infty$. If we choose $K$ as a large number, then it is apparent that all four designs satisfy FLL's testable implication. Hence, we expect no rejection for designs 2-4, albeit they violate the identifying assumptions unless $K$ is set to be relatively small. 

\begin{figure}[!h]
     \centering
        \includegraphics[trim=10 200 10 200,scale=0.65]{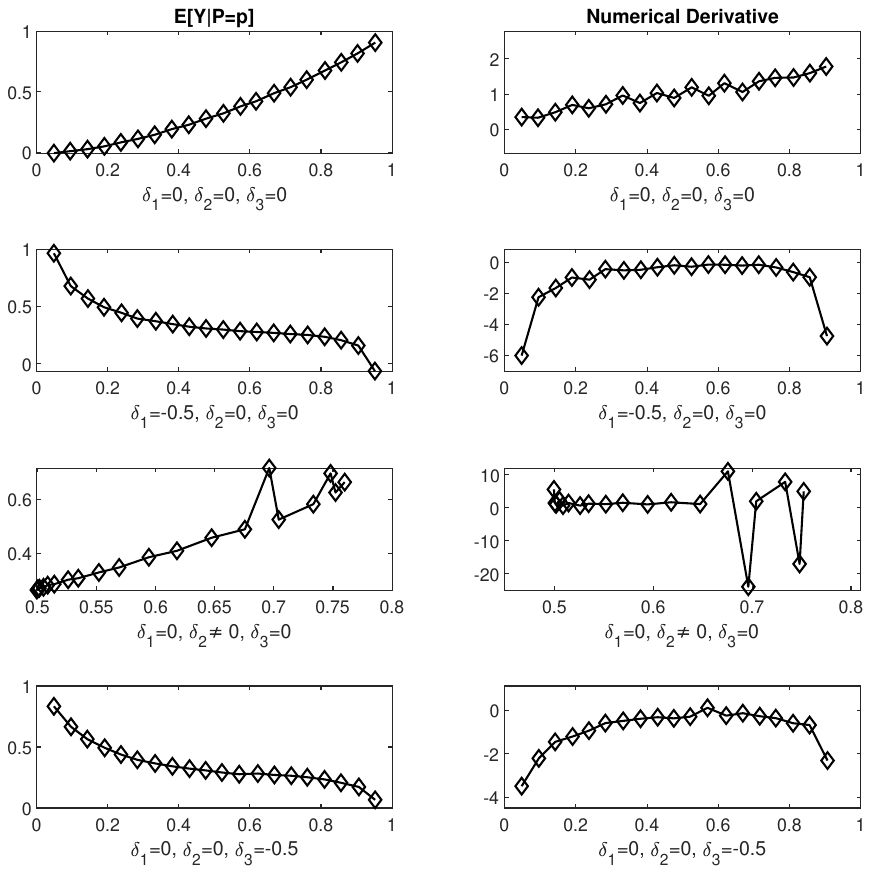} 
        \caption{FLL Testable Restrictions for Different DGPs}
        \label{fig: FLL_no_violation}
\end{figure}

We proceed by implementing our sharp test and FLL's nonparametric test. This comparison is conducted across various parameter values and sample sizes. Specifically, we consider a size design (\textbf{Size} $\delta_1=\delta_2=\delta_3=0$), violation of independence (\textbf{Power1} $\delta_1=-0.5,\delta_2=\delta_3=0$), violation of monotonicity (\textbf{Power2} $\delta_2\neq 0, \delta_1=\delta_3=0$), and violation of exclusion (\textbf{Power3} $\delta_3=-0.5, \delta_1=\delta_2=0$). For each violation, we consider situations with covariates ($\beta_1=\beta_0=1$) or without covariates ($\beta_1=\beta_0=0$ ). When there are covariates, we use \cite{carr2021testing}'s method to control for covariates, as discussed in the previous section. To implement FLL's test, we set $K$ to be the difference between sample maximum ($y_{max}$) and minimum $(y_{min})$: $\Delta_y\equiv y_{max} - y_{min}$. We also consider $K=\frac{\Delta_y}{8}$ and $K=\frac{\Delta_y}{16}$. The results are summarized in \Cref{table: types of violations no x}.

\begin{table}[!h]
\caption{Rejection Rate under Different Types of DGPs}
\begin{center}
\begin{tabular}{lccc|ccc} \label{table: types of violations no x}
& \multicolumn{3}{c}{$\delta_1=\delta_2=\delta_3=0$ (\textbf{Size})}& \multicolumn{3}{c}{$\delta_1=-0.5, \delta_2=\delta_3=0$ (\textbf{Power1})}  \\
Without Covariates&$n=500$&$n=1000$  &$n=2000$&$n=500$&$n=1000$  &$n=2000$ \\\hline\hline
	Sharp Test	&	0.000  &  0.000   & 0.000	&	 0.436   &  0.848  &  0.995		\\
	FLL-nonp, $K=\Delta_y$	&	0.000  &  0.000  &  0.000	&	0.000 &   0.000  &  0.000	\\
	FLL-nonp, $K=\frac{\Delta_y}{8}$	&	0.007  &  0.001  &  0.018	&	0.015  &  0.054 &   0.129	\\
 FLL-nonp, $K=\frac{\Delta_y}{16}$	&	0.064  &  0.284  &  0.719	&	0.101  &  0.376 &   0.839	\\\hline
 &&\\
 & \multicolumn{3}{c}{$\delta_2\neq 0, \delta_1=\delta_3=0$ (\textbf{Power2})}& \multicolumn{3}{c}{$\delta_3=-0.5, \delta_1=\delta_2=0$ (\textbf{Power3})}  \\
Without  Covariates&$n=500$&$n=1000$  &$n=2000$&$n=500$&$n=1000$  &$n=2000$ \\\hline\hline
	Sharp Test	&	0.374  &  0.734   & 0.942	&	 0.183   &  0.503  &  0.902		\\
	FLL-nonp, $K=\Delta_y$	&	0.000 &   0.000  &  0.000	&	0.000 &   0.000  &  0.000	\\
	FLL-nonp, $K=\frac{\Delta_y}{8}$	&	0.015  &  0.037  &  0.079	&	0.005  &  0.004 &   0.008	\\
 FLL-nonp, $K=\frac{\Delta_y}{16}$	&	0.065  &  0.104  &  0.322	&	0.019  &  0.049 &   0.107	\\\hline

 &\\
 && \\
 & \multicolumn{3}{c}{$\delta_1=\delta_2=\delta_3=0$ (\textbf{Size})}& \multicolumn{3}{c}{$\delta_1=-0.5, \delta_2=\delta_3=0$ (\textbf{Power1})}  \\
With Covariates &$n=500$&$n=1000$  &$n=2000$&$n=500$&$n=1000$  &$n=2000$ \\\hline\hline
	Sharp Test	&	0.000  &  0.000   & 0.000	&	 0.424   &  0.821  &  0.993		\\
	FLL-nonp, $K=\Delta_y$	&	0.000  &  0.000  &  0.000	&	0.000 &   0.000  &  0.000	\\
	FLL-nonp, $K=\frac{\Delta_y}{8}$	&	0.003  &  0.000  &  0.000	&	0.029  &  0.018 &   0.041	\\
 FLL-nonp, $K=\frac{\Delta_y}{16}$	&	0.069  &  0.113  &  0.293	&	0.084  &  0.173 &   0.456	\\\hline
 &&\\
 & \multicolumn{3}{c}{$\delta_2\neq 0, \delta_1=\delta_3=0$ (\textbf{Power2})}& \multicolumn{3}{c}{$\delta_3=-0.5, \delta_1=\delta_2=0$ (\textbf{Power3})}  \\
With Covariates &$n=500$&$n=1000$  &$n=2000$&$n=500$&$n=1000$  &$n=2000$ \\\hline\hline
	Sharp Test	&	0.345  &  0.714   & 0.936	&	 0.167   &  0.488  &  0.902		\\
	FLL-nonp, $K=\Delta_y$	&	0.000 &   0.000  &  0.000	&	0.000&   0.000  &  0.000	\\
	FLL-nonp, $K=\frac{\Delta_y}{8}$	&	0.006  &  0.013  &  0.022	&	0.004  &  0.002 &   0.001	\\
 FLL-nonp, $K=\frac{\Delta_y}{16}$	&	0.050  &  0.075  &  0.225	&	0.018  &  0.017 &   0.042	\\\hline
\end{tabular}
\end{center}
\end{table}

Regarding the size property, all tests control the size except the FLL test when $K$ is set to be very small. Our test and the FLL test with $K=\Delta_y$ and $K=\frac{\Delta_y}{8}$ are conservative. When one sets $K=\frac{\Delta_y}{16}$, the rejection probability of FLL's test increases quickly even when all the assumptions are satisfied (the first design). This is unsurprising because a very small $K$ essentially introduced another severe misspecification to the model. However, when examining the power property of the three tests, we see clearly that our test outperforms FLL's tests by a large margin. The proposed sharp test has enough power to detect the violation of any of the three assumptions (independence, exclusion, and monotonicity). In particular, the rejection rates for our sharp test quickly increase with sample size, surpassing 90\% for all cases when the sample size reaches $2000$ (or $100$ cases per judge). Note that in this simulation, the parametric form of the propensity score is correctly specified (except for Power2 when monotonicity is violated); hence, the high power of our test is not because of misspecification of $P(z,\theta_0)$. In contrast, FLL's test has low power performance unless we set $K$ as a small value, which, on the other hand, induces size distortion.

\Cref{table: degrees of violations no x} further examines how the rejection frequency varies as the ``magnitude of violation varies" for independence and exclusion.  For this exercise, we focus on sample size $n=1000$ (50 cases per judge). Not surprisingly, when the magnitude of the violation is small, all tests have low power. However, as the degree of violation increases, the power of our sharp test rises quickly, even quicker than the FLL's nonparametric test with $K=\frac{\Delta_y}{16}$. On the other hand, when $K=\Delta_y$, FLL's nonparametric test does not reject even if the degree of violation is substantial. Again, this table demonstrates that sharp testable implications are desirable in practice.

\begin{table}[!h]
\caption{Rejection Rate under Different Levels of Violations (No Covariates)}
\begin{center}
\begin{tabular}{lcccc} \label{table: degrees of violations no x}
$\delta_2=\delta_3=0$, $n=1000$&$\delta_1=-0.1$&$\delta_1=-0.3$ &$\delta_1=-0.5$&$\delta_1=-0.7$ \\\hline\hline
	Sharp Test	&	0.001  &  0.085   & 0.825	&	 1.000    	\\
	FLL-nonp, $K=\Delta_y$	&	0.000  &  0.000  &  0.000	&	0.000   	\\
	FLL-nonp, $K=\frac{\Delta_y}{8}$	&	0.001  &  0.004  &  0.054	&	  0.911	\\
 FLL-nonp, $K=\frac{\Delta_y}{16}$	&	0.026  &  0.006  &  0.397	&	0.917 \\\hline
 &&\\
$\delta_1=\delta_2=0$, $n=1000$&$\delta_3=-0.1$&$\delta_3=-0.3$ &$\delta_3=-0.5$&$\delta_3=-0.7$ \\\hline\hline
	Sharp Test	&	0.000  &  0.069   & 0.471	&	 0.931   	\\
	FLL-nonp, $K=\Delta_y$	&	0.000 &   0.000  &  0.000	&	0.000   	\\
	FLL-nonp, $K=\frac{\Delta_y}{8}$	&	0.000  &  0.000  &  0.005	&	0.114  	\\
 FLL-nonp, $K=\frac{\Delta_y}{16}$	&	0.027  &  0.002  &  0.032	&	0.798  	\\\hline
\end{tabular}
\end{center}
\end{table}

\subsection{Empirical illustration}

In this subsection, we employ our test to assess the validity of the judge leniency designs using data from \cite{stevenson2018distortion}; see also \cite{cunningham2021causal}, who studies the impact of pretrial detention on conviction. Using Philadelphia court records and leveraging the varying leniency of bail magistrates as an instrumental variable, the author discovers that pretrial detention leads to a 13\% increase in the likelihood of conviction. 

In the Philadelphia court system, following an arrest, individuals are taken to one of seven city police stations for a video conference interview by Pretrial Services, which assesses risk factors and financial details for public defense eligibility. Utilizing this information, Pretrial Services assigns arrestees to a bail recommendation grid. Bail hearings, conducted by magistrates every four hours via video conference, involve a brief process where charges are explained, next court appearances are specified, eligibility for a court-appointed defense attorney is determined, and bail amounts are set based on arrest details, interviews, criminal history, guidelines, and input from representatives. Magistrates hold broad authority to assign bail, which can fall into categories such as release without payment, cash bail with a 10\% deposit, or no bail at all. 

\cite{stevenson2018distortion}'s research design leverages the varying magistrate tendencies to assign affordable bail as an instrument to study detention's impact on case outcomes. To answer the research questions, the author utilizes data from the court records of the Pennsylvania Unified Judicial System, obtained through web scraping of public records in PDF format, which are then transformed for statistical analysis. The dataset encompasses arrests in Philadelphia, where charges were filed between September 13, 2006, and February 18, 2013. The final dataset includes 331,971 cases and eight \emph{randomly} assigned judges, with each observation pertaining to a specific criminal case. As noted in \cite{stevenson2018distortion}, the shift-rotation system at the Philadelphia court forms the basis for such randomness.

\begin{figure} [!h] 
     \centering
     \begin{subfigure}[b]{0.49\textwidth}
         \centering
        \includegraphics[scale=0.7,width=0.8\linewidth]{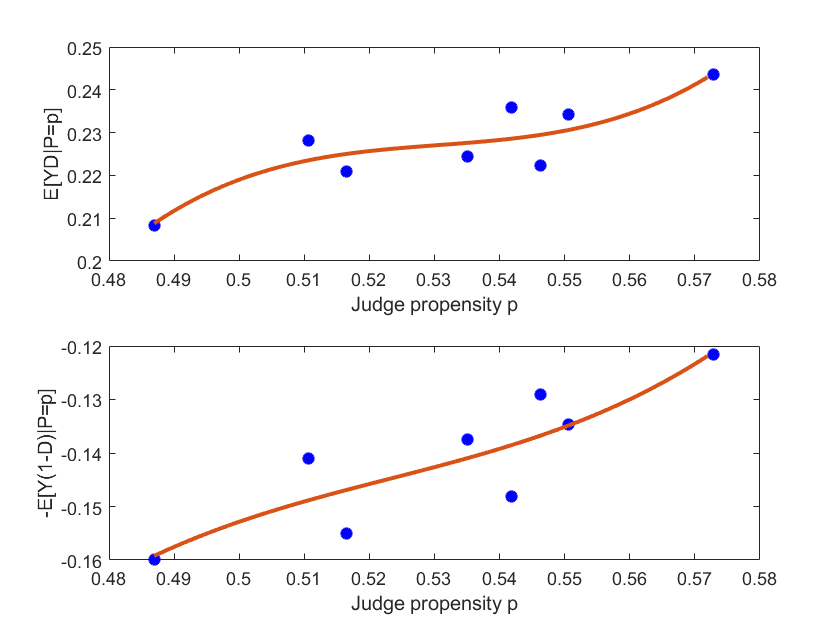} 
        \caption{Aggressive assault}
    \end{subfigure}
    \hfill
   	 \begin{subfigure}[b]{0.49\textwidth}
         \centering
         \includegraphics[scale=0.7,width=0.8\linewidth]{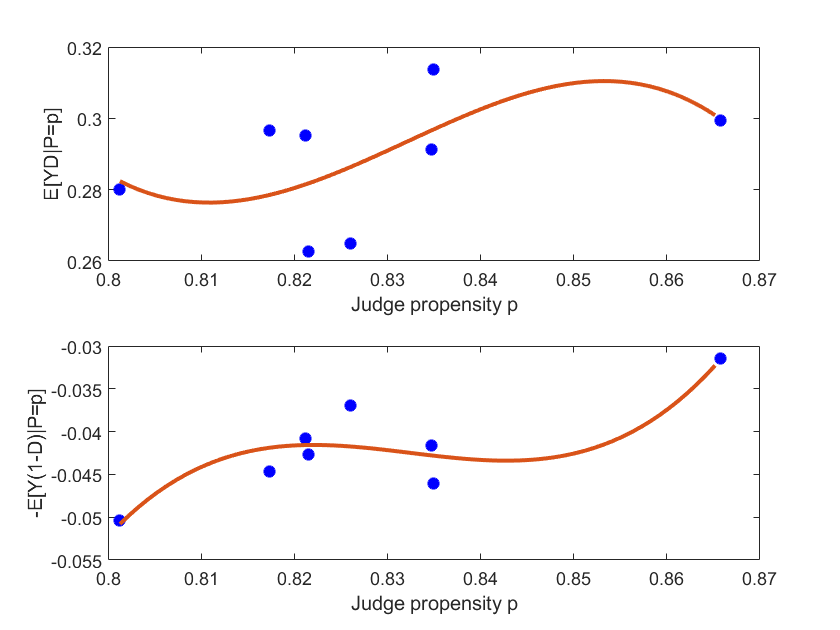} 
         \caption{Robbery}
     \end{subfigure}
     \hfill
     \begin{subfigure}[b]{0.49\textwidth}
         \centering
        \includegraphics[scale=0.7,width=0.8\linewidth]{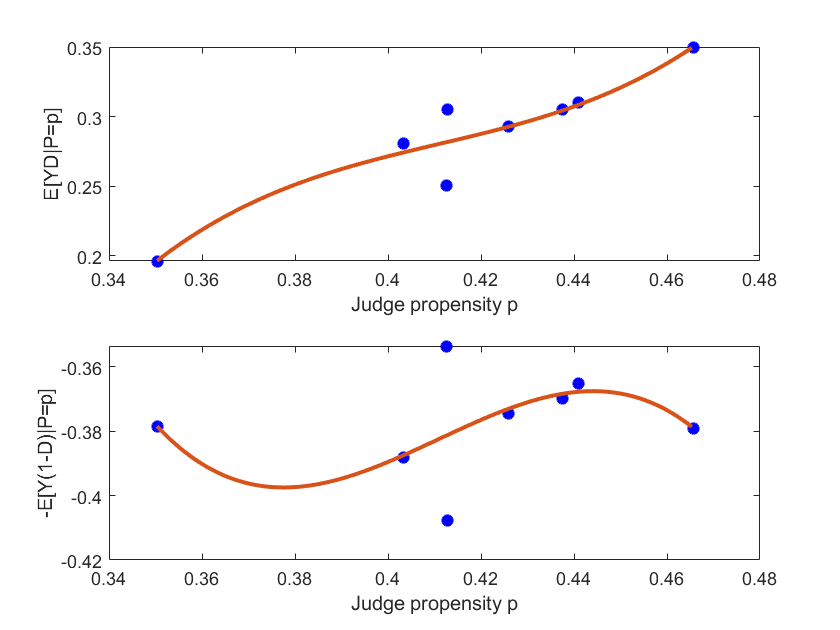} 
        \caption{Drug sell}
    \end{subfigure}
    \hfill
   	 \begin{subfigure}[b]{0.49\textwidth}
         \centering
         \includegraphics[scale=0.7,width=0.8\linewidth]{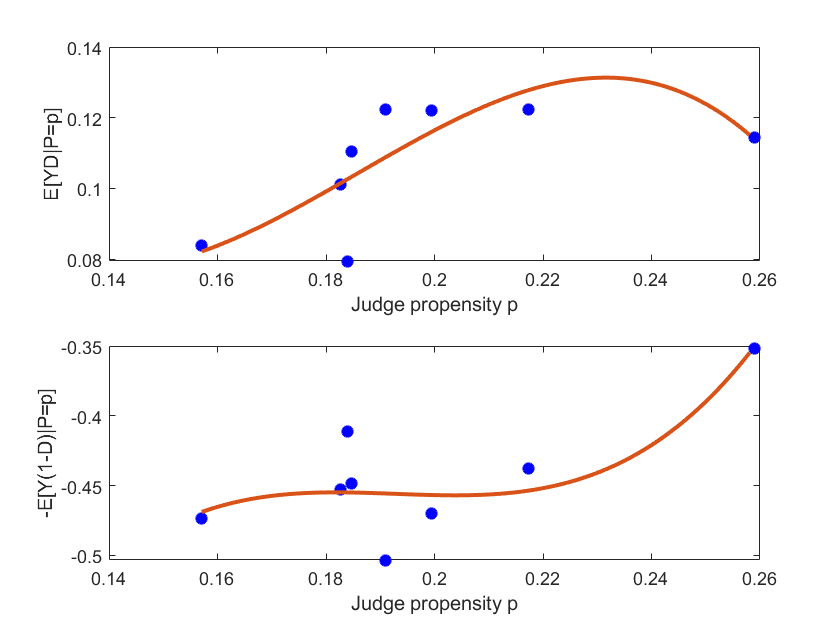} 
         \caption{Drug possession}
     \end{subfigure}
     \hfill
        \caption{Testable restrictions by case types}
        \label{fig5}
\end{figure}

In what follows, we focus on the aggregate dataset (all criminal cases together) and four primary categories of criminal cases in the data: aggressive assault, robbery, drug sale, and drug possession. These four criminal cases we consider in isolation constitute 43\% of the total cases. In \Cref{fig5}, we present two scatter plots for each crime category: $\{(p_j, \mathbb E[YD|P=p_j])\}_{j=1}^8$ and $\{(p_j, \mathbb -E[Y(1-D)|P=p_j])\}_{j=1}^8$, along with a fitted polynomial to illustrate whether the anticipated implications of the judge leniency design framework are satisfied for the considered categories of criminal cases. The graphs indicate $\mathbb E [YD \vert P=p]$ and $\mathbb E [-Y(1-D) \vert P=p]$ are most likely to be non-decreasing for the aggressive assault case.\footnote{Note all the outcome variables are binary. Therefore, the close interval we use for the \Cref{Thm: sharpMTE} is $1\{0<Y\leq 1\}$, which equals to $Y$.} The non-decreasing shape of the functions is unclear for the other types of criminal categories. Although this graphical representation does not constitute a formal test, it offers an intuitive insight. Specifically, it suggests that the assumptions are the least likely to be violated in the aggressive assault case, while the drug possession case shows the highest likelihood of violating the assumptions of the judge leniency design.

We observe a relatively large set of covariates, including fixed effects for year, month, and day of the week. We, consequently, implement the semi-parametric version of our test. For comparison, we also implement FLL's nonparametric and semi-parametric tests.
The results of the three tests are presented in \Cref{table: empirical application} for both the aggregate dataset and separately for each of the four crime categories aforementioned. The nonparametric test introduced by FLL indicates the validity of judge leniency design cannot be rejected either conditioning on each crime category or the aggregate data set at 10\% level, despite that the shape of $\mathbb E [YD \vert P=p]$ and $\mathbb E [-Y(1-D) \vert P=p]$ for the drug possession type suggests the opposite. In contrast, our novel test yields results that align with expectations. For instance, the sharp test does not indicate a rejection of the validity of the judge leniency design assumptions for the aggressive assault. However, for all three other types of offenses, our test rejects the validity of the judge leniency design. Meanwhile, FLL's semi-parametric test rejects the category of aggregate assault.\footnote{For FLL's semi-parametric test, we fit the regression function $\mathbb E[Y|P=p]$ by B-spline with three knots. The results for other numbers of knots are reported in the appendix. The reported p-value is the ``combined p-value'' of the fit component and slope component of the test, and we can see from \Cref{table: FLL p values} in the online appendix that the rejection is mostly generated by the fit component.} These results suggest that using the Wald estimand or the MTE approach for those cases will lead to inconsistent estimates of the causal effects of interest.

Finally, we see no evidence to refute the assumptions underpinning the judge leniency design when applying our sharp test to the aggregate dataset. This outcome may be influenced by the notably high proportion of aggressive assault cases within the dataset compared to other categories. Our result also ascertains that the exclusion restriction or monotonicity can hold for some crime categories but not others, suggesting that controlling the crime type is important in practice. 
\begin{table}[h]
\caption{Testing Judge Leniency Design: p-values}
\begin{center}
\begin{tabular}{lccc} \label{table: empirical application}
&Sharp Test&FLL-Nonp&FLL-Semip \\\hline\hline
	All	&	0.821  &  0.056   & 	0.114 	\\
        Aggressive assault	&	0.913  &  0.996   & 0.015	  	\\
	Robbery &	0.033  &  1.000  &  0.109	 	\\
	Drug sale&	0.005  &  0.116  &  0.180	\\
 Drug possession&	0.000  &  0.929  &  0.610	\\\hline
\end{tabular}
\end{center}
{\footnotesize \textit{Notes}: This table reports the results of the statistical tests using \cite{stevenson2018distortion}'data, including time fixed effects as controls. Specifically, the considered controls are year, month, and day-of-the-week fixed effects. \emph{Sharp Test} stands for our novel semi-parametric test developed in this paper, while \emph{FLL-Nonp} and \emph{FLL-Semip} represent the nonparametric and semi-parametric tests of FLL (three knots B-spline), respectively.}
\end{table}

\section{Salvage the Model under Weaker Assumptions} \label{Section: After the test}

The rejection of the sharp test means that the judge's leniency design assumptions are too stringent for the data. In this case, relaxing some of these assumptions is required to salvage the model. There are different ways to relax a model's assumptions. One way is to maintain the same estimand used in the stringent model and ask under what conditions this estimand can still be interpreted causally. The model relaxation recently entertained by FLL falls into this second approach, providing alternative conditions under which the 2SLS could still have a causal interpretation when \Cref{ass: MTEI,ass: Exclusion,ass: Vyt-Mon} are too stringent for the data. In \Cref{section: average exclusion and monotone}, we revisit the average exclusion assumption proposed by FLL and show it is a special case of a zero-covariance condition: a restriction that may not always be justifiable in all empirical settings. 

There is another approach that focuses on a well-defined policy-relevant parameter and examines how this parameter could be point-identified or set-identified using weaker and more credible assumptions. In such a case, the parameter of interest remains the same, but the (set) estimands may vary depending on the credible assumptions one would be willing to maintain. We will discuss this approach in \Cref{section: partial exclusion and monotone}.\footnote{In this section, we mainly focus on the case in which the exclusion or monotonicity assumption is violated. When the random assignment assumption is violated, one can consider a partial identification, see \cite{mourifie2025layered}.}

\subsection{Average Exclusion and Monotonicity} \label{section: average exclusion and monotone}
We first revisit the average exclusion and monotonicity conditions. For simplicity, suppose $Z$ has finite support as in FLL such that $\mathcal Z=\{1,2,\cdots, J\}$. The general form of the potential outcome model is,
\[
    Y=\tilde{Y}_{1} D + \tilde{Y}_{0} (1-D) , \quad \tilde{Y}_{d}= \sum_{z \in \mathcal Z}Y_{d}(z) 1\{Z=z\} ,\quad D=\sum_{z \in \mathcal Z}D_{z}1\{Z=z\}.
\]

FLL proposes to relax \Cref{ass: Exclusion} and \Cref{ass: Vyt-Mon} with the average exclusion restriction and the average monotonicity assumption, respectively:
\begin{assumption}  \label{ass: Frand}
 Let $\lambda_{z} =\Pr(Z=z)$, $p_{z} = \mathbb{E}[D_{z}]$, $p = \sum_{z \in \mathcal Z} \lambda_{z} p_{z} $, $\bar{D} = \sum_{z \in \mathcal Z} \lambda_{z} D_{z}$, and $\bar{Y}_{d} = \sum_{z \in \mathcal Z} \lambda_{z} Y_{d}(z)$ for $d \in \{0,1\}$. 
 \begin{enumerate}
                \item[(a)] Average exclusion restriction: $$ \mathbb{E} \left[ \sum_{z \in \mathcal Z} \lambda_{z} \left( p_{z} - p \right) \left\{ ( Y_{0}(z) - \bar{Y}_{0} ) (1-D_{z}) + ( Y_{1}(z) - \bar{Y}_{1} ) D_{z} \right\}  \right] =0.$$
        \item[(b)] Average monotonicity: $\omega \equiv \sum_{z \in \mathcal Z} \lambda_{z} \left( p_{z} - p \right) \left( D_{z} - \bar{D} \right) \geq 0$ almost surely. 
    \end{enumerate}
\end{assumption}  
Under \Cref{ass: MTEI,ass: Frand}, FLL's Theorem 3 shows that the 2SLS estimand (of using $P(Z)$ as IV) has a causal interpretation since it can be written as a weighted average of the following treatment effect $\delta = \bar{Y}_{1} - \bar{Y}_{0}$ i.e., 
\begin{equation} \label{eq: 2SlS estimand}
    \frac{\text{Cov}(Y,P(Z))}{\text{Cov}(D,P(Z))} = \mathbb{E}\left[ \frac{\omega}{\mathbb{E}[\omega]} \delta\right].
\end{equation} 
 Note that the $\delta$ in \Cref{eq: 2SlS estimand} is a deterministic function of the collection of potential outcomes $\{Y_{d}(z)\}_{d=0,1,z\in\mathcal Z}$. The 2SLS estimand is causal because it is a weighted average of $\delta$ and the weight $\omega$ is positive by the average monotonicity (\Cref{ass: Frand}-(b)).

\Cref{lem:1} below provides a generalization and more transparent discussion of the FLL's Theorem 3. First, we demonstrate that the average exclusion assumption is essentially equivalent to a zero covariance condition. Second, we show that \Cref{eq: 2SlS estimand} indeed holds for any deterministic function of $\{Y_{d}(z)\}_{d=0,1,z\in\mathcal Z}$, not just for $\delta$. 

To clarify these points, let us define $\alpha_{z} \equiv Y_{1}(z) - Y_{0}(z) $, and $ \tilde{\alpha} = \sum_{z \in \mathcal Z} \alpha_{z} 1\{Z=z\} =Y_1(Z)-Y_0(Z)$. Let $\alpha \equiv h(Y_{1}(1),...,Y_{1}(J),Y_{0}(1),...,Y_{0}(J))$ be an arbitrary measurable deterministic function of the collection of potential outcomes. $\delta$ defined in \Cref{ass: Frand} is a special case when we pick $h(Y_{1}(1),...,Y_{1}(J),Y_{0}(1),...,Y_{0}(J)) = \bar{Y}_{1} - \bar{Y}_{0} $. One could instead be interested in different treatment effects specific to each judge: $\alpha = \alpha_z$, $z=1,2,\cdots, J$. $\alpha$ can also be a quantity without clear economic interpretation such as $\alpha=\sum_{z \in \mathcal Z}zY_{1}(z)$.

\begin{proposition} \label{lem:1} 
\begin{enumerate}
\item [ ]
\item [(a)] Under \Cref{ass: MTEI}, and \Cref{ass: Frand}(b), the following equation holds for any measurable deterministic function $\alpha$ of the collection of potential outcomes:
\begin{equation*} \label{eq3:1}
        \frac{\text{Cov}(Y,P(Z))}{\text{Cov}(D,P(Z))} = \mathbb{E}\left[ \frac{\omega}{\mathbb{E}[\omega]} \alpha \right] + \frac{\text{Cov}\left( (\tilde{\alpha} - \alpha) D + \tilde{Y}_{0},P(Z)\right)}{\mathbb{E}\left[ \omega \right]}
    \end{equation*}
    where $\tilde Y_d = \sum_{z \in \mathcal Z}Y_{d}(z) 1\{Z=z\}$. 
\item [(b)] Under \Cref{ass: MTEI}, for $\alpha= \bar{Y}_{1} - \bar{Y}_{0} = \sum_z \lambda_z (Y_{1}(z)-Y_{0}(z))=\sum_z\lambda_z\alpha_z$ we have:
$$\text{Cov}\left( (\tilde{\alpha} - \alpha) D + \tilde{Y}_{0},P(Z)\right)=\mathbb{E} \left[ \sum_{z \in \mathcal Z} \lambda_{z} \left( p_{z} - p \right) \left\{ ( Y_{0}(z) - \bar{Y}_{0} ) (1-D_{z}) + ( Y_{1}(z) - \bar{Y}_{1} ) D_{z} \right\}  \right].$$
\end{enumerate}
\end{proposition}

The proof for the proposition is collected in \Cref{section: proof of average exclusion and monotone}. Under \Cref{ass: MTEI} (independence), \Cref{lem:1}(b) shows that the average exclusion restriction of FLL is, indeed, a special case of the zero-covariance assumption when $\alpha=\delta$. \Cref{lem:1}(a) further shows that if one targets an arbitrary quantity $\alpha \equiv h(Y_{1}(1),...,Y_{1}(J),Y_{0}(1),...,Y_{0}(J))$, and if one is willing to impose the same zero-covariance assumption on $\alpha$:
\begin{equation}\label{eq:ZC}
\text{Cov}\left( (\tilde{\alpha} - \alpha) D + \tilde{Y}_{0},P(Z)\right)=0,
\end{equation}
then one can always interpret 2SLS estimand as the weighted average of $\alpha$ with positive weights under the average monotonicity \Cref{ass: Frand}(b). This happens because the above zero-covariance condition in \Cref{eq:ZC} is a reduced-form condition, which assumes that the correlation between a reduced-form error (involving the parameter of interest) and the propensity score, i.e. $\text{Cov}\left( Y - \alpha D,P(Z)\right)=0$.

How does one assess the plausibility of the average exclusion condition? FLL provides a heuristic argument.\footnote{\citet[][page 19]{frandsen2023judging}:  ``\textit{Average exclusion can be probed by examining the correlation between judge-level treatment propensity and judge-level averages of alternative channels through which judges may affect outcomes if such channels are observed. Average exclusion
may be more plausible if these correlations are near zero.}"} However, this argument could also be invoked by anyone who wants to impose that \Cref{eq:ZC} holds for other $\alpha\neq \delta$. Also, it is difficult to justify why $\text{Cov}\left( (\tilde{\delta} - \delta) D + \tilde{Y}_{0},P(Z)\right)=0$ but $\text{Cov}\left( (\tilde{\alpha} - \alpha) D + \tilde{Y}_{0},P(Z)\right) \neq 0$ for other $\alpha\neq \delta$. 

Furthermore, it is worth noting that the average exclusion restriction is not invariant to a relabelling of the treatment. In other terms, this assumption may hold if the researcher defines the treatment as $D$ equals $1$ if incarceration and $0$ if not, while it may not hold if the researcher recodes the treatment as $D$ equals $1$ if no incarceration and $0$ if incarceration. Indeed, after a relabelling, the zero-covariance in \Cref{eq:ZC} becomes:
$\text{Cov}\left( (\tilde{\alpha} - \alpha) D + \tilde{Y}_{1},P(Z)\right)=0$. It follows that the average exclusion assumption is invariant to a relabelling if and only if  $\text{Cov}\left(\tilde{Y}_{1},P(Z)\right)=\text{Cov}\left(\tilde{Y}_{0},P(Z)\right)$. 

Despite all those discussed above, we do observe a direct way to assess the validity of \Cref{ass: MTEI,ass: Frand}. In fact, under these assumptions, there is:
\begin{equation*} 
       \left \vert  \frac{\text{Cov}(Y,P(Z))}{\text{Cov}(D,P(Z))}\right \vert 
       \leq \mathbb{E}\left[ \left \vert \frac{\omega}{\mathbb{E}[\omega]} \delta \right \vert \right] \leq 
       \mathbb{E}\left[ \left \vert  \delta \right \vert \right] \leq U-L
\end{equation*}
where $U$ and $L$ are, respectively, the upper and lower bounds of $\mathcal Y$. Therefore, if the support of the outcome is bounded from both above and below, then the absolute value of the 2SLS estimand must also be bounded.  

\subsection{Conditioning on Judge's Characteristics}\label{section: partial exclusion and monotone}
In practice, it is not uncommon for researchers to have good reason to believe the assumptions hold after controlling for the judge's specific characteristics. We explore this idea in this section and demonstrate that it is closely related to the partial exclusion assumption (defined below) and the partial monotonicity assumption made in \cite{mogstad2019identification}. Specifically, we decompose $Z$ into two components: $Z_I$ and $Z_c$, and we assume the monotonicity and exclusion restriction hold conditionally on $Z_c$. Here, $Z_c$ can be a judge's race or political party, and $Z_I$ is a vector of the remaining characteristics.

\begin{assumption}[Partial Exclusion] \label{ass:PExclusion}
Let $Z\equiv(Z_I', Z_c')'$. For $d \in \{0,1\}$, $Y_{d}(z)=Y_d(z_c)$ for all $z \in \mathcal Z$.
\end{assumption}

\begin{assumption}[Partial Monotonicity]\label{ass:PVyt-Mon}
For any $(z_I,z_c)$ and $(z_I',z_c)$ $\in \mathcal Z\times \mathcal Z$ 
either $D(z_I,z_c)\geq D(z_I',z_c)$ for all defendants or $D(z_I,z_c)\leq D(z_I',z_c)$ for all defendants.
\end{assumption}

The partial exclusion assumption relaxes \Cref{ass: Exclusion} and allows the potential outcomes to depend on the subvector $Z_c$. For instance, when the treatment of interest is incarceration, judges could assign and differ in other punishments, such as probation, fines, or sentence length. These other punishments could directly affect potential outcomes, making \Cref{ass: Exclusion} unlikely. Minority judges may be less lenient in their sentence length than their majority counterparts \citep{johnson2014judges}. Beyond the decision to incarcerate, different sentence lengths may have divergent effects on future labor market outcomes. If the sentence length is not observed or controlled, we would expect the potential outcome to depend on whether a judge is a minority judge through this channel. The partial exclusion assumption states that whether and how the judge assigns other types of punishment depends only on a subset of the judge's observable characteristics ($Z_c$), but not on others ($Z_I$). In other words, a defendant will end up with the same pair of potential outcomes $(Y_1(z_c), Y_0(z_c))$ as long as he or she is assigned to judges with the same observed characteristics $Z_c=z_c$. Finally, when the only instrument variable we observe in the data is the identity of the judge $Z_I$, then the partial exclusion assumption is equivalent to the original exclusion \Cref{ass: Exclusion}. 

The partial monotonicity \Cref{ass:PVyt-Mon} was initially introduced in \cite{mogstad2019identification}. It significantly weakens \Cref{ass: Vyt-Mon} since it does not require comparing the level of leniency across judges with different observable characteristics. 
For instance, let $Z_c=(Z^R_c, Z^P_c)$ be composed of the following binary variables: $Z^R_c$ equal to $1$ if the judge is black or Hispanic and 0 if not, while $Z^P_c$ is $1$ if the judge is from the Republican party and $0$ if from the Democratic party. 
Imposing \Cref{ass: Vyt-Mon} means it is not possible to have a black democrat judge be more lenient than a white republican judge for some defendants while being less lenient for other defendants, i.e., these two judges may have different cut-off points, but they rank all the defendants in the same order. Mathematically, we can not have both $\mathbb P(D(z_I,1,0)=1,D(z_I',0,1)=0)>0$ and $\mathbb P(D(z_I',0,1)=1, D(z_I,1,0)=0)>0$. However, there is a large body of empirical evidence of heterogeneity in the ranking of judges' leniency across different types of offense or defendants \citep[see][]{abrams2012judges,stevenson2018distortion}. This is, however, compatible with the partial monotonicity. Its main advantage is that it no longer requires a uniform ranking of defendants across different judges. Judges' rankings are allowed to vary with their characteristics $Z_c$. Applying the result of \cite{vytlacil2002independence}, the partial monotonicity condition can be characterized as a partial single threshold-crossing restriction under the independence assumption \Cref{ass: MTEI}, which we restated below. 

\begin{assumption}[Partial Single Threshold-Crossing]\label{ass: PSTC}
Type $Z=(Z_I,Z_c)$'s judge treatment assignment mechanism is governed by the following threshold crossing model $D_z=1\{\nu(Z_I,Z_c)\geq U_{Z_c}\}$ for a measurable function $\nu$, where the distribution of $U_{z_c}$ is absolutely continuous for all $z_c \in \mathcal Z_c$.
\end{assumption}

Under \Cref{ass: MTEI,ass: PSTC}, we can apply the standard normalization,
\[
D(z_I,z_c) = 1\left\{F_{U_{z_c}|Z_c}(\nu(z_I,z_c)|z_c)\geq F_{U_{z_c}|Z_c}(U_{z_c}|z_c)\right\} \equiv 1\left\{P(z_I,z_c)\geq V_{z_c}\right\}, 
\]
where $F_{U_{z_c}}(\cdot)$ is the distribution function of $U_{z_c}$, $P(z_I,z_c)$ is identified from the observed $(D,Z)$ by $P(z_I,z_c) \equiv \mathbb P(D=1|Z_I=z_I, Z_c=z_c)$. Note by construction, $V_{z_c}$ follows $Uniform[0,1]$ distribution because the distribution of $U_{z_c}$ is absolute continuous; also, $V_{z_c}$ is independent with $(Z_I,Z_c)$. 

The key difference between the STC and the Partial STC is even though $V_{z_c}$ follows $Uniform[0,1]$ distribution, each defendant does not face a single $V$. Instead, he or she faces a collection of $\{V_{z_c},z_c\in\mathcal Z_c\}$. This unobserved latent variable is now different for judges with distinct observable characteristics. The partial STC has a natural interpretation as an extension of the Roy model \citep[][]{canay2024use}. We can interpret $P(z_I,z_c) $ as the perceived gain of incarcerating a defendant by a type $z=(z_I,z_c)$ judge, and  $V_{z_c}$ as the expected cost (but unobserved to the econometrician) of incarcerating a defendant. The particularity of the partial STC is that the expected cost can vary across judges with distinct observable characteristics $z_c$, but is fixed within judges with the same $z_c$. In the standard monotonicity assumption, the cost $V$ would be the same regardless of the characteristics $(z_I,z_c)$. For the same reason, the partial STC is also more reasonable in settings where decision-makers (judges) differ in their preferences and skills \citep{chan2022selection}.

Here, we provide an example of eight judges deciding whether to incarcerate a given defendant to elucidate further the richer heterogeneity enabled by the partial monotonicity (or, equivalently, the partial STC) assumption. We consider the two observable characteristics of the judges introduced earlier, $Z_c \equiv (Z_c^{R}, Z_c^{P}) \in \{0,1\} \times \{0,1\}$. These two binary observable characteristics result in four types of judges. The eight judges are evenly allocated across these four types.

\begin{figure} [!htbp]  
\centering
 \includegraphics[scale=0.8,width=0.7\linewidth]{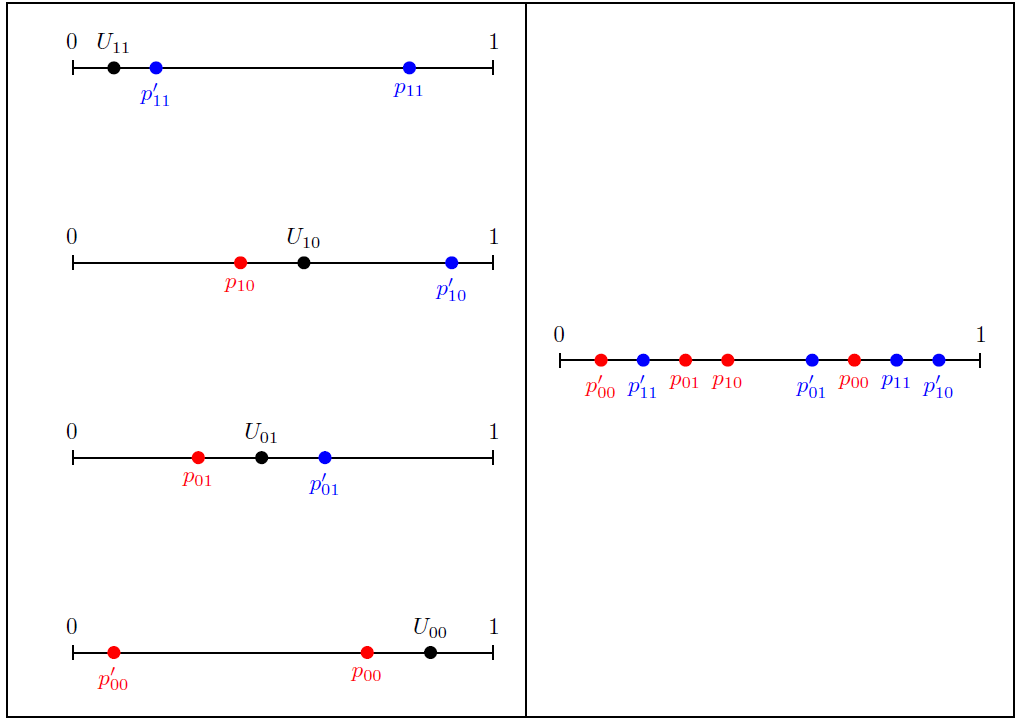} 
\caption{Monotonicity in Judge IV and Conditional Judge IV designs}
\label{condJudge2}
\end{figure}
 The left rectangle of \Cref{condJudge2} shows the benefit and the expected cost of incarcerating the defendant in a separate unit segment for each judge. For example, $p_{11}$ and $p_{11}'$ are the benefits of the two black democratic judges with type $Z_c=(1,1)$ to incarcerate the defendant. The right rectangle of \Cref{condJudge2} plots the benefit numbers of all eight judges on the same unit segment. Similarly, $U_{11}$ represents the expected cost of incarcerating the defendant by a black democratic judge: they share the same expected cost or skills. A judge incarcerates the defendant when the corresponding benefit is higher than the expected cost of incarceration. In \Cref{condJudge2}, the judges who incarcerate the defendant are blue-colored, while those who release the defendant are red-colored. 

The behavior of the eight judges does not violate \Cref{ass:PVyt-Mon} or \Cref{ass: PSTC}. However, the standard monotonicity \Cref{ass: Vyt-Mon} is clearly violated (right rectangle of \Cref{condJudge2}). Indeed, the judge with propensity score $p_{11}^{\prime}$ incarcerates the defendant (blue-colored), whereas judges with higher propensity scores $p_{01}$, $p_{10}$, or $p_{00}$ do not incarcerate the defendant (red-colored). Note that \Cref{ass: Vyt-Mon} would not be violated for this group of judges only under one of these two conditions: (i) all four $V_{z_c}$ are greater than the maximum of the eight propensities or smaller than the minimum of all eight properties. In other words,  when all judges make the same decision regarding this defendant, or (ii) judges who do not incarcerate the defendant must have lower benefit scores than judges who incarcerate the defendant. Moreover, one of these two conditions must hold for all defendants when we impose \Cref{ass: Vyt-Mon} (or \Cref{ass: STC}). 

However, \Cref{ass:PVyt-Mon} (or \Cref{ass: PSTC}) does not require such a binding restriction. In particular, under the partial monotonicity assumption, defendants are allowed to be defiers across judges with distinct observed characteristics $Z_C$. For instance, in \Cref{condJudge2} and using propensity scores to identify judges, the defendant is a $p_{11}^{\prime}-p_{01}$ defier, a $p_{11}^{\prime}-p_{10}$ defier, a $p_{11}^{\prime}-p_{00}$ defier, and a $p_{01}^{\prime}-p_{00}$ defier.

\Cref{ass: MTEI,ass:PExclusion,ass: PSTC} are weaker than 
\Cref{ass: MTEI,ass: Exclusion,ass: Vyt-Mon}. We show that under these weaker conditions, it is still possible to identify meaningful treatment effect parameters.

\begin{theorem}[Identification under Partial exclusion and monotonicity] \label{thm: newIdent}
 If \Cref{ass: MTEI,ass:PExclusion,ass: PSTC} hold, then:
 \begin{enumerate}

     \item[(i)] (Identification of the LATE). Let $\mathcal P_{z_c}$ be the support of $P(Z_I,Z_c)$ conditioning on $Z_c=z_c$. Then for any pair  $(p,p') \in \mathcal P_{z_c} \times \mathcal P_{z_c} $ such that $p < p'$ we have the following identification results:
\begin{multline*}
    \frac{\mathbb E [g(Y)\vert P=p', Z_c=z_c]-\mathbb E [g(Y) \vert P=p, Z_c=z_c]}{p'-p}\\ =\mathbb E [g(Y_1(z_c))-g(Y_0(z_c))\vert 1\{p <  V_{z_c} \leq p'\}].
\end{multline*}
\item[(ii)] (Identification of the MTE). For any $p \in \mathcal P_{z_c}$ such that $\mathbb E [g(Y)\vert P=\cdot, Z_c=z_c]$ is continuously differentiable in the neighborhood of $p$, then, 
\begin{eqnarray*}
    \frac{\partial \mathbb E [g(Y)\vert P=t, Z_c=z_c]}{\partial t}\big |_{t=p}=\mathbb E [g(Y_1(z_c))-g(Y_0(z_c))\vert V_{z_c}=p].
\end{eqnarray*}

     \item [(iii)] (Testable restrictions). For any fixed $z_c \in \mathcal Z_{c}$,  $\mathbb P(y<Y \leq y' , D=1 \vert P=p, Z_c=z_c)$ and $-\mathbb P(y<Y \leq y' , D=0 \vert P=p, Z_c=z_c)$ are non-decreasing in $p$ for all $p \in \mathcal P_{Z_c}$ and any $y, y' \in \mathcal Y$.
 \end{enumerate}
 
\end{theorem}

The proof of \Cref{thm: newIdent} is similar to \Cref{Thm: sharpMTE} after conditioning on $Z_c=z_c$ and therefore omitted. The identification results stated in \Cref{thm: newIdent} (i)-(ii) demonstrate that whenever there are two judges with distinct $Z_I$ but share the same observed characteristics $Z_c=z_c$, the conditional Wald estimand identifies the LATE provided the propensity scores for these two judges are different. Moreover, when the distribution of $Z_I|Z_c=z_c$ allows one to take the derivative of $\mathbb E [g(Y)\vert P=\cdot, Z_c=z_c]$, the conditional LIV estimand identifies the MTE. This identification result is a local version of the standard LATE and MTE identification. 

\Cref{thm: newIdent} (iii) presents the testable implications of the weaker monotonicity and exclusion assumptions. The testable implications in \Cref{thm: newIdent} (iii) are weaker than those in \Cref{Thm: sharpMTE} (i). To see this, let us consider the same example of the eight judges discussed above, where the outcome of interest is recidivism ($Y \in \{0,1\}$). We consider the same two observable characteristics of the judges, $Z_c \equiv (Z_c^{R}, Z_c^{P}) \in \{0,1\} \times \{0,1\}$. Let $ \theta^{d}(p) = \mathbb P(Y=0 , D=d \vert P=p) $ for $d \in \{0,1\}$. In this simple case, the sharp testable implications under the standard judge leniency design, i.e. \Cref{ass: MTEI,ass: Exclusion,ass: Vyt-Mon} are:
\begin{align*}
     &\theta^{1}(p_{00}^{\prime}) \leq  \theta^{1}(p_{11}^{\prime}) \leq  \theta^{1}(p_{01}) \leq  \theta^{1}(p_{10}) \leq  \theta^{1}(p_{01}^{\prime}) \leq  \theta^{1}(p_{00}) \leq \theta^{1}(p_{11}) \leq   \theta^{1}(p_{10}^{\prime})\\
     &\theta^{0}(p_{00}^{\prime}) \geq  \theta^{0}(p_{11}^{\prime}) \geq  \theta^{0}(p_{01}) \geq  \theta^{0}(p_{10}) \geq  \theta^{0}(p_{01}^{\prime}) \geq  \theta^{0}(p_{00}) \geq \theta^{0}(p_{11}) \geq   \theta^{0}(p_{10}^{\prime}),
\end{align*}
which is a total of fourteen inequalities. However, when invoking our weaker set of assumptions, we have only eight inequalities that characterize the sharp testable implications:
\begin{align*}
     &\theta^{1}(p_{11}^{\prime})  \leq  \theta^{1}(p_{11}), \quad
 \theta^{1}(p_{10})  \leq  \theta^{1}(p_{10}^{\prime}) , \quad \theta^{1}(p_{01})  \leq \theta^{1}(p_{01}^{\prime})  ,  \quad\theta^{1}(p_{00}^{\prime})  \leq  \theta^{1}(p_{00})\\
 &\theta^{0}(p_{11}^{\prime})  \geq  \theta^{0}(p_{11}) , \quad \theta^{0}(p_{10})  \geq  \theta^{0}(p_{10}^{\prime}) ,  \quad\theta^{0}(p_{01})  \geq \theta^{0}(p_{01}^{\prime}),\quad \theta^{0}(p_{00}^{\prime})  \geq  \theta^{0}(p_{00}).
\end{align*}
The comparison of the testable implications in \Cref{Thm: sharpMTE,thm: newIdent} confirms that the judge leniency design is more stringent than the conditional judge leniency design. Hence, whenever the standard judge leniency design is rejected, the researcher may rely on its relaxed versions as long as the testable implications derived in \Cref{thm: newIdent} are satisfied.

\section{Conclusion}
In this paper, we derive the sharp testable implications for identifying assumptions for the judge's leniency design in a general framework where the instruments can be either discrete or continuous and propose a consistent test for the implications. Our simulation study and empirical results highlight the importance of considering sharp implications for a better use of information in the data. While we focus on the primary application of testing the validity of judge leniency design, our method can be readily applied to a broad range of other applications. 

\clearpage
\appendix
\begin{center}
{\Large APPENDIX}
\par\end{center}
\small
\section{Implementation of the test} \label{app: implementation}
In this section, we describe the details of calculating the test statistics for \Cref{algorithm: critical value}. Let $\{Y_i, Z_i, D_i\}_{i=1}^n$ be a random sample and $P(D_i=1|Z_i)=P(Z_i,\theta_0)$ be the propensity known to $\theta_0$. Note that when $Z$ is the judge's identity and if the number of defendants for each judge diverges to infinity, we can simply use the frequency estimator $\hat P_i=\frac{\sum_{k=1}^n D_k 1\{Z_k=Z_i\}}{\sum_{k=1}^n 1\{Z_k=Z_i\}}$. Therefore, in the appendix sections, we focus on the case in which $Z$ is continuous to simplify notation. 
\subsection{Constructing $\hat\nu_d(\ell)$}
First, when $Z$ is continuous, we estimate $\theta_0$ by MLE, 
\begin{align}
\hat{\theta}&=\argmax_{\theta\in\Theta} \frac{1}{n}\sum_{i=1}^n \log f(Y_i,D_i,Z_i,X_i,\theta) \nonumber\\
&\equiv \argmax_{\theta\in \Theta} \frac{1}{n}\sum_{i=1}^n D_i \log P(Z_i,\theta)+(1-D_i) \log (1-P(Z_i,\theta)).\label{eq: MLE without x}
\end{align}
where $P(z,\theta)$ is parameterized and depends on $z$ through $z'\theta$. For example, $P(z,\theta) = \Phi(z'\theta_z)$ for Probit or $P(z,\theta) = \frac{exp(z'\theta_z)}{1+\exp(z'\theta_z)}$ for Logit. 

Next, note that 
\begin{align*}
&\nu_1(y,r_y, p_1,p_2,r_p,\theta_0)=m_{1}(y, r_y,p_2,r_p,\theta_0)\cdot  
w(p_1,r_p,\theta_0)- m_{1}(y, r_y,p_1,r_p,\theta_0)\cdot  
w(p_2,r_p,\theta_0),\\
&\nu_0(y,r_y, p_1,p_2,r_p,\theta_0)=m_{0}(y, r_y,p_2,r_p,\theta_0)\cdot  
w(p_1,r_p,\theta_0)- m_{0}(y, r_y,p_1,r_p,\theta_0)\cdot  
w(p_2,r_p,\theta_0),
\end{align*}
where 
\begin{align} 
&m_{1}(y, r_y,p,r_p,\theta)=\mathbb{E}[D 1(y \leq Y\leq y+r_y)  1(p \leq P(Z,\theta)\leq p+r_p)],\label{eq: m1 function}\\
&m_{0}(y, r_y,p,r_p,\theta)=\mathbb{E}[(D-1) 1(y \leq Y\leq y+r_y)  1(p \leq P(Z,\theta)\leq p+r_p)],\label{eq: m0 function}\\
&w(p,r_p,\theta)=\mathbb{E}[ 1(p \leq P(Z,\theta)\leq p+r_p)].\label{eq: w function}
\end{align}
We can estimate $m_d(y, r_y,p,r_p,\theta)$ and $w(p,r_p,\theta)$ by sample analogs and $\theta$ be replaced by its MLE $\hat\theta$:
\begin{align} 
&\hat m_{d}(y, r_y,p,r_p,\hat \theta)=\frac{1}{n}\sum_{i=1}^n m_{di}(y, r_y,p,r_p,\hat \theta),\quad d=0,1 \label{eq: m functoin estimates} \\
&\hat w(p,r_p,\hat \theta)=\frac{1}{n}\sum_{i=1}^nw_i(p,r_p,\hat \theta).\label{eq: w functoin estimates}
\end{align}
with 
\begin{align*}
&m_{1i}(y, r_y,p,r_p,\theta)=D_i 1(y \leq Y_i\leq y+r_y)  1(p \leq P(Z_i,\theta)\leq p+r_p),\\
&m_{0i}(y, r_y,p,r_p,\theta)=(D_i-1) 1(y \leq Y_i\leq y+r_y)  1(p \leq P(Z_i,\theta)\leq p+r_p),\\
&w_i(p,r_p,\theta)= 1(p \leq P(Z_i,\theta)\leq p+r_p).
\end{align*}

Then, for a given $\ell=(y,r_y,p_1,p_2,r_p)'$, we can estimate $\nu_1(\ell)$ and $\nu_0(\ell)$ by
\begin{align}
    \label{eq: nu1estimation}&\hat\nu_1(\ell)=\hat m_{1}(y, r_y,p_2,r_p,\hat \theta)\cdot  
\hat w(p_1,r_p,\hat \theta)- \hat m_{1}(y, r_y,p_1,r_p,\hat \theta)\cdot  
\hat w(p_2,r_p,\hat \theta),\\
\label{eq: nu0estimation}&\hat\nu_0(\ell)=\hat m_{0}(y, r_y,p_2,r_p,\hat \theta)\cdot  
\hat w(p_1,r_p,\hat \theta)- \hat m_{0}(y, r_y,p_1,r_p,\hat \theta)\cdot  
\hat w(p_2,r_p,\hat \theta).
\end{align}

\subsection{Constructing $\hat\nu_d^b(\ell)$}\label{app: multiplier boostrap}
In this appendix, we show how to construct the bootstrap estimates $\hat\nu_d^b(\ell)$. For bootstrap iteration $b$, let $\{W_1^b,W_2^b,\cdots,W_n^b\}$ be a sequence of i.i.d. random variables with both mean and variance equal to one. For instance, we can choose standard normal. Let $\hat{\theta}^b$ be the MLE based on the $b$-th bootstrapped sample: 
\begin{multline*}
\hat{\theta}^b=\argmax_{\theta\in\Theta} \frac{1}{n}\sum_{i=1}^n W_i^b \log f(Y_i,D_i;Z_i\theta)\\
\equiv \argmax_{\theta\in \Theta} \frac{1}{n}\sum_{i=1}^n W_i^b \left\{D_i \log P(Z_i,\theta)+(1-D_i) \log (1-P(Z_i,\theta))\right\},
\end{multline*}
and the estimated propensity score for the $b$-th bootstrap as
\begin{equation}\label{eq: boot MLE}
    \hat P_i^b = P(Z_i,\hat\theta^b)
\end{equation}
We define the weighted bootstrapped estimators for $m_{1}(y, r_y,p,r_p,\theta_0)$, $m_{0}(y, r_y,p,r_p,\theta_0)$ and $w(p,r_p,\theta_0)$ be
\begin{align*}
&\widehat{m}^{b}_{1}(y, r_y,p,r_p,\hat{\theta}^{b})
=\frac{1}{n}\sum_{i=1}^n W_i^b\cdot m_{1i}(y, r_y,p,r_p,\hat{\theta}^{b})\Big/ \frac{1}{n}\sum_{i=1}^n W_i^b,\\
&\widehat{m}^{b}_{0}(y, r_y,p,r_p,\hat{\theta}^{b})=\frac{1}{n}\sum_{i=1}^n W_i^b\cdot m_{0i}(y, r_y,p,r_p,\hat{\theta}^{b})\Big/ \frac{1}{n}\sum_{i=1}^n W_i^b,\\
&\widehat{w}^{b}(p,r_p,\hat{\theta}^{b}) = \frac{1}{n}\sum_{i=1}^n W_i^b\cdot w_i(p,r_p,\hat{\theta}^{b})\Big/ \frac{1}{n}\sum_{i=1}^n W_i^b,
\end{align*}
Finally, for a given $\ell=(y,r_y,p_1,p_2,r_p)'$, we can construct $\hat\nu_d^b(\ell)$ for the $b$-th bootstrap iteration 
\begin{align}
    \label{eq: nu1estimation boot}&\hat\nu_1^b(\ell)=\hat m_{1}^b(y, r_y,p_2,r_p,\hat \theta)\cdot  
\hat w^b(p_1,r_p,\hat \theta^b)- \hat m_{1}^b(y, r_y,p_1,r_p,\hat \theta^b)\cdot  
\hat w^b(p_2,r_p,\hat \theta^b),\\
\label{eq: nu0estimation boot}&\hat\nu_0^b(\ell)=\hat m_{0}^b(y, r_y,p_2,r_p,\hat \theta^b)\cdot  
\hat w^b(p_1,r_p,\hat \theta^b)- \hat m_{0}^b(y, r_y,p_1,r_p,\hat \theta^b)\cdot  
\hat w^b(p_2,r_p,\hat \theta^b).
\end{align}
 
\section{Proof of Main Results} \label{app: main proof}
\subsection{Proof of \Cref{Thm: sharpMTE}} \label{proof: sharpMTE}

\proof \Cref{Thm: sharpMTE}-(i) is a direct application of \cite{heckman2005structural}'s testable implications where $g(Y)=1\{Y  \in (y,y']\}$ for $y \leq y'$. We focus on part (ii). 

We define some notation. Let $\mathcal{L}(\mathcal P)$ be the set of limit points of $\mathcal P$, $\mathcal{L}^o(\mathcal P)$ be a set of interior point of $\mathcal P$, and $\mathcal{C}(\mathcal P)$ be the closure of $\mathcal P$. Furthermore, let $I(\mathcal P)=\mathcal{C}(\mathcal P)/\mathcal{L}^o(\mathcal P)$ be the complement of $\mathcal{L}^o(\mathcal P)$ in the closure of $\mathcal P$. So $I(\mathcal P)$ also contains isolation points. Note that $\mathcal{L}^o(\mathcal P)$ can be written as a union of countable or finite exclusive open intervals: $\mathcal{L}^o(\mathcal P)=\cup_{j=1}^J (a_j,b_j)$, where $ (a_j,b_j)\subseteq \mathcal P$, $b_j<a_{j+1}$, and $J$ can be infinity. Let  $\Omega(\mathcal P)$ be a collection of intervals belonging to $(0,1]$ defined  as follows:
\[\Omega(\mathcal P)\equiv \big\{(p, p']: p, p' \in I(\mathcal P) \cup\{0,1\} \text{ and }  (p,p')\cap \mathcal P=\emptyset \big \}.\]
So the interior of each interval does not intersect with $\mathcal P$. $\Omega(\mathcal P)$ contains a generic element $(c_k,d_k]$, where $c_k$, $d_k\in I(\mathcal P)$, $d_k\leq c_{k+1}$, $k=1,2,\cdots, K$ with $K$ possibly equals to $\infty$, depending on how many isolation points there are in $\mathcal P$. Note that with above notation, for any $v \in (0,1]$, $v$ must belongs to one of the following categories:  (i) an element of  $\mathcal{L}^o(\mathcal P)$ so that $v\in  (a_j,b_j)$ for some $j$, (ii)  $v \in \mathcal{L}(\mathcal P) / \mathcal{L}^o(\mathcal P)$, and (iii) there exist an integer $k$ such that $v \in (c_k,d_k]$. The following figure illustrates the partition of the unit interval. 

\begin{figure}[h]\label{construction}
	\centering
		\includegraphics[trim=100 655 100 50,clip=true,scale=0.8]{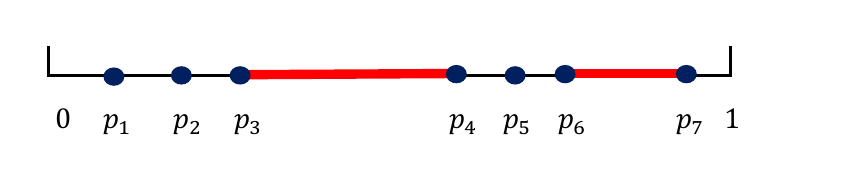}
		\caption{An illustration: $\mathcal P=\{p_1,p_2,p_5\}\cup [p_3,p_4]\cup [p_6,p_7]$, $\mathcal{L}^o(\mathcal P)=(p_3,p_4)\cup (p_6,p_7)$, and $\Omega(\mathcal P)=\{(0,p_1], (p_1,p_2],(p_4,p_5],(p_5,p_6],(p_7,1]\}$. }
\end{figure}

We will assume that $\mathbb P(y<Y \leq y', D=1\vert P=p)$ and $\mathbb P(y<Y \leq y', D=0\vert P=p)$ are continuously differentiable over $\mathcal L^o$ as a regularity condition under which the local instrumental variable (LIV) estimand is well defined. 

First, we construct  $\tilde V$ and $\tilde D$ as follows:  
\begin{equation*}
\mathbb P(\tilde V \leq t \vert P=p)=t,\forall (t,p)\in [0,1]\times \mathcal P, \text{ and } \tilde D=1\{P(Z)\geq \tilde V\}.\\
\end{equation*} 
By construction, \Cref{ass: STC} is satisfied. Next, we propose the following distribution for $\tilde Y_1 \vert \tilde V, P$. For any arbitrary $ p \in \mathcal P$ and $v\in(0,1]$, we define 
\begin{eqnarray*} 
\mathbb P(\tilde{Y}_1 \leq y \vert \tilde V=v,  P=p)&=&
\begin{cases}
  \frac{\partial }{\partial t} \mathbb P(Y \leq y, D=1\vert P=t)|_{t=v} \;\; \text{ if } v \in \mathcal{L}^o(\mathcal P)  \\
  \lim_{\tilde v\rightarrow v}\frac{\partial }{\partial t} \mathbb P(Y \leq y, D=1\vert P=t) |_{t=\tilde v}\;\; \text{ if } v \in \mathcal{L}(\mathcal P) / \mathcal{L}^o(\mathcal P)  \\
   \frac{\mathbb P(Y \leq y, D=1\vert P=d_k)-\mathbb P(Y \leq y, D=1\vert P=c_k)}{d_k-c_k} \;\; \text{ if } v \notin L(P) \text{ but } v \in (c_k,d_k]  \in \Omega(\mathcal P).  \\
\end{cases}
\end{eqnarray*}

\begin{eqnarray*} 
\mathbb P(\tilde{Y}_0 \leq y \vert \tilde V=v,  P=p)&=&
\begin{cases}
  -\frac{\partial }{\partial v} \mathbb P(Y \leq y, D=0\vert P=t)|_{t=v} \;\; \text{ if } v \in \mathcal{L}^o(\mathcal P)  \\
  -\lim_{\tilde v\rightarrow v}\frac{\partial }{\partial v} \mathbb P(Y \leq y, D=0\vert P=t)|_{t=\tilde v} \;\; \text{ if } v \in \mathcal{L}^o(\mathcal P)  \\
   \frac{\mathbb P(Y \leq y, D=0\vert P=c_k)-\mathbb P(Y \leq y, D=0\vert P=d_k)}{d_k-c_k} \;\; \text{ if } v \notin L^o(P) \text{ but } v \in  (c_k,d_k]  \in \Omega(\mathcal P).  \\
\end{cases}
\end{eqnarray*}
Note that the conditioning on $\tilde V=v$ and $P=p$, the distribution of $\tilde Y_1$ does not depend on $p$. Hence, \Cref{ass: MTEI} is satisfied by construction. 

We now show that the distribution function constructed above is well defined. We focus on $\mathbb P(\tilde{Y}_1 \leq y \vert \tilde V=v,  P=p)$ and the verification for $\mathbb P(\tilde{Y}_0 \leq y \vert \tilde V=v,  P=p)$ is analogous. Let $\underline y$ and $\overline{y}$ be the lower and upper bounds of the support of $Y$, respectively. 

\begin{enumerate}

\item $ \mathbb P(\tilde{Y}_1 < \underline y-\epsilon \vert \tilde V=v,  P=p)=0$ for all $v \in [0,1]$ and for any arbitrarily small $\epsilon>0$. To see this, suppose $v\notin \mathcal{L}(\mathcal P)$, then there exists $(c_k,d_k]\in \Omega(\mathcal P)$ such that $v\in (c_k,d_k]$, therefore,
\begin{align*}
&\mathbb P(\tilde{Y}_1 \leq  \underline y -\epsilon\vert \tilde V=v,  P=p) \\=& \frac{\mathbb P(Y \leq  \underline y-\epsilon, D=1\vert P=d_k)-\mathbb P(Y \leq  \underline y-\epsilon, D=1\vert P=c_k)}{d_k-c_k} = \frac{0-0}{d_k-c_k}=0.
\end{align*}
On the other hand, if $v\in \mathcal{L}^o(\mathcal P)$, then $\mathbb P(Y \leq\underline y -\epsilon, D=1\vert P=\tilde v)=0$ for all $\tilde v$ in a small neighborhood of $v$, which implies $\frac{\partial }{\partial v} \mathbb P(Y \leq \underline y -\epsilon, D=1\vert P=v) =0$. The case that $v \in \mathcal{L}^o(\mathcal P)  $ follows straightforwardly. 
\item $ \mathbb P(\tilde{Y}_1 \leq \overline y \vert \tilde V=v,  P=p)=1$. First, if $v\in \mathcal{L}^o(\mathcal P)$, then 
\[\mathbb P(Y \leq\overline y , D=1\vert P=v)=\mathbb P( D=1\vert P=v)=v\Rightarrow \frac{\partial }{\partial v} \mathbb P(Y \leq \overline y, D=1\vert P=v) =1.\] 
On the other hand, if $v\notin \mathcal{L}(\mathcal P)$, then
\begin{align*}
&\mathbb P(\tilde{Y}_1 \leq  \overline y \vert \tilde V=v,  P=p) = \frac{\mathbb P(Y \leq  \overline y, D=1\vert P=d_k)-\mathbb P(Y \leq  \overline y, D=1\vert P=c_k)}{p'-p} = \frac{d_k-c_k}{d_k-c_k}=1.
\end{align*}
 \item $\mathbb P(\tilde Y_1\leq y\vert \tilde V=v,  P=p)$ is nondecreasing in $y$. For $y<y'$ we have 
 \begin{multline*} 
 \mathbb P(\tilde{Y}_1 \leq y' \vert \tilde V=v,  P=p)-\mathbb P(\tilde{Y}_1 \leq y \vert \tilde V=v,  P=p)\\=
\begin{cases}
  \frac{\partial }{\partial t} \mathbb P(y<Y \leq y', D=1\vert P=t)|_{t=v}\geq 0  \;\; \text{ if } v \in \mathcal{L}^o(\mathcal P),  \\
    \lim_{\tilde v\rightarrow v}\frac{\partial }{\partial t} \mathbb P(y<Y \leq y, D=1\vert P=t) |_{t=\tilde v} \geq 0\;\; \text{ if } v \in \mathcal{L}(\mathcal P) / \mathcal{L}^o(\mathcal P)  \\
   \frac{\mathbb P(y<Y \leq y', D=1\vert P=d_k)-\mathbb P(y<Y \leq y', D=1\vert P=c_k)}{d_k-c_k}\geq 0 \text{ if } v \notin L^o(P) \text{ but } v \in [c_k,d_k]  \in \Omega(\mathcal P),  
\end{cases}
\end{multline*}
 where the last inequalities hold whenever the testable implications hold, i.e. $\mathbb P(y<Y \leq y', D=1\vert P=p)$ is a non-decreasing function for all $p \in \mathcal P$ and all $y<y'$, and by the continuous differentiability of $\mathbb P(y<Y \leq y', D=1\vert P=p)$ over $\mathcal{L}(\mathcal P)$. 
\end{enumerate}

Finally, we show that $(\tilde{V},\tilde{Y}_d,P(Z))$, $d \in \{0,1\}$ is observationally equivalent to $(V,Y_d, P(Z))$ $d \in \{0,1\}$. For this, we show that the conditioning distribution of $(\tilde Y,\tilde D)$ given $P(Z)$ is the same as the conditioning of $(Y,D)$ given $P(Z)$. Take an arbitrary $p\in \mathcal P$.

Suppose first $p\notin \mathcal{L}^o(\mathcal P)$, then $(0,p]$ can be expressed as unions of exclusive intervals $\left(\cup_{j=1}^{J^*} (a_j,b_j) \right)\cup \left(\cup _{k=1}^{K^*} (c_k,d_k]\right)$ for some $J^*$ and $K^*$, where $(a_j,b_j)s$ are connected subsets of $\mathcal P$. Therefore, 
\begin{multline*}
\mathbb P(\tilde{Y} \leq y, \tilde{D}=1 \vert P=p)=\mathbb P(\tilde{Y}_1 \leq y, \tilde{V} \leq p \vert P=p)=\int_{0}^p \mathbb P(\tilde Y_1\leq y|\tilde V=v,P=p)d v \\
=\sum_{j=1}^{J^*}\int_{a_j}^{b_j}\mathbb P(\tilde{Y}_1 \leq y \vert \tilde V=v,  P=p)dv+ \sum_{k=1}^{K^*}\int_{c_k}^{d_k}\mathbb P(\tilde{Y}_1 \leq y \vert \tilde V=v,  P=p)dv\\
=\sum_{j=1}^{J^*}\left(\mathbb P(Y \leq y, D=1\vert P=b_j)-\mathbb P(Y \leq y, D=1\vert P=a_{j})\right) 
\\+ \sum_{k=1}^{K^*}\left(\mathbb P(Y \leq y, D=1\vert P=d_k)-\mathbb P(Y \leq y, D=1\vert P=c_k)\right) \\
=\mathbb P(Y \leq y, D=1\vert P=p)-\mathbb P(Y \leq y, D=1\vert P=0)=\mathbb P(Y \leq y, D=1\vert P=p),
\end{multline*}
where the first equality is by construction that $\tilde V$ satisfies \Cref{ass: STC}, the third equality holds because $(0,p]$ can be expressed as unions of exclusive intervals $\left(\cup_{j=1}^{J^*} (a_j,b_j) \right)\cup \left(\cup _{k=1}^{K^*} (c_k,d_k]\right)$, the fourth equality is obtained by inserting the constructed counterfactural distributions, and the last one holds because $\mathbb P(Y \leq y, D=1\vert P=0)=0$.

Suppose that $p\in (a_{j^*},b_{j^*})\subseteq  \mathcal L^0(\mathcal P)$ for some $j^*$, then the right hand side equals to 
\begin{multline*}
\mathbb P(\tilde{Y} \leq y, \tilde{D}=1 \vert P=p)= \mathbb P(\tilde{Y}_1 \leq y, \tilde{V} \leq p \vert P=p)=\int_{0}^p \mathbb P(\tilde Y_1\leq y|\tilde V=v,P=p)d v \\
=\int_{0}^{a_{j^*}}\mathbb P(\tilde{Y}_1 \leq y \vert \tilde V=v,  P=p)dv+ \int_{a_{j^*}}^{p}\mathbb P(\tilde{Y}_1 \leq y \vert \tilde V=v,  P=p)dv\\
=\mathbb P(Y \leq y, D=1\vert P=a_{j^*})+ \int_{a_{j^*}}^{p}\frac{\partial }{\partial v} \mathbb P(Y \leq y, D=1\vert P=v)dv\\
=\mathbb P(Y \leq y, D=1\vert P=a_{j^*})+  \mathbb P(Y \leq y, D=1\vert P=p)- \mathbb P(Y \leq y, D=1\vert P=a_{j^*})\\
=\mathbb P(Y \leq y, D=1\vert P=p),
\end{multline*}
where the $\int_{0}^{a_{j^*}}\mathbb P(\tilde{Y}_1 \leq y \vert \tilde V=v,  P=p)dv=\mathbb P(Y \leq y, D=1\vert P=a_{j^*})$ holds by the above argument and the fifth equality holds by inserting the constructed counterfactural distributions. This completes the proof.\qed

\subsection{Proof of \Cref{thm: validity}} \label{proof: validity}
We begin by listing a few regularity conditions for the proof of \Cref{thm: validity}. Again, when $Z$ is the judge's identity, we use the frequency estimator $\hat P_i=\frac{\sum_{k=1}^n D_k 1\{Z_k=Z_i\}}{\sum_{k=1}^n 1\{Z_k=Z_i\}}$ for the propensity score. For its root-n-consistency, we only need $\sum_{i=1}^n 1\{Z_k=j\}\rightarrow \infty$ for each judge $j$, and i.i.d. of $(Y_i,D_i,X_i)$ among defendants conditioning on judges. So \Cref{ass: iid,ass: continuity,ass: theta influence,ass: wb theta influence} are mostly for the case of continuous instrument $Z$. 

\begin{assumption}\label{ass: iid}
    The observations $\{(Y_i, D_i, Z_i, X_i)\}_{i=1}^n$ are i.i.d.\ across $i$.  
\end{assumption}

For notational simplicity, \Cref{ass: iid} assumes that all cases are mutually independent, which is equivalent to assuming that each judge handles exactly one case. All inference results can be extended straightforwardly to settings where judges handle multiple cases (with varying case counts across judges) by accounting for clustering at the judge level.

\begin{assumption} \label{ass: continuity} We impose the following smoothness conditions: 
\begin{enumerate}
    \item  The conditional density of $(Y,D)$ given $P(Z,\theta_0)=p$, denoted by $f_{Y,D|P}(y,d|p)$, is Lipschitz continuous both in $p$ on $\mathcal P$ and in $y$ on $\mathcal Y$ for $d=0,1$.
    \item For all $z\in\mathcal Z$, $P(z,\theta)$ is continuously differentiable in $\theta$ at $\theta_0$ with bounded derivatives. 
\end{enumerate}
\end{assumption}

Note that \Cref{ass: continuity}-(1) does not exclude the case of discrete propensity score. When $P$ is discrete and $\mathcal P$ contains finite many distinguished elements, any convergent sequence in $\mathcal P$ must be a constant sequence eventually, and in that case \Cref{ass: continuity}-(1) holds automatically. \Cref{ass: continuity}-(1)  implies that the functions $m_d$ and $\omega$, defined in \Cref{eq: m1 function,eq: m0 function,eq: w function}, are continuous functions of $\ell$.  \Cref{ass: continuity}-(2) implies that the class of functions  $\{1(p \leq P(Z,\theta)\leq p+r_p) : \theta\in\Theta, p\in [0,1], r_p\in[0,1]\}$  is a Vapnik-Chervonenkis (VC) class of function.

\begin{assumption}\label{ass: theta influence}
    The parameter space $\Theta$ for $\theta_0$ is compact, and $\theta_0$ is in the interior of $\Theta$. The estimator $\hat\theta$ admits an influence function of the following form,
    \begin{equation} \label{eq: theta influence}
            \sqrt{n} (\hat\theta-\theta_0) = \frac{1}{\sqrt{n}}\sum_{i=1}^n s(D_i,Z_i,\theta_0) + o_p(1),
    \end{equation}
    where $s(\cdot,\cdot,\cdot)$ is measurable, satisfying $\mathbb{E}[s(D_i,Z_i,\theta_0)]=0$,  $\mathbb{E}[\sup_{\theta}|s(D_i,Z_i,\theta)|]<\infty$, and $V(\sup_{\theta}|s(D_i,Z_i,\theta)|)<\infty$.
\end{assumption}
\Cref{ass: theta influence} is satisfied for common maximum likelihood estimators and parametric binary response models. For example, if one estimates $\theta_0$ by Probit model $D_i = 1[Z_i'\theta_0\geq V_i]$, with $V_i\sim N(0,1)$, then
\[
s(D_i,Z_i,\theta_0) =  \frac{\phi((2D_i-1)Z_i'\theta_0)}{\Phi((2D_i-1)Z_i'\theta_0)}Z_i.
\]
 If the Logit model is used, then
\[
s(D_i,Z_i,\theta_0) = \left(D_i-\frac{\exp(Z_i'\theta_0)}{1+\exp(Z_i'\theta_0)}\right) Z_i.
\]

\begin{assumption}\label{ass: bootstrap weights}
    $\{W_i\}_{i=1}^n$ is a sequence of i.i.d.\ pseudo random variables that is independent of the sample path with $E[W_i]=1$ and $Var[W_i]=1$.  
\end{assumption}

\begin{assumption}\label{ass: wb theta influence}
    The estimator $\hat\theta^{b}$ satisfies that 
    \begin{equation} \label{eq: wb theta influence}
            \sqrt{n} (\hat\theta^{b}-\hat\theta) = \frac{1}{\sqrt{n}}\sum_{i=1}^n (W_i-1)\cdot s(D_i,Z_i,\theta_0) + o_p(1),
    \end{equation}
    where $s_\theta(\cdot)$ is the same as in Assumption \ref{ass: theta influence}.
\end{assumption}
\Cref{ass: wb theta influence} is satisfied under our weighted bootstrap procedure.

The proof of Theorem \Cref{thm: validity} follows from the same arguments as Theorems 5.1 and 5.2 of \cite{hsu2017consistent} once \Cref{lemma: m and w influence,lemma: wb influence nu,lemma: influence nu,lemma: wb variance} are established, as detailed in \Cref{app: theorem 2 lemmas}.

\subsection{Proofs for \Cref{lem:1}} \label{section: proof of average exclusion and monotone}
\textbf{Part (a)}. Note that, 
\begin{align*}
    \text{Cov}(Y,P(Z)) =& \text{Cov}\left(\tilde{Y}_{1} D + \tilde{Y}_{0} (1-D),P(Z) \right) \\
    = & \text{Cov}\left(\tilde{\alpha} D + \tilde{Y}_{0},P(Z) \right) \\
    = & \text{Cov}\left(\alpha D + (\tilde{\alpha} - \alpha) D + \tilde{Y}_{0},P(Z) \right) \\
    = & \text{Cov}(\alpha D ,P(Z)) + \text{Cov}\left( (\tilde{\alpha} - \alpha) D + \tilde{Y}_{0},P(Z) \right).
    \end{align*}
    The first term on the right hand side can be written as
\begin{align*}
    \text{Cov}(\alpha D ,P(Z)) =& \mathbb{E}\left[ \alpha D (P(Z) - p) \right]  \\
    = & \sum_{z=1}^{J} \mathbb{E}\left[ \alpha D_{z} (p_{z} - p) \mid Z=z \right]  \lambda_{z}  \\
    = &  \mathbb{E}\left[ \alpha \sum_{z=1}^{J} D_{z} (p_{z} - p) \lambda_{z} \right] =\mathbb{E}\left[ \alpha \omega \right],
    \end{align*}
where the conditioning variable $Z=z$ is removed by the independence \Cref{ass: MTEI}. This shows that 
\[
\text{Cov}(Y,P(Z)) = \mathbb{E}\left[ \alpha \omega \right] + \text{Cov}\left( (\tilde{\alpha} - \alpha) D + \tilde{Y}_{0},P(Z) \right).
\]
Next, it is easy to verify that 
\[
    \text{Cov}(D ,P(Z)) =  \mathbb{E}\left[ D (P(Z) - p) \right]= \mathbb{E}\left[  \omega \right].
\]
Finally, the proof is completed by taking the ratio of $\text{Cov}(Y ,P(Z))$ and $\text{Cov}(D ,P(Z))$ (which is possible as long as the instrument is relevant).

\textbf{Part (b).}
\begin{align*}
    \text{Cov}\left( (\tilde{\alpha} - \alpha) D + \tilde{Y}_{0},P(Z) \right) =&  \mathbb{E}\left[ \left( (\tilde{\alpha} - \alpha) D + \tilde{Y}_{0} \right) (P(Z) - p) \right]  \\
    = & \sum_{z=1}^{J} \mathbb{E}\left[ \left( (\alpha_{z} - \alpha) D_{z} + Y_{0z} \right) (p_{z} - p) \mid Z=z \right] \lambda_{z} \\
    = & \sum_{z=1}^{J} \mathbb{E}\left[ \lambda_{z} (p_{z} - p)  \left( (\alpha_{z} - \alpha) D_{z} + Y_{0z} \right) \right] \\
    = & \sum_{z=1}^{J} \mathbb{E}\left[ \lambda_{z} (p_{z} - p)  \left( Y_{1z} D_{z} + Y_{0z} (1-D_{z}) -   \alpha D_{z} \right) \right] \\ 
    = & \scriptsize{\sum_{z=1}^{J} \mathbb{E}\left[ \lambda_{z} (p_{z} - p)  \left( Y_{1z} D_{z} + Y_{0z} (1-D_{z}) -   \left( \bar{Y}_{1} D_{z} + \bar{Y}_{0} (1-D_{z}) - \bar{Y}_{0}  \right) \right) \right]} \\ 
    = & \sum_{z=1}^{J} \mathbb{E}\left[ \lambda_{z} (p_{z} - p)  \left( (Y_{1z} - \bar{Y}_{1}) D_{z} + (Y_{0z} - \bar{Y}_{0} ) (1-D_{z}) + \bar{Y}_{0}  \right) \right] \\
    =&  \mathbb{E}\left[ \sum_{z=1}^{J} \lambda_{z} (p_{z} - p)  \left( (Y_{1z} - \bar{Y}_{1}) D_{z} + (Y_{0z} - \bar{Y}_{0} ) (1-D_{z})  \right) \right],
\end{align*}
where the third equality is by the independence\Cref{ass: MTEI}, the fifth is by substituting for $\alpha = \bar{Y}_{1} - \bar{Y}_{0} $, and the last equality holds because $ \mathbb{E}\left[ \sum_{z=1}^{J} \lambda_{z} (p_{z} - p)  \bar{Y}_{0}  \right] =0 $. \qed

\subsection{Detailed derivation for \Cref{example: FFL fails}} \label{section: numerical example}
For the ease of reading, we restate the DGP below. Consider the potential outcome model:
    \[
  \begin{cases}
  Y = Y_{1}D + Y_{0}(1-D),  \\
  D=1\left\{P\geq V\right\}.
\end{cases}
    \]
We assume $V$ is independent of $(Y_1,Y_0,P)$. However, $(Y_1,Y_0)$ and $P$ are dependent:
    \[
      \begin{cases}
  Y_1|P=\tilde p\sim \text{ degenerate at 1 }, &\text{ if } \tilde p<\frac{1}{2}  \\
  Y_1|P=\tilde p\sim Bernoulli(\tilde p),&\text{ if } \tilde p\geq \frac{1}{2}
\end{cases}
    \]
        \[
      \begin{cases}
  Y_0|P=\tilde p\sim \text{ degenerate at 0 }, &\text{ if } \tilde p<\frac{1}{2}  \\
  Y_0|P=\tilde p\sim Bernoulli(\tilde p),&\text{ if } \tilde p\geq \frac{1}{2}
\end{cases}
    \]
    
    We first check inequality (\ref{eq:SF1}). Let $p'>p>\frac{1}{2}$, and use the condition that $V$ is independent of $(Y_1,Y_0,P)$, we have
    \begin{multline*}
            W(g(YD),p,p') =\frac{\mathbb E[YD|P=p']- \mathbb E[YD|P=p]}{p'-p} \\ =\frac{\mathbb E[Y_1|P=p']p'- \mathbb E[Y_1|P=p]p}{p'-p} = \frac{p'^2-p^2}{p'-p} =p'+p > 1\equiv U_g.
    \end{multline*}

    Therefore, condition (\ref{eq:SF1}) is violated. 
    
    Next, we check condition (\ref{eq: Frand-Test}). Based on the relative positoin of $p'$, $p$, and $\frac{1}{2}$, we verify it by four cases. 
    
    (i) Suppose first $p'>\frac{1}{2}>p$,
    \begin{multline*}
    W(g(Y),p,p') =\frac{\mathbb E[Y_1D+(1-D)Y_0|P=p']- \mathbb E[Y_1D+(1-D)Y_0|P=p]}{p'-p}  \\=\frac{\mathbb E[Y_1|P=p']p'- \mathbb E[Y_1|P=p]p + \mathbb E[Y_0|P=p'](1-p')-\mathbb E[Y_0|P=p](1-p)}{p'-p} \\
     = \frac{p'^2-p + \mathbb E[Y_0|P=p'](1-p')-\mathbb E[Y_0|P=p](1-p)}{p'-p}\\
     = \frac{p'^2-p +p'(1-p')}{p'-p} =1 =U_g-L_g,
    \end{multline*}
    where $ \mathbb E[Y_0|P=p]=0$ because $Y_0$ is degenerate at $0$ when conditioning on $P=p<\frac{1}{2}$, and $\mathbb E[Y_0|P=p']=p'$ because $Y_0\sim Bernoulli(p')$ when conditioning on $P=p'\geq \frac{1}{2}$. 

    (ii) Suppose $p>\frac{1}{2}>p'$,
    \begin{multline*}
    W(g(Y),p,p') = \frac{\mathbb E[Y_1|P=p']p'- \mathbb E[Y_1|P=p]p + \mathbb E[Y_0|P=p'](1-p')-\mathbb E[Y_0|P=p](1-p)}{p'-p} \\
     = \frac{p'-p^2 + \mathbb E[Y_0|P=p'](1-p')-\mathbb E[Y_0|P=p](1-p)}{p'-p}\\
     = \frac{p'-p^2 -p(1-p)}{p'-p} =1 =U_g-L_g.
    \end{multline*}
    
    (iii) If $\frac{1}{2}>p'>p$, then
        \begin{multline*}
    W(g(Y),p,p') = \frac{\mathbb E[Y_1|P=p']p'- \mathbb E[Y_1|P=p]p + \mathbb E[Y_0|P=p'](1-p')-\mathbb E[Y_0|P=p](1-p)}{p'-p} \\
     = \frac{p'-p }{p'-p}=1 =U_g-L_g,
    \end{multline*}
    because in this case $Y_1$ and $Y_0$ are degenerate at $1$ and $0$, respectively. 
    
    (iv) If $p'>p>\frac{1}{2}$, then
        \begin{multline*}
    W(g(Y),p,p') = \frac{\mathbb E[Y_1|P=p']p'- \mathbb E[Y_1|P=p]p + \mathbb E[Y_0|P=p'](1-p')-\mathbb E[Y_0|P=p](1-p)}{p'-p} \\
     = \frac{p'^2-p^2 + p'(1-p')-p(1-p)}{p'-p}
     = \frac{p'-p}{p'-p} =1 =U_g-L_g,
    \end{multline*}
    because in this case both $Y_1$ and $Y_0$ follows Bernoulli distribution. 
    
    Combining (i)--(iv), we can conclude that condition (\ref{eq: Frand-Test}) always holds and has no power to detect the violation. 

    On the other hand, our testable implication can capture such a violation. Consider 
    \[
    \mathbb E[YD|P=p] = \mathbb E[Y_1|P=p]p = \begin{cases}
  p \;\; \text{ if } p <\frac{1}{2},  \\
 p^2\;\; \text{ if } p\geq \frac{1}{2}.   \\
\end{cases}.
    \]
    It is apparent that $\mathbb E[YD|P=p]$ is not a monotone function of $p$, and therefore violates our testable implication. 

\section{A Finite sample test} \label{section: finite sample test}
This appendix section considers the case with a finite number of $J$ judges, $j=1,2,\cdots, J$, and judge $j$ handles a finite number of $n_j$ defendants. For notation simplicity, we assume $n_j=n_{j'}=n^*$, so that the total number of defendants $n=Jn^*$, but our test can be straightforwardly extended to allow for heterogeneous $n_j$. For defendant $i$, let $Z_i\in\{1,2,\cdots, J\}$ be the identity of the judge who handles his/her case. Let $p_{j}$ be the propensity score or stringency measure of judge $j$, defined as
\[
p_{j} = P(D_i=1|Z_i=j)
\]
We assume a judge treats all his/her defendants independently. In this section, we consider the case where $Y$ is a binary variable, as in \cite{frandsen2023judging}. Let $W^{1}_i=Y_i D_i$ and $W^{0}_i=-Y_i(1-D_i)$, and define $q_j^{1} =\mathbb E[W_i^1|Z_i=j]= \mathbb P(W^{1}_i=1|Z_i=j)$ and $q_j^{0} = \mathbb E[W_i^0|Z_i=j]=-\mathbb P(W^{0}_i=-1|Z_i=j)$. Note that because $D_i\geq W_i^1\geq 0$ and $0\geq W_i^0\geq (D_i-1)$, we have $0\leq q_j^1\leq p_j$ and $0\leq -q_j^0\leq p_j$.

The judge leniency design would imply that  $(p_j- p_{j'})(q_j^{d} - q_{j'}^{d})\geq 0 \text{ for all } j,j'\in\mathcal J\text{ and }d\in\{0,1\}$, where $\mathcal J\equiv \{1,2,\cdots, J\}$. With this notation, we rewrite the null hypothesis as
\begin{equation}\label{eq: finite sample null hypothesis}
H_0:  (p_j- p_{j'})(q_j^{d} - q_{j'}^{d})\geq 0 \text{ for all } j,j'\in\mathcal J\text{ and }d\in\{0,1\}
\end{equation} 

To implement a test for $H_0$, we first consider a test $\phi^{j,j'}$ for the null hypothesis of a given pair $(j,j')$ at $\tilde \alpha$ level, that is, 
\[
\mathbb P(\phi^{j,j'}=1|H_0)\leq \tilde\alpha,
\]
Then, we can define the overall test $\phi$ such that $\phi=0$ if $\phi^{j,j'}=0$ for all pairs $(j,j')$, and $\phi=1$ otherwise. That is, we reject $H_0$ if we reject at least one out of $J(J-1)/2$ pairs. If we choose $\tilde \alpha = \frac{2\alpha}{J(J-1)}$, then, we can ensure that 
\[
\mathbb P(\phi=1|H_0)=\mathbb P(\cup_{j>j'}\{\phi^{j,j'}=1\}|H_0)\leq \sum_{j>j'}\mathbb P(\phi^{j,j'}=1|H_0) \leq \frac{J(J-1)}{2} \tilde \alpha=\alpha.
\]

Now we construct $\phi^{j,j'}$ for the pair $(j,j')$. Define $\delta_p=p_j-p_{j'}$, $\delta_q^d=q_j^d-q_{j'}^d$. The relevant null hypothesis is $H^{j,j'}_0:\delta_p\delta_q^d\geq 0$ for $d=0,1$. The idea of constructing $\phi^{j,j'}$ is as follows. We first construct the least favorable confidence interval for $\delta_p$ and $\delta_q^d$, $d=0,1$. Then, suppose we observe that the upper bound of the confidence interval for $\delta_p$ is below zero, while the lower bound of the confidence interval for $\delta_q^d$ is above zero. In that case, we consider this as evidence against the null hypothesis that $\delta_p$ and $\delta_q^d$ have to have the same sign. Similarly, we also reject when the lower bound of the confidence interval for $\delta_p$ is above zero while the upper bound for $\delta_q^d$ is below zero. 

Let $\tilde{\tilde\alpha}=\frac{\tilde\alpha}{4}$. Let $\hat\delta_p=\hat p_j -\hat p_{j'}$ and $\hat\delta_q^d=\hat q^d_j -\hat q^d_{j'}$ be estimators for $\delta_p$ and $\delta^d_q$, respectively, where
\[
\hat p_j = \frac{\sum_{i=1}^N D_i 1\{Z_i=j\}}{n^*},\quad \hat q_j^d = \frac{\sum_{i=1}^N W^d_i1\{Z_i=j\}}{n^*}.
\]

Let $\hat c_p$ be the smallest support point of $\hat\delta_p$ such that $\mathbb P(\hat\delta_p>\hat c_p|p_j=p_{j'}=0.5)\leq \tilde{\tilde\alpha}$. Note that the distribution of $\hat\delta_p$ is symmetric around zero under $p_j=p_{j'}=0.5$ because defendants handled by judges $j$ and $j'$ are independent, then we would know that $-\hat c_p$ is the largest support point of $\hat\delta_p$ such that $\mathbb P(\hat\delta_p<-\hat c_p|p_j=p_{j'}=0.5)\leq \tilde{\tilde\alpha}$. Clearly, $\hat c_p$ is known and can be tabulated for each $n$ and $\tilde{\tilde\alpha}$ by simulation (When $n_j$ and $n_{j'}$ are different, we can simulate the $\tilde{\tilde\alpha}$ and $1-\tilde{\tilde\alpha}$ quantiles too). Another observation is that for all $\tilde{\tilde\alpha}<0.5$, 
\[
\mathbb P(\hat\delta_p<-\hat c_p|p_j=p_{j'}=p)\leq \mathbb P(\hat\delta_p<-\hat c_p|p_j=p_{j'}=0.5)=\tilde{\tilde\alpha},\quad \forall p\in(0,1). 
\]
That is, the distribution of $\hat\delta_p$ is most dispersed when $p_j=p_{j'}=0.5$. For instance, if $p_j=p_{j'}=1$, then $\hat\delta_p\equiv 0$ is degenerate. Let $\widehat{CL}_{p,U}=\hat\delta_p+\hat c_p$ and $\widehat{CL}_{p,L}=\hat\delta_p-\hat c_p$, then define 
\begin{equation} \label{eq: finite sample CS for p}
    \widehat{CS}_p\equiv [\widehat{CS}_{p,L}, \widehat{CS}_{p,U}].
\end{equation}
It is easy to verify that $\widehat{CS}_p$ is a valid $1-2\tilde{\tilde\alpha}$ level confidence set for $\delta_p$ among the models with $p_j=p_{j'}$. 

Similarly, we define $\hat c_q^d$ be the largest support point of $\hat\delta_q^d$ such that $\mathbb P(\hat\delta_q^d>\hat c_q^d|p_j=p_{j'}=0.5)\leq \tilde{\tilde\alpha}$, and define its $1-2\tilde{\tilde\alpha}$ level confidence set as 
\begin{equation}\label{eq: finite sample CS for q}
    \widehat{CS}_q^d\equiv [\widehat{CS}_{q,L}^d, \widehat{CS}_{q,U}^d],
\end{equation}
where $\widehat{CS}_{q,L}^d=\hat \delta_q^d - \hat c_q^d$ and $\widehat{CS}_{q,U}^d=\hat \delta_q^d + \hat c_q^d$. 

Now we are ready to define $\phi^{j,j'}$. We reject $H^{j,j'}_0$ when any of the following events happen:
\[
\{\widehat{CS}_{p,U}<0,\widehat{CS}_{q,L}^1>0\},\{\widehat{CS}_{p,U}<0,\widehat{CS}_{q,L}^0>0\},\{\widehat{CS}_{p,L}>0,\widehat{CS}_{q,U}^1<0\},\{\widehat{CS}_{p,L}>0,\widehat{CS}_{q,U}^1<0\}
\]
We first verify that $\mathbb P(\phi^{j,j'}=1|H_0^{j,j'})\leq 4\tilde{\tilde\alpha}=\tilde\alpha$. Note that, 
\begin{multline}\label{eq: finite sample reject rate 1}
    \mathbb P(\phi^{j,j'}=1|H_0^{j,j'}) \leq \mathbb P(\{\widehat{CS}_{p,U}<0,\widehat{CS}_{q,L}^1>0\}|H_0^{j,j'}) + \mathbb P(\{\widehat{CS}_{p,U}<0,\widehat{CS}_{q,L}^0>0\}|H_0^{j,j'})\\+ \mathbb P(\{\widehat{CS}_{p,L}>0,\widehat{CS}_{q,U}^1<0\}|H_0^{j,j'})+ \mathbb P(\{\widehat{CS}_{p,L}>0,\widehat{CS}_{q,U}^1<0\}|H_0^{j,j'}).
\end{multline}
Consider the first term on the right-hand side of \Cref{eq: finite sample reject rate 1}. We have, 
\begin{equation} \label{eq: finite sample reject rate 2}
     \mathbb P(\{\widehat{CS}_{p,U}<0,\widehat{CS}_{q,L}^1>0\}|H_0^{j,j'})\leq \min\{ \mathbb P(\{\widehat{CS}_{p,U}<0\}|H_0^{j,j'}), \mathbb P(\widehat{CS}_{q,L}^1>0\}|H_0^{j,j'})\}.
\end{equation}
If $H_0^{j,j'}$ is such that $\delta_p>0$ and $\delta_q^1>0$, then, 
\begin{multline} \label{eq: finite sample reject rate 3}
\min\{ \mathbb P(\{\widehat{CS}_{p,U}<0\}|H_0^{j,j'}), \mathbb P(\widehat{CS}_{q,L}^1>0\}|H_0^{j,j'})\} \leq \mathbb P(\{\widehat{CS}_{p,U} <0\}|\delta_p>0,\delta_q^1>0) \\  \leq \mathbb P(\{\widehat{CS}_{p,U} <0\}|\delta_p=0) = \mathbb P(\{\hat\delta_p <-\hat c_p\}|\delta_p=0) = \tilde{\tilde\alpha}. 
\end{multline}
where the first equality hold trivially, the first inequality and second equality are by the properties of the confidence interval $\widehat{CS}_p$. 

Similarly, if $H_0^{j,j'}$ is such that $\delta_p<0$ and $\delta_q^1<0$, then, 
\begin{multline} \label{eq: finite sample reject rate 4}
\min\{ \mathbb P(\{\widehat{CS}_{p,U}<0\}|\delta_p<0,\delta_q^1<0), \mathbb P(\widehat{CS}_{q,L}^1>0\}|H_0^{j,j'})\} \leq \mathbb P(\{\widehat{CS}^1_{q,L} >0\}|\delta_p<0,\delta_q^1<0) \\  \leq \mathbb P(\{\widehat{CS}^1_{q,L} >0\}|\delta_q^1=0) = \mathbb P(\{\hat\delta_q >\hat c_q^1\}|\delta_q^1=0) = \tilde{\tilde\alpha}. 
\end{multline}
Therefore, we can conclude that the first right-hand side term of \Cref{eq: finite sample reject rate 1} satisfies 
\[
 \mathbb P(\{\widehat{CS}_{p,U}<0,\widehat{CS}_{q,L}^1>0\}|H_0^{j,j'})\leq \tilde{\tilde\alpha}. 
\]
Applying the same derivations to the remaining three right-hand side terms, we can conclude that
\[
 \mathbb P(\phi^{j,j'}=1|H_0^{j,j'}) \leq 4\tilde{\tilde\alpha} = \tilde\alpha. 
\]

We summarize the procedure below. 
\begin{algo}\label{algorithm: finite sample test} Finte Sample Test. 

   \begin{enumerate}\setlength{\itemsep}{8pt}
   \item Let $J$ be the number of judges and $\alpha$ be the prechosen significance level. Set $\tilde \alpha = \frac{2\alpha}{J(J-1)}$ and $\tilde{\tilde\alpha} = \frac{\tilde\alpha}{4}$.
    \item For each judge $j$, calculate $\hat\delta_p=\hat p_j -\hat p_{j'}$, and $\hat\delta_q^d=\hat q^d_j -\hat q^d_{j'}$, where $(\hat p_j,\hat q_j^d)$ are sample frequency estimators for $(p_j,q_j^d)$.

\item Let $B$ be a large integer (can be millions). For each $b=1,2,\cdots, B$, draw two independent random samples of Bernoulli(0.5) random variables, each with sample size $n^*$. Calculate $\Delta_b $ as the difference of the average of the two samples for iteration $b$. Let $\hat c$ be the smallest point from $\{-1,-\frac{n^*-1}{n^*},\cdots,-\frac{1}{n^*},0,\frac{1}{n^*},\cdots, \frac{n^*-1}{n^*},1\}$ such that $\frac{1}{B}\sum_{b=1}^B1\{\Delta_b >\hat c\})\leq \tilde{\tilde\alpha}$.
\item Set $\hat c_p=\hat c_q^1 =\hat c_q^0 = \hat c$. 
\item Calculate the confidence sets according to \Cref{eq: finite sample CS for p,eq: finite sample CS for q}.
\item For a given pair $(j,j')$, set $\phi^{j,j'}=1$ if any of the following events happen: $\{\widehat{CS}_{p,U}<0,\widehat{CS}_{q,L}^1>0\},\{\widehat{CS}_{p,U}<0,\widehat{CS}_{q,L}^0>0\},\{\widehat{CS}_{p,L}>0,\widehat{CS}_{q,U}^1<0\},\{\widehat{CS}_{p,L}>0,\widehat{CS}_{q,U}^1<0\}$. 
\item Reject the null hypothesis if $\phi^{j,j'}=1$ for at least one pair $(j,j')$. 
    \end{enumerate}
\end{algo}

We report the rejection probability of the finite sample for the design in \Cref{section: continuous outcome simulation}, where we set $\delta_3=-0.5$. We can see that the power is lower than the asymptotic test that we reported in \Cref{table: types of violations no x}, particularly when the violation is relatively mild. This is not surprising because the asymptotic test is based on the assumption that the propensity score is consistently estimated, and therefore, we can consistently estimate the ranking of the propensity score. In contrast, for the finite sample test, the ranking of the propensity score is unknown. Another loss of power is that we only consider one type of interval that $1\{Y\geq 0.5\}$ here, whereas the asymptotic test considers all possible intervals of the form $1\{y\leq Y<y'\}$.  However, we still observe that for each given sample size, the rejection frequencies increase as the magnitude of the violation increases, as well as with the number of cases per judge.

To conclude this section, we want to emphasize that while there is a potential loss of power for our test, we trade this off for a substantial computational advantage. As discussed in \citet[][Supplementary material, page 9]{frandsen2023judging}, implementing a finite sample test can be quite computationally challenging when involving large-dimensional nonlinear optimization.\footnote{For this reason, we do not offer a simulation comparison with FLL's finite sample test, for which FLL does not provide a complete simulation study either.} On the contrary, our test requires little more than drawing Bernoulli random numbers and is very easy to implement. It thus serves as a useful complement to the existing literature.

\begin{figure}
     \centering
        \includegraphics[trim=10 200 10 200,scale=0.6]{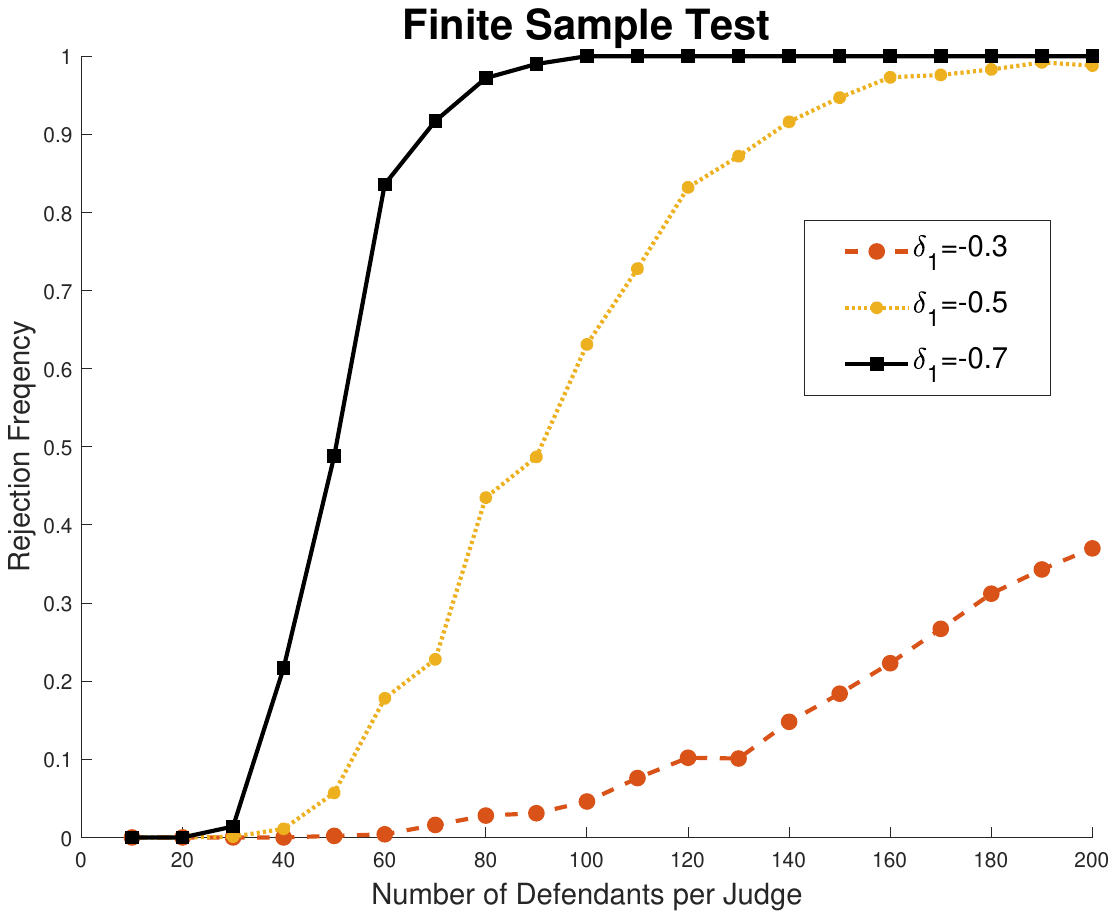} 
        \caption{Finite Sample Test Results}
        \label{fig: FLL_no_violation}
\end{figure}

\section{Lemmas and Intermediary Results}\label{app: lemmas}

\subsection{Lemmas for the proof of \Cref{thm: validity}}\label{app: theorem 2 lemmas}
This section collects useful Lemmas, intermediary results, and additional assumptions for establishing the asymptotic results in \Cref{thm: validity}.

\begin{lemma}\label{lemma: m and w influence}
    Suppose \Cref{ass: theta influence,ass: continuity} are satisfied, then uniformly in $\ell\in\mathcal L$, 
\begin{align}\label{eq: m1 hat influence}
&\sqrt{n}(\hat{m}_{1}(y, r_y,p,r_p,\hat{\theta})-{m}_{1}(y, r_y,p,r_p,\theta_0))\nonumber\\
=&\frac{1}{\sqrt{n}}\sum_{i=1}^n \phi_{m_1,i}(y, r_y,p,r_p,\theta_0)+o_p(1)\nonumber\\
\equiv&\frac{1}{\sqrt{n}}\sum_{i=1}^n \left( m_{1,i}(y, r_y,p,r_p,\theta_0)-{m}_{1}(y, r_y,p,r_p,\theta_0)+\triangledown_\theta m_{1}(y, r_y,p,r_p,\theta_0)\cdot  s(D_i,Z_i,\theta_0)\right) +o_p(1).
\end{align}
\begin{align}
&\sqrt{n}(\hat{m}_{0}(y, r_y,p,r_p,\hat{\theta})-{m}_{0}(y, r_y,p,r_p,\theta_0))\nonumber\\
=&\frac{1}{\sqrt{n}}\sum_{i=1}^n \phi_{m_0,i}(y, r_y,p,r_p,\theta_0)+o_p(1)\nonumber\\
\equiv& \frac{1}{\sqrt{n}}\sum_{i=1}^n \left( m_{0,i}(y, r_y,p,r_p,\theta_0)-{m}_{0}(y, r_y,p,r_p,\theta_0)+\triangledown_\theta m_{0}(y, r_y,p,r_p,\theta_0)\cdot  s(D_i,Z_i,\theta_0)\right) +o_p(1),\label{eq: m0 hat influence}
\end{align}
\begin{align}
&\sqrt{n}(\hat{w}(p,r_p,\hat{\theta})-{w}(p,r_p,\theta_0))\nonumber\\
=&\frac{1}{\sqrt{n}}\sum_{i=1}^n \phi_{w,i}(p,r_p,\theta_0)+o_p(1)\nonumber\\
\equiv&\frac{1}{\sqrt{n}}\sum_{i=1}^n \left( w_{i}(p,r_p,\theta_0)-{w}(p,r_p,\theta_0)+\triangledown_\theta w(p,r_p,\theta_0)\cdot  s(D_i,Z_i,\theta_0)\right) +o_p(1)\label{eq: w hat influence}
\end{align} 
where functions $m_d$ and $w$ are defined in \Cref{eq: m1 function,eq: m0 function,eq: w function} and
\begin{align*}
&m_{1i}(y, r_y,p,r_p,\theta)=D_i 1(y \leq Y_i\leq y+r_y)  1(p \leq P(Z_i,\theta)\leq p+r_p),\\
&m_{0i}(y, r_y,p,r_p,\theta)=(D_i-1) 1(y \leq Y_i\leq y+r_y)  1(p \leq P(Z_i,\theta)\leq p+r_p),\\
&w_i(p,r_p,\theta)= 1(p \leq P(Z_i,\theta)\leq p+r_p).
\end{align*}

\proof Let $f_P(p)$ denote the density function of $P(Z;\theta_0)$. Following \cite{hsu2021inference}, we calculate the derivatives for $m_d(y, r_y,p,r_p,\cdot)$ and $w(p,r_p,\cdot)$ as:
\begin{multline*} 
\triangledown_\theta m_{1}(y, r_y,p,r_p,\theta_0)= \mathbb{E}[D 1(y \leq Y\leq y+r_y)| P(Z,\theta_0)=p]\cdot f_P(p)
\mathbb{E}[\triangledown_\theta  P(Z,\theta_0)| P(Z,\theta_0)=p]\\
-\mathbb{E}[D 1(y \leq Y\leq y+r_y)| P(Z,\theta_0)=p+r_p]\cdot f_P(p+r_p)
\mathbb{E}[\triangledown_\theta  P(Z,\theta_0)| P(Z,\theta_0)=p+r_p],
\end{multline*}
\begin{multline*} 
\triangledown_\theta m_{0}(y, r_y,p,r_p,\theta_0)=\mathbb{E}[(D-1) 1(y \leq Y\leq y+r_y)| P(Z,\theta_0)=p]\cdot f_P(p)
\mathbb{E}[\triangledown_\theta  P(Z,\theta_0)| P(Z,\theta_0)=p]\\
-\mathbb{E}[(D-1) 1(y \leq Y\leq y+r_y)| P(Z,\theta_0)=p+r_p]\cdot f_P(p+r_p)
\mathbb{E}[\triangledown_\theta  P(Z,\theta_0)| P(Z,\theta_0)=p+r_p],
\end{multline*}
\begin{multline*} 
\triangledown_\theta w(p,r_p,\theta_0)=  f_P(p)
\mathbb{E}[\triangledown_\theta  P(Z,\theta_0)| P(Z,\theta_0)=p]-f_P(p+r_p)
\mathbb{E}[\triangledown_\theta  P(Z,\theta_0)| P(Z,\theta_0)=p+r_p].
\end{multline*}

Now we prove \Cref{eq: m1 hat influence}, the results for \Cref{eq: m0 hat influence,eq: w hat influence} are similar. Note that
\begin{align} \label{eq: m1 decompose}
    &\sqrt{n}(\hat{m}_{1}(y, r_y,p,r_p,\hat{\theta})-{m}_{1}(y, r_y,p,r_p,\theta_0)) \nonumber\\
    =& \sqrt{n}(\hat{m}_{1}(y, r_y,p,r_p,\hat{\theta})-{m}_{1}(y, r_y,p,r_p,\hat\theta)) + \sqrt{n}({m}_{1}(y, r_y,p,r_p,\hat\theta)-{m}_{1}(y, r_y,p,r_p,\theta_0)) \nonumber\\
    =& \sqrt{n}(\hat{m}_{1}(y, r_y,p,r_p,\hat{\theta})-{m}_{1}(y, r_y,p,r_p,\hat\theta)) + \triangledown_\theta m_{1}(y, r_y,p,r_p,\theta_0)' \sqrt{n}(\hat\theta -\theta_0) + o(\sqrt{n}\|\hat\theta-\theta_0\|)\nonumber\\
     =& \sqrt{n}(\hat{m}_{1}(y, r_y,p,r_p,\hat{\theta})-{m}_{1}(y, r_y,p,r_p,\hat\theta)) +\frac{1}{\sqrt{n}}\sum_{i=1}^n \triangledown_\theta m_{1}(y, r_y,p,r_p,\theta_0) s(D_i,Z_i,\theta_0) + o_p(1)
\end{align}
where the second equality holds because $m_1(\ell,\theta)$ is continuously differentiable in $\theta$ under \Cref{ass: continuity}-(2), and the third equality is due to \Cref{ass: theta influence}. 

Let $\hat{\mathbb G}_{m_1}(\theta,\ell)\equiv \sqrt{n}(\hat{m}_{1}(y, r_y,p,r_p,{\theta})-{m}_{1}(y, r_y,p,r_p,\theta))$, $\theta\in\Theta,\ell\in\mathcal L$. It remains to show that $\sup_{\ell\in\mathcal L}|\hat{\mathbb G}_{m_1}(\hat\theta,\ell) - \hat{\mathbb G}_{m_1}(\theta_0,\ell)|=o_p(1)$. 

By \Cref{ass: continuity}-(ii), the class of functions  $\{1(p \leq P(Z,\theta)\leq p+r_p) : \theta\in\Theta, p\in [0,1], r_p\in[0,1]\}$  is a Vapnik-Chervonenkis (VC) class of function. Therefore, the class of functions $\{1\{y\leq Y\leq y+r_y\}\times 1(p \leq P(Z,\theta):\theta\in\Theta,p\in [0,1], r_p\in[0,1],r_y\in[0,1]\}$ is also VC class. Hence, the process $\hat{\mathbb G}_{m_1}$ is stochastically equicontinuous with respect to $(\theta,\ell)$. Note $\hat\theta\convp \theta_0$, then there exist $\delta_n\downarrow 0$ such that with probability approaching one, $(\hat \theta,\ell) \in B((\theta_0,\ell),\delta_n)$, where $B((\theta_0,\ell),\delta_n) $ is a ball in $\Theta\times \mathcal L$ centered at $(\theta_0,\ell)$ with radius $\delta_n$. Therefore, 
\begin{align}
&\sup_{\ell\in\mathcal L}|\sqrt{n}(\hat{m}_{1}(y, r_y,p,r_p,\hat{\theta})-{m}_{1}(y, r_y,p,r_p,\hat\theta)) - \sqrt{n}(\hat{m}_{1}(y, r_y,p,r_p,{\theta_0})-{m}_{1}(y, r_y,p,r_p,\theta_0))| \nonumber\\
 =& \sup_{\ell\in\mathcal L}|\hat{\mathbb G}_{m_1}(\hat\theta,\ell) - \hat{\mathbb G}_{m_1}(\theta_0,\ell)|\nonumber\\
\leq& \sup_{\theta_0\in\Theta,\ell\in\mathcal L} \sup_{(\theta',\ell')\in B((\theta_0,\ell),\delta_n)} |\hat{\mathbb G}_{m_1}(\theta',\ell') - \hat{\mathbb G}_{m_1}(\theta_0,\ell)| =o_p(1).\label{eq: m1 equicontinuous}
\end{align}
where the last equality is by the stochastic equicontinuity of the process $\hat{\mathbb G}_{m_1}$. Combine both \Cref{eq: m1 equicontinuous,eq: m1 decompose}, the result then follows. 
\qed
\end{lemma}

\begin{lemma} \label{lemma: influence nu}
Suppose \Cref{ass: MTEI,ass: STC,ass: Vyt-Mon,ass: Exclusion,ass: theta influence,ass: continuity} are satisfied, then uniform in $\ell$, 
\begin{align} 
&\sqrt{n}(\hat\nu_1(y,r_y, p_1,p_2,r_p,\hat{\theta})-\nu_1(y,r_y, p_1,p_2,r_p,\theta_0))
=\frac{1}{\sqrt{n}}\sum_{i=1}^n \phi_{\nu_1,i}(y, r_y,p_1,p_2,r_p,\theta_0)+o_p(1),\label{eq: influence nu1}\\ \label{eq: influence nu0}
&\sqrt{n}(\hat\nu_0(y,r_y, p_1,p_2,r_p,\hat{\theta})-\nu_0(y,r_y, p_1,p_2,r_p,\theta_0))
=\frac{1}{\sqrt{n}}\sum_{i=1}^n \phi_{\nu_0,i}(y, r_y,p_1,p_2,r_p,\theta_0)+o_p(1),
\end{align}
where
\begin{align*}
&\phi_{\nu_1,i}(y, r_y,p_1,p_2,r_p,\theta_0)
=w(p_1,r_p,\theta_0)\cdot \phi_{m_1,i}(y, r_y,p_2,r_p,\theta_0)+
m_{1}(y, r_y,p_2,r_p,\theta_0)\cdot  \phi_{w,i}(p_1,r_p,\theta_0)\\
&-w(p_2,r_p,\theta_0)\cdot \phi_{m_1,i}(y, r_y,p_1,r_p,\theta_0)-
m_{1}(y, r_y,p_1,r_p,\theta_0)\cdot  \phi_{w,i}(p_2,r_p,\theta_0),\\
&\phi_{\nu_0,i}(y, r_y,p_1,p_2,,r_p,\theta_0)
=w(p_1,r_p,\theta_0)\cdot \phi_{m_0,i}(y, r_y,p_2,r_p,\theta_0)+
m_{0}(y, r_y,p_2,r_p,\theta_0)\cdot  \phi_{w,i}(p_1,r_p,\theta_0)\\
&-w(p_2,r_p,\theta_0)\cdot \phi_{m_0,i}(y, r_y,p_1,r_p,\theta_0)-
m_{0}(y, r_y,p_1,r_p,\theta_0)\cdot  \phi_{w,i}(p_2,r_p,\theta_0).
\end{align*}
Furthermore, 
\begin{align*}
\sqrt{n}(\widehat{\nu}_1(\cdot,\widehat{\theta})-\nu_1(\cdot,\theta_0))
\Rightarrow \Phi_{\nu_1} (\cdot),~~~~
\sqrt{n}(\widehat{\nu}_0(\cdot,\widehat{\theta})-\nu_0(\cdot,\theta_0))
\Rightarrow \Phi_{\nu_0} (\cdot),
\end{align*}
where $\Phi_{\nu_1} (\cdot)$ and $\Phi_{\nu_0} (\cdot)$ are Gaussian processes with variance-covariance 
kernel generated by $\phi_{\nu_1}(\cdot,\theta_0)$ and $\phi_{\nu_0}(\cdot,\theta_0)$, respectively. 

\proof We show \Cref{eq: influence nu1}. \Cref{eq: influence nu0} holds analogously. Recall 
\[
   \hat\nu_1(\ell)=\hat m_{1}(y, r_y,p_2,r_p,\hat \theta)\cdot  
\hat w(p_1,r_p,\hat \theta)- \hat m_{1}(y, r_y,p_1,r_p,\hat \theta)\cdot  
\hat w(p_2,r_p,\hat \theta)
\]
To save space, for generic $\ell$, we write $\hat m_1(\hat\theta)\equiv \hat m_1(\ell,\hat\theta)$ and $ \hat w(\hat \theta)\equiv  \hat w(\ell,\hat \theta)$. Similarly, $m_1(\theta_0)\equiv m_1(\ell,\theta_0)$ and $w(\theta_0)\equiv w(\ell,\theta_0)$. Then, 
\begin{align*}
    \hat m_{1}(\hat \theta) \hat w(\hat \theta) -&m_1(\theta_0)w(\theta_0 )
     = ( \hat m_{1}(\hat \theta)-m_{1}(\theta_0) + m_{1}(\theta_0)) (\hat w(\hat \theta) - w (\theta_0)+ w(\theta_0))-m_1(\theta_0)w(\theta_0 )\\
      =& ( \hat m_{1}(\hat \theta)-m_{1}(\theta_0)) w(\theta_0) + (\hat w(\hat \theta) - w (\theta_0))m_{1}(\theta_0)  + ( \hat m_{1}(\hat \theta)-m_{1}(\theta_0))(\hat w(\hat \theta) - w (\theta_0))\\
       =& ( \hat m_{1}(\hat \theta)-m_{1}(\theta_0)) w(\theta_0) + (\hat w(\hat \theta) - w (\theta_0))m_{1}(\theta_0)  + o_p\left(\frac{1}{\sqrt{n}}\right),
\end{align*}
where the last equality is because $\hat m_{1}(\hat \theta)-m_{1}(\theta_0)=O_p(1/\sqrt{n})$ and $\hat w(\hat \theta) - w (\theta_0)=O_p(1/\sqrt{n})$ by \Cref{lemma: m and w influence}. Then we have 
\begin{align*}
     \hat\nu_1(\ell) -\nu_1(\ell) =&w(p_1,r_p,\theta_0)\cdot( \hat m_1(y, r_y,p_2,r_p,\hat\theta) - m_1(y, r_y,p_2,r_p,\theta_0))\\&~~  +
m_{1}(y, r_y,p_2,r_p,\theta_0)\cdot  (\hat w(p_1,r_p,\hat\theta)-w(p_1,r_p,\theta_0))\\
&~~-w(p_2,r_p,\theta_0)\cdot (\hat m_1(y, r_y,p_1,r_p,\hat\theta)-m_1(y, r_y,p_1,r_p,\theta_0))\\&~~  -
m_{1}(y, r_y,p_1,r_p,\theta_0)\cdot (\hat w(p_2,r_p,\hat\theta) - w(p_2,r_p,\theta_0))+ o_p\left(\frac{1}{\sqrt{n}}\right).
\end{align*}
\Cref{eq: influence nu1} then follows by inserting \Cref{eq: m0 hat influence,eq: m1 hat influence,eq: w hat influence} to the above equation. 

Finally, under \Cref{ass: continuity}, each element of $\triangledown_\theta m_{1}(y, r_y,p,r_p,\theta_0)$ is Lipschitz continuous in $y$, $r_y$, $p$, $r_p$ and it implies that $\{\partial m_{1}(y, r_y,p,r_p,\theta_0)/\partial \theta_j : (y, r_y,p,r_p) \in [0,1]^4\}$ is a VC class of functions for each $j$.  Similarly, each element of $\triangledown_\theta w(p,r_p,\theta_0)$ is Lipschitz continuous in $p$, $r_p$.  It follows that $\{\phi_{m_1}(y, r_y,p,r_p,\theta_0):  (y, r_y,p,r_p) \in [0,1]^4 \}$, $\{\phi_{m_0}(y, r_y,p,r_p,\theta_0):  (y, r_y,p,r_p) \in [0,1]^4 \}$ and $\{\phi_{w}(p,r_p,\theta_0):  (p,r_p) \in [0,1]^2 \}$ are all VC classes of functions.  weak convergence follows from the fact that $\{\phi_{\nu_0}(y, r_y,p_1,p_2,,r_p,\theta_0):  (y, r_y,p_1,p_2,r_p) \in [0,1]^5 \}$ and $\{\phi_{\nu_0}(y, r_y,p_1,p_2,,r_p,\theta_0):  (y, r_y,p_1,p_2,r_p) \in [0,1]^5 \}$ are both VC classes of functions. Therefore, we have 
\begin{align*}
\sqrt{n}(\widehat{\nu}_1(\cdot,\widehat{\theta})-\nu_1(\cdot,\theta_0))
\Rightarrow \Phi_{\nu_1} (\cdot),~~~
\sqrt{n}(\widehat{\nu}_0(\cdot,\widehat{\theta})-\nu_0(\cdot,\theta_0))
\Rightarrow \Phi_{\nu_0} (\cdot).
\end{align*}\qed
\end{lemma}

\begin{lemma} \label{lemma: wb influence nu}
Suppose \Cref{ass: MTEI,ass: STC,ass: Vyt-Mon,ass: Exclusion,ass: theta influence,ass: wb theta influence,ass: continuity} are satisfied, then uniform in $\ell$ over $\mathcal L$,
\begin{align} 
&\sqrt{n}(\hat\nu_1^{b}(y,r_y, p_1,p_2,r_p,\hat{\theta}^{b})-\hat\nu_1(y,r_y, p_1,p_2,r_p,\hat{\theta}))\nonumber\\
=&\frac{1}{\sqrt{n}}\sum_{i=1}^n (W_i-1)\phi_{\nu_1,i}(y, r_y,p_1,p_2,r_p,\theta_0)+o_p(1),\label{eq: wb influence nu1}\\ 
&\sqrt{n}(\hat\nu_0^{b}(y,r_y, p_1,p_2,r_p,\hat{\theta}^{b})-\hat\nu_0(y,r_y, p_1,p_2,r_p,\hat{\theta}))\nonumber\\
=&\frac{1}{\sqrt{n}}\sum_{i=1}^n (W_i-1)\phi_{\nu_0,i}(y, r_y,p_1,p_2,r_p,\theta_0)+o_p(1),\label{eq: wb influence nu0}
\end{align}
where
$\phi_{\nu_1,i}(y, r_y,p_1,p_2,r_p,\theta_0)$ and 
$\phi_{\nu_0,i}(y, r_y,p_1,p_2,,r_p,\theta_0)$ are the same as in \Cref{lemma: influence nu}.
\end{lemma}

The proof to \Cref{lemma: wb influence nu} is similar to \Cref{lemma: influence nu} and is therefore omitted. 

\begin{lemma} \label{lemma: wb variance}
Suppose \Cref{ass: MTEI,ass: STC,ass: Vyt-Mon,ass: Exclusion,ass: theta influence,ass: wb theta influence,ass: continuity} are satisfied, then $\hat{\sigma}^2_{d}(\ell)$ defined in (\ref{eq: variance estimator}) satisfies that for $d=0,1$, 
$\sup_{\ell}|\hat{\sigma}^2_{d}(\ell)-{\sigma}^2_{d}(\ell)|=o_p(1)$.

\proof Recall that for a given $\ell\in\mathcal L$,
\begin{equation*}
 \hat{\sigma}^2_{d}(\ell)
 =\frac{n}{B}\sum_{b=1}^B \big(\hat{\nu}^{b}_d(\ell)-
 \overline{\hat{\nu}^{b}_d}(\ell)\big)^2,\text{ where }\overline{\hat{\nu}}^{b}_d(\ell)=\frac{1}{B}\sum_{b=1}^B
\hat{\nu}^{b}_d(\ell).
\end{equation*}
It can be written as
\begin{equation}\label{eq: variance decompose}
\hat{\sigma}^2_{d}(\ell) = \frac{n}{B}\sum_{b=1}^B \big(\hat{\nu}^{b}_d(\ell)-
 \hat{\nu}_d(\ell)\big)^2 + 2 \frac{n}{B}\sum_{b=1}^B \big(\hat{\nu}^{b}_d(\ell)-
 \hat{\nu}_d(\ell)\big) \big (\hat{\nu}_d(\ell) -\overline{\hat{\nu}^{b}_d}(\ell) \big ) + \frac{n}{B}\sum_{b=1}^B \big (\hat{\nu}_d(\ell) -\overline{\hat{\nu}^{b}_d}(\ell) \big )^2
\end{equation}

We first consider the second term on the right-hand side of \Cref{eq: variance decompose}. Let $\bar W_i = \frac{1}{B} \sum_{b=1}^BW_i^b$, Using \Cref{lemma: wb influence nu}, we know that for a given $b=1,2,\cdots,B$, and uniformly over $\ell\in\mathcal L$,
\[
\hat{\nu}^{b}_d(\ell)-  \hat{\nu}_d(\ell) = \frac{1}{n}\sum_{i=1}^n(W_i^b-1)\phi_{\nu_d,i}(\ell,\theta_0) + o_p(1).
\] 
So it can be written as
\begin{align*}
    &\frac{n}{B}\sum_{b=1}^B \big(\hat{\nu}^{b}_d(\ell)-
 \hat{\nu}_d(\ell)\big) \big (\hat{\nu}_d(\ell) -\overline{\hat{\nu}^{b}_d}(\ell) \big )\\
 = &\frac{1}{B} \frac{1}{n} \sum_{b=1}^B \big( \sum_{i=1}^n(W_i^b-1)\phi_{\nu_d,i}(\ell,\theta_0) \big)\big( \sum_{i=1}^n(\bar W_i-1)\phi_{\nu_d,i}(\ell,\theta_0) \big)+o_p(1)\\
  =& \frac{1}{B} \frac{1}{n}  \sum_{b=1}^B  \sum_{i=1}^n(W_i^b-1)(\bar W_i-1) \phi^2_{\nu_d,i}(\ell,\theta_0)  + \frac{1}{B} \frac{1}{n}  \sum_{b=1}^B \sum_{i\neq j}^n (W_i^b-1)(\bar W_j-1)\phi_{\nu_d,i}(\ell,\theta_0) \phi_{\nu_d,j}(\ell,\theta_0) +o_p(1) \\
   = &\frac{1}{B^2} \frac{1}{n}  \sum_{b=1}^B  \sum_{i=1}^n(W_i^b-1)^2 \phi^2_{\nu_d,i}(\ell,\theta_0)  + \frac{1}{B^2} \frac{1}{n}  \sum_{b=1}^B \sum_{b'\neq b}^B \sum_{i=1}^n (W_i^b-1)( W_j^{b'}-1) \phi^2_{\nu_d,i}(\ell,\theta_0)\\&~~+ \frac{1}{B} \frac{1}{n}  \sum_{b=1}^B \sum_{i\neq j}^n (W_i^b-1)(\bar W_j-1)\phi_{\nu_d,i}(\ell,\theta_0) \phi_{\nu_d,j}(\ell,\theta_0) +o_p(1) 
\end{align*}
The first right-hand side term is of order $\frac{1}{B}$ and is negligible as $B\rightarrow \infty$. The second term on the right-hand side is negligible because $E[(W_i^b-1)( W_i^{b'}-1)|(Y,D,Z)]=0$ as long as $b\neq b'$. The third term on the right-hand side is negligible because $E[(W_i^b-1)( W_j^b-1)|(Y,D,Z)]=0$ as long as $i\neq j$. For similarly reasoning, the third right-hand side term of \Cref{eq: variance decompose} is also negligible as $B\rightarrow \infty$. 

Now consider the first term on the right-hand side of \Cref{eq: variance decompose}. Uniformly over $\ell$, 
\begin{align*}
 &\frac{n}{B}\sum_{b=1}^B \big(\hat{\nu}^{b}_d(\ell)- \hat{\nu}_d(\ell)\big)^2= \frac{1}{B} \frac{1}{n} \sum_{b=1}^B \big( \sum_{i=1}^n(W_i^b-1)\phi_{\nu_d,i}(\ell,\theta_0) \big)^2 +o_p(1) \\
  =& \frac{1}{B} \frac{1}{n}  \sum_{b=1}^B  \sum_{i=1}^n(W_i^b-1)^2 \phi^2_{\nu_d,i}(\ell,\theta_0)  + \frac{1}{B} \frac{1}{n}  \sum_{b=1}^B \sum_{i=1}^n \sum_{j\neq i}^n (W_i^b-1)(W_j^b-1)\phi_{\nu_d,i}(\ell,\theta_0) \phi_{\nu_d,j}(\ell,\theta_0) +o_p(1).
\end{align*}
Conditioning on the sample, because $W_i^b$ are i.i.d. across $b$ and $i$, has expectation and variance equal to one, we know $E[(W_i^b-1)(W_j^b-1)|(Y,D,Z)]=0$ and $E[(W_i^b-1)^2|(Y,D,Z)]=1$. As $B\rightarrow \infty$, the right-hand side converges in probability (with respect to the distribution of $\{W^b\}_{b=1}^B$) to $\frac{1}{n}  \sum_{i=1}^n \phi^2_{\nu_d,i}(\ell,\theta_0) + o_p(1)$, which in turn converges to $\sigma^2_{d}(\ell)$ uniformly over $\ell$ as $n\rightarrow \infty$. \qed

\end{lemma}

 \subsection{The influence function with covariate case} \label{app: influence covariate}

In this subsection, we derive the influence function for estimating $\nu_d(\ell)$ in the presence of covariates. First, we estimate $\theta_0\equiv (\theta_{0z},\theta_{0x})$ by MLE, 
\begin{align}
\hat{\theta}&=\argmax_{\theta\in\Theta} \frac{1}{n}\sum_{i=1}^n \log f(Y_i,D_i,Z_i,X_i,\theta) \nonumber\\
&\equiv \argmax_{\theta\in \Theta} \frac{1}{n}\sum_{i=1}^n D_i \log P(Z_i,X_i,\theta)+(1-D_i) \log (1-P(Z_i,X_i,\theta)).\label{eq: MLE with X}
\end{align}
where $P(z,x,\theta)$ is parameterized and depends on $(z,x)$ and $\theta\equiv (\theta_z',\theta_x')'$ through $z'\theta_z+x'\theta_x$. For example, $P(z,x,\theta) = \Phi(z'\theta_z+x'\theta_x)$ for Probit or $P(z,x,\theta) = \frac{exp(z'\theta_z+x'\theta_x)}{1+\exp(z'\theta_z+x'\theta_x)}$ for Logit.

As in \Cref{app: theorem 2 lemmas}, we make the following assumptions. 

\begin{assumption} \label{ass: vc class with x} Assuming following conditions hold
\begin{enumerate}
\item   The conditional density of $(Y,X,D)$ given $P(Z,X,\theta_0)=p$, denoted by $f_{Y,X,D|P}(y,x,d|p)$, is Lipschitz continuous in $(y,x,p)$ over the joint support of $(Y,X,P)$ for $d=0,1$.
    \item For all $z\in\mathcal Z$ and $x\in\mathcal X$, $P(z,x,\theta)$ is continuously differentiable in $\theta$ at $\theta_0$ with bounded derivatives. 
\end{enumerate}
\end{assumption}

\begin{assumption}\label{ass: theta beta influence}
    The estimator $\hat\theta$, $\hat{\beta}_1$, $\hat{\beta}_0$ admits an influence function of the following form,
\begin{align} \label{eq: theta beta influence}
          &\sqrt{n} (\hat\theta-\theta_0) = \frac{1}{\sqrt{n}}\sum_{i=1}^n s_{\theta_0}(D_i,Z_i,X_i,\theta_0) + o_p(1),\\
            &\sqrt{n} (\hat\beta_1-\beta_1) = \frac{1}{\sqrt{n}}\sum_{i=1}^n s_{\beta_1}(D_i,Y_i, Z_i,X_i,\beta_1) + o_p(1),\label{eq: beta1 influence}\\
        &\sqrt{n} (\hat\beta_0-\beta_0) = \frac{1}{\sqrt{n}}\sum_{i=1}^n s_{\beta_0}(D_i,Y_i, Z_i,X_i,\beta_0) + o_p(1),\label{eq: beta0 influence}
    \end{align}
    where $s_{\theta_0}(\cdot)$,
   $s_{\beta_1}(\cdot)$ and $s_{\beta_0}(\cdot)$  are measurable, satisfying 
   $E[s_{\theta_0}(D_i,Z_i,X_i,\theta_0)]=0$,$E[s_{\beta_1}(D_i,Y_i, Z_i,X_i,\beta_1)]=0$,
   $E[s_{\beta_0}(D_i,Y_i, Z_i,X_i,\beta_0)]=0$, $E[\sup_{\theta}\|s_{\theta_0}(D_i,Z_i,\theta)\|^{2+\delta}]<\infty$,
   $E[\sup_{\beta}\|s_{\beta_1}(D_i,Y_i, Z_i,X_i,\beta)\|^{2+\delta}]<\infty$, and  $E[\sup_{\beta}\|s_{\beta_0}(D_i,Y_i, Z_i,X_i,\beta)\|^{2+\delta}]<\infty$ for some $\delta>0$.
\end{assumption}

Note that under similar conditions as in Section 4 of Hsu, Liao and Lin (2022, Econometric Reviews), (\ref{eq: beta1 influence}) and (\ref{eq: beta0 influence}) would hold.
We define the following quantities for generic $(y, r_y,p,r_p,b, \theta)$:
\begin{align} \label{eq: md functions}
&m_{1}(y, r_y,p,r_p,b, \theta)=\mathbb{E}[D 1(y \leq Y-X'b\leq y+r_y)  1(p \leq P(Z,X,\theta)\leq p+r_p)],\\
&m_{0}(y, r_y,p,r_p,b,\theta)=\mathbb{E}[(D-1) 1(y \leq Y-X'b\leq y+r_y)  1(p \leq P(Z,X,\theta)\leq p+r_p)],\\ 
&w(p,r_p,\theta)=\mathbb{E}[ 1(p \leq P(Z,X,\theta)\leq p+r_p)].
\end{align}
Let $f_P(p)$ denote the density function of $P(Z,X,\theta_0)\equiv \mathbb P(D=1|X,Z;\theta_0)$. Following the calculation in \cite{hsu2021inference}, we can analogously obtain the derivatives with respect to $\theta$, evaluating at the true parameter values $(\beta_1,\beta_0,\theta_0)$ as
\begin{align*}
&\triangledown_\theta m_{1}(y, r_y,p,r_p,\beta_1,\theta_0)\\
= &\mathbb{E}[D 1(y \leq Y-X'\beta_1\leq y+r_y)| P(Z,X,\theta_0)=p]\cdot f_P(p)
\mathbb{E}[\triangledown_\theta  P(Z,X,\theta_0)| P(Z,X,\theta_0)=p]\\
&-\mathbb{E}[D 1(y \leq Y-X'\beta_1\leq y+r_y)| P(Z,X,\theta_0)=p+r_p]\cdot f_P(p+r_p)
\mathbb{E}[\triangledown_\theta  P(Z,X,\theta_0)| P(Z,X,\theta_0)=p+r_p],\\
&\triangledown_\theta m_{0}(y, r_y,p,r_p,\beta_0,\theta_0)\\
= &\mathbb{E}[(D-1) 1(y \leq Y-X'\beta_0\leq y+r_y)| P(Z,X,\theta_0)=p]\cdot f_P(p)
\mathbb{E}[\triangledown_\theta  P(Z,X,\theta_0)| P(Z,X,\theta_0)=p]\\
&-\mathbb{E}[(D-1) 1(y \leq Y-X'\beta_0\leq y+r_y)| P(Z,X,\theta_0)=p+r_p]\cdot f_P(p+r_p)
\mathbb{E}[\triangledown_\theta  P(Z,X,\theta_0)| P(Z,X,\theta_0)=p+r_p],\\
&\triangledown_\theta w(p,r_p,\theta_0)\\
= & f_P(p)
\mathbb{E}[\triangledown_\theta  P(Z,X,\theta_0)| P(Z,X,\theta_0)=p]-f_P(p+r_p)
\mathbb{E}[\triangledown_\theta  P(Z,X,\theta_0)| P(Z,X,\theta_0)=p+r_p].
\end{align*}
In addition, let $f_{u_d|zxd}(y|z,x,d)$ denote the conditional density of $U_d$ conditional on $(Z,X,D)=(z,x,d)$, then the derivatives with respect to $\beta$, evaluating at the true parameter values $(\beta_1,\beta_0,\theta_0)$ are
\begin{align*}
&\triangledown_\beta m_{1}(y, r_y,p,r_p,\beta_1,\theta_0)\\
=&\mathbb{E}[P(Z,X,\theta_0) (f_{u_1|zxd}(y+r_y|Z,X,1)-
f_{u_1|zxd}(y|Z,X,1) X\cdot  1(p \leq P(Z,X,\theta)\leq p+r_p)] ],\\
&\triangledown_\beta m_{0}(y, r_y,p,r_p,\beta_0,\theta_0)\\
=&\mathbb{E}[(1-P(Z,X,\theta_0)) (f_{u_0|zxd}(y+r_y|Z,X,0)-
f_{u_0|zxd}(y|Z,X,0)X \cdot  1(p \leq P(Z,X,\theta)\leq p+r_p)] ].
\end{align*}

Let the estimators for $m_{1}(y, r_y,p,r_p,\beta,\theta)$, $m_{0}(y, r_y,p,r_p,\beta,\theta)$ and $w(p,r_p,\theta)$ be
\begin{align*}
&\hat{m}_{1}(y, r_y,p,r_p,\beta,\theta)=\frac{1}{n}\sum_{i=1}^n m_{1,i}(y, r_y,p,r_p,\beta,\theta),\\
&\hat{m}_{0}(y, r_y,p,r_p,\beta,\theta)=\frac{1}{n}\sum_{i=1}^n m_{0,i}(y, r_y,p,r_p,\beta,\theta),\\
&\hat{w}(p,r_p,\theta) = \frac{1}{n}\sum_{i=1}^n w_i(p,r_p,\theta).
\end{align*}
where  
\begin{align*}
&m_{1,i}(y, r_y,p,r_p,\beta,\theta)=
D_i 1(y \leq Y_i-X_i\beta \leq y+r_y)  1(p \leq P(Z_i,X_i,\theta)\leq p+r_p),\\
&m_{0,i}(y, r_y,p,r_p,\beta,\theta)=
(1-D_i) 1(y \leq Y_i-X_i\beta \leq y+r_y)  1(p \leq P(Z_i,X_i,\theta)\leq p+r_p),\\
&w_i(p,r_p,\theta)= 1(p \leq P(Z_i,X_i,\theta)\leq p+r_p),
\end{align*}
and
\begin{align*}
&\sqrt{n}(\hat{m}_{1}(y, r_y,p,r_p,\hat{\beta}_1,\hat{\theta})-{m}_{1}(y, r_y,p,r_p,{\beta_1},\theta_0))\\
=&\frac{1}{\sqrt{n}}\sum_{i=1}^n m_{1,i}(y, r_y,p,r_p,\beta_1,\theta_0)-{m}_{1}(y, r_y,p,r_p,\beta_1,\theta_0)
+\triangledown_\theta m_{1}(y, r_y,p,r_p,\beta_1,\theta_0)\cdot  s(D_i,Z_i,X_i,\theta_0)\\
&~~~~~+\triangledown_\beta m_{1}(y, r_y,p,r_p,\beta_1,\theta_0)\cdot  s_{\beta_1}(D_i,Y_i,Z_i,X_i,\beta_1)  +o_p(1)\\
\equiv& \frac{1}{\sqrt{n}}\sum_{i=1}^n \phi_{m_1,i}(y, r_y,p,r_p,\beta_1,\theta_0)+o_p(1),
\end{align*}
\begin{align*}
&\sqrt{n}(\hat{m}_{0}(y, r_y,p,r_p,\hat{\beta}_0,\hat{\theta})-{m}_{0}(y, r_y,p,r_p,\beta_0,\theta_0))\\
=&\frac{1}{\sqrt{n}}\sum_{i=1}^n m_{0,i}(y, r_y,p,r_p,\beta_0,\theta_0)-{m}_{0}(y, r_y,p,r_p,\beta_0,\theta_0)
+\triangledown_\theta m_{0}(y, r_y,p,r_p,\beta_0,\theta_0)\cdot  s(D_i,Z_i,X_i,\theta_0)
\\
&~~~~~+\triangledown_\beta m_{0}(y, r_y,p,r_p,\beta_0,\theta_0)\cdot  s_{\beta_0}(D_i,Y_i,Z_i,X_i,\beta_0)  +o_p(1)\\
\equiv &\frac{1}{\sqrt{n}}\sum_{i=1}^n \phi_{m_0,i}(y, r_y,p,r_p,\theta_0)+o_p(1),
\end{align*}
\begin{align*}
&\sqrt{n}(\hat{w}(p,r_p,\hat{\theta})-{w}(p,r_p,\theta_0))\\
=&\frac{1}{\sqrt{n}}\sum_{i=1}^n w_{i}(p,r_p,\theta_0)-{w}(p,r_p,\theta_0)
+\triangledown_\theta w(p,r_p,\theta_0)\cdot  s(D_i,Z_i,X_i,\theta_0) +o_p(1)\\
\equiv& \frac{1}{\sqrt{n}}\sum_{i=1}^n \phi_{w,i}(p,r_p,\theta_0)+o_p(1).
\end{align*} 

By \Cref{ass: vc class with x}, all elements of $\triangledown_\theta m_{1}(y, r_y,p,r_p,\beta_1,\theta_0)$,
$\triangledown_\beta m_{1}(y, r_y,p,r_p,\beta_1,\theta_0)$,
 $\triangledown_\theta m_{0}(y, r_y,p,r_p,\beta_0,\theta_0)$, and $\triangledown_\beta m_{0}(y, r_y,p,r_p,\beta_0,\theta_0)$, are Lipschitz continuous in $y$, $r_y$, $p$, $r_p$, and each element of $\triangledown_\theta w(p,r_p,\theta_0)$ is Lipschitz continuous in $p$, $r_p$. It follows that 
$\{\phi_{m_1}(y, r_y,p,r_p,\beta_1\theta_0):  (y, r_y,p,r_p) \in [0,1]^4 \}$, $\{\phi_{m_0}(y, r_y,p,r_p,\beta_0,\theta_0):  (y, r_y,p,r_p) \in [0,1]^4 \}$ and $\{\phi_{w}(p,r_p,\theta_0):  (p,r_p) \in [0,1]^2 \}$ are all VC classes of functions. 
Finally, let  
\begin{align*}
&\nu_1(y,r_y, p_1,p_2,r_p,\beta_1,\theta_0)=m_{1}(y, r_y,p_2,r_p,\beta_1,\theta_0)\cdot  
w(p_1,r_p,\theta_0)- m_{1}(y, r_y,p_1,r_p,\beta_1,\theta_0)\cdot  
w(p_2,r_p,\theta_0),\\
&\nu_0(y,r_y, p_1,p_2,r_p,\beta_1,\beta_0,\theta_0)=m_{0}(y, r_y,p_2,r_p,\beta_0,\theta_0)\cdot  
w(p_1,r_p,\theta_0)- m_{0}(y, r_y,p_1,r_p,\beta_0,\theta_0)\cdot  
w(p_2,r_p,\theta_0),\\
&\hat{\nu}_1(y,r_y, p_1,p_2,r_p,\hat{\beta}_1,\hat{\theta})=\hat{m}_{1}(y, r_y,p_2,r_p,\hat{\beta}_1,\hat{\theta})\cdot  
\hat{w}(p_1,r_p,\hat{\theta})- \hat{m}_{1}(y, r_y,p_1,r_p,\hat{\beta}_1,\hat{\theta})\cdot  
\hat{w}(p_2,r_p,\hat{\theta}),\\
&\hat{\nu}_0(y,r_y, p_1,p_2,r_p,\hat{\beta}_0,\hat{\theta})=\hat{m}_{0}(y, r_y,p_2,r_p,\hat{\beta}_0,\hat{\theta})\cdot  
\hat{w}(p_1,r_p,\hat{\theta})- \hat{m}_{0}(y, r_y,p_1,r_p,\hat{\beta}_0,\hat{\theta})\cdot  
\hat{w}(p_2,r_p,\hat{\theta}).
\end{align*}

\begin{lemma}
Suppose \Cref{ass: MTEI,ass: STC,ass: Vyt-Mon,ass: Exclusion,ass: theta beta influence,ass: functional form x,ass: vc class with x} are satisfied, then, 
\begin{align} \label{eq: influence beta nu1}
&\sqrt{n}(\hat\nu_1(y,r_y, p_1,p_2,r_p,\hat{\beta}_1,\hat{\theta})-\nu_1(y,r_y, p_1,p_2,r_p,{\beta}_1,\theta_0))
=\frac{1}{\sqrt{n}}\sum_{i=1}^n \phi_{\nu_1,i}(y, r_y,p_1,p_2,r_p,{\beta}_1,\theta_0)+o_p(1),\\ \label{eq: influence beta nu0}
&\sqrt{n}(\hat\nu_0(y,r_y, p_1,p_2,r_p,\hat{\beta}_0,\hat{\theta})-\nu_0(y,r_y, p_1,p_2,r_p,{\beta}_0,\theta_0))
=\frac{1}{\sqrt{n}}\sum_{i=1}^n \phi_{\nu_0,i}(y, r_y,p_1,p_2,r_p,{\beta}_0,\theta_0)+o_p(1),
\end{align}
where
\begin{align*}
&\phi_{\nu_1,i}(y, r_y,p_1,p_2,r_p,{\beta}_1,\theta_0)
\\
=&w(p_1,r_p,\theta_0)\cdot \phi_{m_1,i}(y, r_y,p_2,r_p,{\beta}_1,\theta_0)+
m_{1}(y, r_y,p_2,r_p,{\beta}_1,\theta_0)\cdot  \phi_{w,i}(p_1,r_p,\theta_0)\\
&~~-w(p_2,r_p,\theta_0)\cdot \phi_{m_1,i}(y, r_y,p_1,r_p,{\beta}_1,\theta_0)+
m_{1}(y, r_y,p_1,r_p,{\beta}_1,\theta_0)\cdot  \phi_{w,i}(p_2,r_p,\theta_0),\\
&\phi_{\nu_0,i}(y, r_y,p_1,p_2,r_p,{\beta}_0,\theta_0)
\\
=&w(p_1,r_p,\theta_0)\cdot \phi_{m_0,i}(y, r_y,p_2,r_p,{\beta}_0,\theta_0)+
m_{0}(y, r_y,p_2,r_p,{\beta}_0,\theta_0)\cdot  \phi_{w,i}(p_1,r_p,\theta_0)\\
&~~-w(p_2,r_p,\theta_0)\cdot \phi_{m_0,i}(y, r_y,p_1,r_p,{\beta}_0,\theta_0)+
m_{0}(y, r_y,p_1,r_p,{\beta}_0,\theta_0)\cdot  \phi_{w,i}(p_2,r_p,\theta_0).
\end{align*}
\end{lemma}

The proofs are similar to those in \Cref{app: theorem 2 lemmas}, so we omit the details.
\end{spacing}

\section{Additional Empirical Results}
\begin{table}[ht]
\tiny
\caption{FLL Semi-parametric test, B-Spline}
\begin{tabular}{l|ccc|ccc|ccc|ccc} \label{table: FLL p values}
 &\multicolumn{3}{c}{2 Knots}&\multicolumn{3}{c}{3 Knots}& \multicolumn{3}{c}{4 Knots}&\multicolumn{3}{c}{5 Knots}\\
 &$p_f$  &$p_s$&combined  &$p_f$  &$p_s$&combined &$p_f$  &$p_s$&combined  & $p_f$  &$p_s$&combined \\  \hline\hline
All	&	0.13	&	1.00	&	0.25	&	0.06	&	1.00	&	0.11	&	0.02	&	1.00	&	0.04	&	na	&	1.00	&	1.00	\\
Aggressive Assault	&	0.02	&	1.00	&	0.04	&	0.01	&	0.91	&	0.02	&	0.00	&	0.96	&	0.00	&	na	&	1.00	&	1.00	\\
Robbery	&	0.13	&	0.73	&	0.25	&	0.06	&	0.99	&	0.11	&	0.02	&	0.44	&	0.04	&	na	&	0.45	&	0.90	\\
Drug Sale	&	0.18	&	0.83	&	0.36	&	0.09	&	0.33	&	0.18	&	0.03	&	0.55	&	0.06	&	na	&	0.73	&	1.00	\\
Drug Possession	&	0.45	&	0.82	&	0.89	&	0.31	&	1.00	&	0.61	&	0.14	&	0.99	&	0.27	&	na	&	0.98	&	1.00	
\\\hline\hline
\end{tabular}
\end{table}

\clearpage
\bibliographystyle{econometrica}
\bibliography{Bib_MTE}

@article{heckman2005structural,
  title={Structural equations, treatment effects, and econometric policy evaluation 1},
  author={Heckman, James J and Vytlacil, Edward},
  journal={Econometrica},
  volume={73},
  number={3},
  pages={669--738},
  year={2005},
  publisher={Wiley Online Library}}

@article{AndrewsShi2013,
	Author = {Andrews, Donald W. K. and Shi, Xiaoxia},
	Doi = {10.3982/ECTA9370},
	Issn = {1468-0262},
	Journal = {Econometrica},
	Number = {2},
	Pages = {609--666},
	Publisher = {Blackwell Publishing Ltd},
	Title = {Inference Based on Conditional Moment Inequalities},
	Url = {http://dx.doi.org/10.3982/ECTA9370},
	Volume = {81},
	Year = {2013}}

@article{vytlacil2002independence,
	Author = {Vytlacil, Edward},
	Journal = {Econometrica},
	Number = {1},
	Pages = {331--341},
	Publisher = {Wiley Online Library},
	Title = {Independence, Monotonicity, and Latent Index Models: An Equivalence Result},
	Volume = {70},
	Year = {2002}}

@article{MW2017,
author = {Ismael Mourifi\'e and Yuanyuan Wan},
title = {Testing Local Average Treatment Effect Assumptions},
journal = {The Review of Economics and Statistics},
volume = {99},
number = {2},
pages = {305-313},
year = {2017}
}

@article{kitagawa2015,
	author = {Kitagawa, Toru},
	title = { A Test for Instrument Validity},
	journal = {Econometrica},
        number = {5},
       volume = {83},
	pages = {2043--2063},
	year = {2015}
	}

@article{huber2015testing,
  title={Testing instrument validity for LATE identification based on inequality moment constraints},
  author={Huber, Martin and Mellace, Giovanni},
  journal={Review of Economics and Statistics},
  volume={97},
  number={2},
  pages={398--411},
  year={2015},
  publisher={MIT Press}
}

@article{Brinch2017,
author = {Christian N. Brinch and Magne Mogstad and Matthew Wiswall},
title = {Beyond LATE with a Discrete Instrument},
journal = {Journal of Political Economy},
volume = {125},
number = {4},
pages = {985-1039},
year = {2017}
}

@ARTICLE{CarneiroLee2009,
title = {Estimating distributions of potential outcomes using local instrumental variables with an application to changes in college enrollment and wage inequality},
author = {Carneiro, Pedro and Lee, Sokbae (Simon)},
year = {2009},
journal = {Journal of Econometrics},
volume = {149},
number = {2},
pages = {191-208}
}

@article{carneiro2010evaluating,
  title={Evaluating marginal policy changes and the average effect of treatment for individuals at the margin},
  author={Carneiro, Pedro and Heckman, James J and Vytlacil, Edward},
  journal={Econometrica},
  volume={78},
  number={1},
  pages={377--394},
  year={2010},
  publisher={Wiley Online Library}
}

@article{hsu2019testing,
  title={Testing generalized regression monotonicity},
  author={Hsu, Yu-Chin and Liu, Chu-An and Shi, Xiaoxia},
  journal={Econometric Theory},
  volume={35},
  number={6},
  pages={1146--1200},
  year={2019},
  publisher={Cambridge University Press}
}

@article{aizer2015juvenile,
  title={Juvenile incarceration, human capital, and future crime: Evidence from randomly assigned judges},
  author={Aizer, Anna and Doyle Jr, Joseph J},
  journal={The Quarterly Journal of Economics},
  volume={130},
  number={2},
  pages={759--803},
  year={2015},
  publisher={MIT Press}
}

@article{doyle2015measuring,
  title={Measuring returns to hospital care: Evidence from ambulance referral patterns},
  author={Doyle Jr, Joseph J and Graves, John A and Gruber, Jonathan and Kleiner, Samuel A},
  journal={Journal of Political Economy},
  volume={123},
  number={1},
  pages={170--214},
  year={2015},
  publisher={University of Chicago Press Chicago, IL}
}

@article{farre2020patent,
  title={What is a patent worth? Evidence from the US patent “lottery”},
  author={Farre-Mensa, Joan and Hegde, Deepak and Ljungqvist, Alexander},
  journal={The Journal of Finance},
  volume={75},
  number={2},
  pages={639--682},
  year={2020},
  publisher={Wiley Online Library}
}

@article{dobbie2017consumer,
  title={Consumer bankruptcy and financial health},
  author={Dobbie, Will and Goldsmith-Pinkham, Paul and Yang, Crystal S},
  journal={Review of Economics and Statistics},
  volume={99},
  number={5},
  pages={853--869},
  year={2017},
  publisher={MIT Press One Rogers Street, Cambridge, MA 02142-1209, USA journals-info~…}
}

@article{gross2022temporary,
  title={Temporary stays and persistent gains: The causal effects of foster care},
  author={Gross, Max and Baron, E Jason},
  journal={American Economic Journal: Applied Economics},
  volume={14},
  number={2},
  pages={170--199},
  year={2022},
  publisher={American Economic Association 2014 Broadway, Suite 305, Nashville, TN 37203-2425}
}

@book{cunningham2021causal,
  title={Causal inference: The mixtape},
  author={Cunningham, Scott},
  year={2021},
  publisher={Yale university press}
}

@article{di2013criminal,
  title={Criminal recidivism after prison and electronic monitoring},
  author={Di Tella, Rafael and Schargrodsky, Ernesto},
  journal={Journal of Political Economy},
  volume={121},
  number={1},
  pages={28--73},
  year={2013},
  publisher={University of Chicago Press Chicago, IL}
}

@article{mueller2015criminal,
  title={The criminal and labor market impacts of incarceration},
  author={Mueller-Smith, Michael},
  journal={Unpublished Working Paper},
  volume={18},
  year={2015}
}

@techreport{dobbie2018intergenerational,
  title={The intergenerational effects of parental incarceration},
  author={Dobbie, Will and Gr{\"o}nqvist, Hans and Niknami, Susan and Palme, M{\aa}rten and Priks, Mikael},
  year={2018},
  institution={National Bureau of Economic Research}
}

@article{kling2006incarceration,
  title={Incarceration length, employment, and earnings},
  author={Kling, Jeffrey R},
  journal={American Economic Review},
  volume={96},
  number={3},
  pages={863--876},
  year={2006}
}

@article{mogstad2019identification,
  title={Identification of causal effects with multiple instruments: Problems and some solutions},
  author={Mogstad, Magne and Torgovitsky, Alexander and Walters, Christopher},
  journal={NBER Working Paper},
  number={w25691},
  year={2019}
}

@article{norris2021effects,
  title={The effects of parental and sibling incarceration: Evidence from ohio},
  author={Norris, Samuel and Pecenco, Matthew and Weaver, Jeffrey},
  journal={American Economic Review},
  volume={111},
  number={9},
  pages={2926--2963},
  year={2021},
  publisher={American Economic Association 2014 Broadway, Suite 305, Nashville, TN 37203}
}

@techreport{bhuller2018incarceration,
  title={Incarceration spillovers in criminal and family networks},
  author={Bhuller, Manudeep and Dahl, Gordon B and L{\o}ken, Katrine V and Mogstad, Magne},
  year={2018},
  institution={National Bureau of Economic Research}
}

@article{frandsen2023judging,
  title={Judging Judge Fixed Effects},
  author={Frandsen, Brigham and Lefgren, Lars and Leslie, Emily},
  journal={American Economic Review},
  volume={113},
  number={1},
  pages={253--77},
  year={2023}
}

@article{canay2024use,
  title={On the use of outcome tests for detecting bias in decision making},
  author={Canay, Ivan A and Mogstad, Magne and Mountjoy, Jack},
  journal={Review of Economic Studies},
  volume={91},
  number={4},
  pages={2135--2167},
  year={2024},
  publisher={Oxford University Press UK}
}

@article{johnson2014judges,
  title={Judges on trial: A reexamination of judicial race and gender effects across modes of conviction},
  author={Johnson, Brian D},
  journal={Criminal Justice Policy Review},
  volume={25},
  number={2},
  pages={159--184},
  year={2014},
  publisher={Sage Publications Sage CA: Los Angeles, CA}
}

@article{chan2022selection,
  title={Selection with variation in diagnostic skill: Evidence from radiologists},
  author={Chan, David C and Gentzkow, Matthew and Yu, Chuan},
  journal={The Quarterly Journal of Economics},
  volume={137},
  number={2},
  pages={729--783},
  year={2022},
  publisher={Oxford University Press}
}

@article{stevenson2018distortion,
  title={Distortion of justice: How the inability to pay bail affects case outcomes},
  author={Stevenson, Megan T},
  journal={The Journal of Law, Economics, and Organization},
  volume={34},
  number={4},
  pages={511--542},
  year={2018},
  publisher={Oxford University Press}
}

@article{abrams2012judges,
  title={Do judges vary in their treatment of race?},
  author={Abrams, David S and Bertrand, Marianne and Mullainathan, Sendhil},
  journal={The Journal of Legal Studies},
  volume={41},
  number={2},
  pages={347--383},
  year={2012},
  publisher={University of Chicago Press Chicago, IL}
}

@article{hsu2021inference,
  title={Inference for ROC curves based on estimated predictive indices},
  author={Hsu, Yu-Chin and Lieli, Robert P},
  journal={arXiv preprint arXiv:2112.01772},
  year={2021}
}

@article{carr2021testing,
  title={Testing instrument validity with covariates},
  author={Carr, Thomas and Kitagawa, Toru},
  journal={arXiv preprint arXiv:2112.08092},
  year={2021}
}

@article{hsu2017consistent,
  title={Consistent tests for conditional treatment effects},
  author={Hsu, Yu-Chin},
  journal={The econometrics journal},
  volume={20},
  number={1},
  pages={1--22},
  year={2017},
  publisher={Oxford University Press Oxford, UK}
}

@techreport{kowalski2016doing,
  title={Doing more when you're running late: Applying marginal treatment effect methods to examine treatment effect heterogeneity in experiments},
  author={Kowalski, Amanda E},
  year={2016},
  institution={National Bureau of Economic Research}
}

@article{carneiro2011estimating,
  title={Estimating marginal returns to education},
  author={Carneiro, Pedro and Heckman, James J and Vytlacil, Edward J},
  journal={American Economic Review},
  volume={101},
  number={6},
  pages={2754--2781},
  year={2011},
  publisher={American Economic Association}
}

@article{sun2023instrument,
  title={Instrument validity for heterogeneous causal effects},
  author={Sun, Zhenting},
  journal={Journal of Econometrics},
  volume={237},
  number={2},
  pages={105523},
  year={2023},
  publisher={Elsevier}
}

@article{jochmans2023many,
  title={Many (Weak) Judges in Judge-Leniency Designs},
  author={Jochmans, Koen},
  year={2023},
  publisher={TSE Working Paper}
}

@article{sithole2024locally,
  title={A Locally Robust Semiparametric Approach to Examiner IV Designs},
  author={Sithole, Lonjezo},
  journal={arXiv preprint arXiv:2404.19144},
  year={2024}
}

@article{ren2024extrapolating,
  title={Extrapolating LATE with Weak IVs},
  author={Ren, Muyang},
  journal={working paper},
  year={2024}
}

@article{yap2024inference,
  title={Inference with Many Weak Instruments and Heterogeneity},
  author={Yap, Luther},
  journal={arXiv preprint arXiv:2408.11193},
  year={2024}
}

@article{mourifie2025layered,
  title={Layered policy analysis in program evaluation using the marginal treatment effect},
  author={Mourifi{\'e}, Ismael and Wan, Yuanyuan},
  journal={Journal of Econometrics},
  volume={251},
  pages={106060},
  year={2025},
  publisher={Elsevier}
}

@article{li2024discordant,
  title={Discordant relaxations of misspecified models},
  author={Li, Lixiong and K{\'e}dagni, D{\'e}sir{\'e} and Mourifi{\'e}, Isma{\"e}l},
  journal={Quantitative Economics},
  volume={15},
  number={2},
  pages={331--379},
  year={2024},
  publisher={Wiley Online Library}
}
\end{document}